\documentclass[english]{article}
\usepackage{graphicx}
\usepackage{amssymb,amsmath}
\usepackage[hypcap,font=footnotesize,labelfont=bf,singlelinecheck=false]{caption}
\usepackage{hyperref}
\usepackage{cleveref}
\usepackage{url}
\usepackage{fancyhdr}
\usepackage{bbm}
\usepackage[numbers]{natbib}
\makeatletter
\def\url@leostyle{%
  \@ifundefined{selectfont}{\def\UrlFont{\sf}}{\def\UrlFont{\small\ttfamily}}}
\makeatother
\crefformat{subsection}{\S#2#1#3}
\crefmultiformat{subsection}{\S\S#2#1#3}{and~#2#1#3}{, #2#1#3}{, and~#2#1#3}

\begin{document}

\title{Efficient valuation method for the SABR model%
\footnote{The views represented herein are the author's own views and do not necessarily 
represent the views of Morgan Stanley or its affiliates, 
and are not a product of Morgan Stanley Research}%
}

\author{Hyukjae Park\footnote{Morgan Stanley, E-mail: Hyukjae.Park@morganstanley.com}}

\date{November 8th, 2013%
\footnote{The original version of the paper was published on August 1st, 20013. In this version, we have added more numerical test results.}%
}
\maketitle

\begin{abstract}
In this article, we show how the scaling symmetry of the SABR model can be utilized
to efficiently price European options. 
For special kinds of payoffs, the complexity of the problem is reduced by one dimension. 
For more generic payoffs, instead of solving the $1+2$ dimensional SABR PDE, 
it is sufficient to solve $N_V$ uncoupled $1+1$ dimensional PDE's,
where $N_V$ is the number of points used to discretize one dimension.
Furthermore, the symmetry argument enables us to obtain prices of multiple options, 
whose payoffs are related to each other by convolutions, by valuing one of them.
The results of the method are compared with the Monte Carlo simulation.
\end{abstract}

\section{SABR Model}

The SABR model\cite{hagan:02,hagan:05,antonov:12} is one of the most commonly used models to price European
swaptions. It is a two factor model with stochastic volatility and
is described by a pair of coupled SDE's:
\begin{equation}
\begin{aligned}
\mathrm{d}F_t &= \alpha_t F_t^\beta \, \mathrm{d} B_{1,t}\\
\mathrm{d}\alpha_t &= \nu \alpha_t \, \mathrm{d} B_{2,t}
\end{aligned}
\end{equation}
where
\begin{itemize}
\item $F_t$ is the forward swap rate
\item $\alpha_t$ is the stochastic volatility 
\item $B_{1,t}$ and $B_{2,t}$ are the Brownian motions with correlation
$\rho$
\item $0 \le \beta \le 1$
\end{itemize}
The above equations are taken from the forward swap annuity measure%
\footnote{Throughout the article, we will only work in this measure and
all payoffs and valuations are expressed in the unit of the measure's num\'eraire, namely, 
the forward swap annuity.},
where $F_t$ is a martingale. When $\beta < 1$,
$F_t$ can reach $0$ with non-zero probability. Once it hits $0$,
$F_t$ must stay there to maintain its martingale property. 
It has the absorbing boundary condition at $\{F_t = 0\}$.
$\alpha_t$ is the log-normal process and can be easily solved:
\begin{equation}
\alpha_t= \alpha_0 \exp\left[\nu B_{2,t}-\frac{1}{2}\nu^2 t\right]
\end{equation}

In this article, we will only consider European options whose payoffs
at the maturity are functions of $F_t$ and $\alpha_t$. With this
restriction, the problems become Markov and the prices of the options
can be obtained by solving the Kolmogorov backward equation, we call
SABR PDE. This PDE is $1+2$ dimensional: 1 dimension for time and 
2 dimensions for Markovian state variables: $F_t$ and $\alpha_t$.
As mentioned earlier, it has the absorbing boundary condition at $\{F_t=0\}$.
Unfortunately, no-closed form solutions to this equation
are known except for some special cases. Instead, asymptotic
expansions in $\nu^2 t$ can be computed analytically
and they are commonly used to price European swaptions%
\cite{hagan:02, hagan:05, berestycki:04, henry-labordere:08, obloj:08, paulot:09,  antonov:12}.
Note that
these are asymptotic expansions. Hence, their convergence is not guaranteed.
One can easily see that, with any finite value of $\nu^2 t$, the
probability of $F_t$ hitting $0$ is non-zero, which cannot
be obtained from the series expansion in $\nu^2 t$ around $0$.

When $\rho=0$, there are semi-analytic solutions to the PDE\cite{islah:09, antonov:12}. Conditioned
on a path of $\alpha_t$, $F_t$ becomes a time-changed CEV process.
Closed-form solutions to the CEV PDE are known since a simple change
of variable would transform this PDE into the CIR PDE\cite{antonov:12}. The solutions
will have dependency in the path of $\alpha_t$ only through the elapsed time, $\int^T_0 \alpha_t^2 \,\mathrm{d}t$.
The distribution of this integral is also known semi-analytically%
\cite{antonov:12,yor:92,linetsky:04,dufresne:89,dufresne:90,carmona:97,donati-martin:01}. 
Therefore, the unconditional solutions to the SABR PDE
will be given as integrals of the CEV solutions over this distribution.

\section{Symmetry Argument}
In this section, we will introduce the scaling symmetry of the SABR model and 
examine its consequences in pricing European options. Special attention will be paid to swaption cases.

\subsection{Scaling Symmetry}\label{sec:scalingSymmetry}

The SABR SDE's are invariant under the following scaling transformation,
\begin{equation}
\begin{aligned}
F_t &\rightarrow \lambda F_t\\
\alpha_t &\rightarrow \lambda^{1-\beta} \alpha_t
\end{aligned}\label{eq:symmetry}
\end{equation}
for any $\lambda >0$. Instead of $F_t$ and $\alpha_t$, it is convenient
to work with a different set of variables where one of the variables in the set is invariant under the scaling transformation.
This will make it easy to analyze scaling properties of solutions. 
We will consider 2 such sets. The first is given below:
\begin{equation}
\begin{aligned}
Z_t &= \frac{F_t^{1-\beta}}{ \alpha_t} \\
X_t &= \log F_t
\end{aligned}\label{eq:newVariablesZX}
\end{equation}
Under the scaling transformation, $Z_t$ is invariant while $X_t \rightarrow X_t + \eta$
where $\eta = \log\lambda$. 
Note that when $\beta < 1$, the change of variables from $F_t$ and $\alpha_t$ to $Z_t$ and $X_t$ becomes singular 
at locus $\{F_t = 0\}$.
This means that not all solutions of the SABR PDE can be expressed as a function of $Z_t$ and $X_t$.
For most of our applications, payoff functions are independent of $\alpha_T$ when $F_T = 0$. 
This, together with the absorbing boundary condition at $\{F_t = 0 \}$ guarantees that
solutions are also independent of $\alpha_t$ when $F_t = 0$. In such cases, variables $Z_t$ and $X_t$ can be
used to express solutions.

The second set of variables we are going to use is
\begin{equation}
\begin{aligned}
W_t &= F_t \alpha_t^{-\frac{1}{1-\beta}}\\
Y_t & = \log\alpha_t
\end{aligned}\label{eq:newVariablesWY}
\end{equation}
$W_t$ is invariant under the scaling and $Y_t \rightarrow Y_t + (1-\beta)\eta$.
For a fixed set of model parameters, 
these variables are good to use for all values of $F_t$ and $\alpha_t$, and, unlike $Z_t$ and $X_t$, there will
be no restriction on kinds of payoff functions for which these variables can be utilized.
However, the change of variables becomes singular as $\beta$ approaches to $1$. 
This causes large numerical errors in solutions when $\beta$ is too close to $1$.

To deal with these two sets of variables uniformly, we will use following notation. 
$U_t$ and $V_t$ will denote the variables. $U_t$ will be invariant and
$V_t \rightarrow V_t + c \eta$ under the scaling transformation.
For the first set of variables, $U_t = Z_t$, $V_t = X_t$ and $c = 1$. For the second set, $U_t = W_t$, $V_t = Y_t$, and $c = (1-\beta)$.

\subsection{Special Payoff}\label{sec:specialPayoff}

The scaling symmetry can be utilized to reduce complexity of solving the SABR
PDE. To see this, let's consider an European option whose payoff at
maturity $T$ is $f(U_T)\exp(k V_T)$ for some function 
$f$\footnote{$F_T^m \alpha_T^n$ is an example of such payoffs.}.
The value of this option at time $t$
is the expected value of the payoff conditioned on $\mathfrak{F}_{t}$,
a filtration generated by all information available by time $t$:
\begin{equation}
P_f(U_t,V_t,t,T)=\mathbb{E}\left[f(U_T)\exp(k V_T)\big|\mathfrak{F}_t\right]
\end{equation}
Here, we have used the fact that the problem at hand is Markovian
and the solution has dependency on $\mathfrak{F}_t$ only through $U_t$
and $V_t$. Using the symmetry, one can show
\begin{equation}
P_f(U_t,V_t,t,T)=\exp(k c \eta)P_f(U_t,V_t-c \eta,t,T)
\end{equation}
for any $\eta$. In particular, one can choose $\eta=\frac{V_t}{c}$ and conclude
\begin{equation}
\begin{aligned}
P_f(U_t,V_t,t,T) &= \exp(k V_t)P_f(U_t,0,t,T)\\
&= \exp\!\left(k V_t\right) p(U_t,t)
\end{aligned}
\end{equation}
for some function $p$. Note that $P_f(U_t,V_t,t,T)$ is a solution
to the SABR PDE, which is $1+2$ 
dimensional\footnote{One dimension for time and other 2 dimensions for Markovian state
variables, $U_t$ and $V_t$.}.
By applying it to the SABR PDE, we obtain another
PDE that $p(U_t,t)$ satisfies:
\begin{equation}\label{eq:pdeRealPower}
\left(
\frac{\partial}{\partial t} 
+ \frac{1}{2}\sigma_{UU}^2\frac{\partial^2}{\partial u^2}
+ \left( \mu_U+ \sigma_{UV}^2 k \right) \frac{\partial}{\partial u}
+ \frac{1}{2} \sigma_{VV}^2 k^2 + \mu_V k
\right) p(u,t) = 0 
\end{equation}
where $\sigma_{UU}^2, \sigma_{UV}^2, \sigma_{VV}^2, \mu_U$, and $\mu_V$ are functions of $u$ only
and their
functional forms depend on
the choice of variables $U_t$ and $V_t$.

For variables $Z_t$ and $X_t$, the corresponding functions are
\begin{equation}
\begin{aligned}
\sigma_{ZZ}^2 &= (1-\beta)^2 - 2 \rho \nu (1-\beta) z + \nu^2 z^2\\
\sigma_{ZX}^2 &= \frac{1-\beta}{z}- \rho \nu\\
\sigma_{XX}^2 &= \frac{1}{z^2}\\
\mu_Z &= \nu^2 z -  \rho \nu ( 1-\beta)- \frac{\beta(1-\beta)}{2z}\\
\mu_X &= -\frac{1}{2}\sigma_{XX}^2\\
\end{aligned}
\end{equation}
Note that the above PDE%
\footnote{After completion of this research, the author became aware that a
similar PDE for payoff $F_T^m$ has been derived in \cite{islah:11}. Instead of the symmetry argument used here, 
\citeauthor{islah:11} noted that the SDE for $Z_t$
is uncoupled from $X_t$ and used the measure change to derive the PDE.} 
has a singularity at $z = 0$. This is due to the singularity of the change of variables 
we discussed in \cref{sec:scalingSymmetry}.
For this PDE to work, $f(z)$ should go to $0$ fast enough as $z\rightarrow 0$.

For variables $W_t$ and $Y_t$, 
\begin{equation}
\begin{aligned}
\sigma_{WW}^2 &= w^{2\beta} - 2 \frac{\rho \nu }{1-\beta} w^{1+\beta} + \left(\frac{\nu}{1-\beta}\right)^2 w^2\\
\sigma_{WY}^2 &= \rho w^\beta - \frac{\nu}{1-\beta}w \\
\sigma_{YY}^2 &= 1\\
\mu_W &= \frac{\nu^2(2-\beta)}{2(1-\beta)^2}w - \frac{\rho\nu}{1-\beta} w^\beta\\
\mu_Y &= -\frac{1}{2}\sigma_{YY}^2\\
\end{aligned}
\end{equation}

The PDE in \cref{eq:pdeRealPower} is $1+1$ dimensional
since $p(U_t,t)$ does not depend on $V_t$. $p(U_t,t)$ can be obtained by solving this
PDE with terminal condition $p(U_T, T) = f(U_T)$. Hence, using the
symmetry, we have reduced the complexity of the problem by one dimension.

\subsection{Generic Payoff}\label{sec:genericPayoff}

The payoffs we have considered so far are rather limited. For more
generic payoffs, the benefit of the symmetry is much more subtle.
Consider a generic payoff function $f(U_T,V_T)$. Using the Fourier transform along
$V_T$ direction, we can decompose it as follows:
\begin{equation}
f(U_T,V_T)=\int_{-\infty}^\infty a_k(U_T) \exp(i k V_T) \, \mathrm{d}k
\end{equation}
where 
\begin{equation}
a_k(U_T)=\int_{-\infty}^\infty f(U_T,v) \exp(-i k v ) \,\frac{\mathrm{d}v}{2\pi}
\end{equation}
The value of the option at time $t$, $P_f(U_t,V_t,t,T)$, is
given by the expected value of the payoff. With sufficiently regular
$f$, we can interchange the integration and the expectation:
\begin{equation}
\begin{aligned}
P_f(U_t,V_t,t,T) &= \mathbb{E}\left[f(U_T,V_T) \big| \mathfrak{F}_t\right]\\
&= \int_{-\infty}^\infty\mathbb{E}\left[ a_k(U_T)\exp(i k V_T) \big| \mathfrak{F}_t	\right]\mathrm{d}k
\end{aligned}
\end{equation}
Now, the same symmetry argument in \cref{sec:specialPayoff} applies and we can 
use this to separate $V_t$ dependency:
\begin{equation}
P_f(U_t,V_t,t,T)=\int_{-\infty}^\infty \exp(i k V_t)p_k(U_t,t)\,\mathrm{d}k
\end{equation}
for some functions $p_k(U_t,t)$'s. As before, $p_k(U_t,t)$'s
are solutions of $1+1$ dimensional PDE's that are all uncoupled from
each other:
\begin{equation}\label{eq:pdeImaginaryPower}
\left(
\frac{\partial}{\partial t} 
+ \frac{1}{2}\sigma_{UU}^2\frac{\partial^2}{\partial u^2}
+ \left( \mu_U+ i\sigma_{UV}^2 k \right) \frac{\partial}{\partial u}
- \frac{1}{2} \sigma_{VV}^2 k^2 + i\mu_V k
\right) p_k(u,t) = 0 
\end{equation}
These PDE's are obtained by replacing $k$ in \cref{eq:pdeRealPower} with $ik$.
$p_k(U_t,t)$'s can be obtained by solving the PDE's with terminal
condition $p_k(U_T,T) = a_k(U_T)$.

In practice, we cannot solve these PDE's analytically. Instead, we
discretize variables and resort to numerical techniques. For our discussion,
it is enough to discretize along the $V_t$ direction only. After
discretization, we impose the periodic boundary
condition in the $V_t$ direction on the payoff function and decompose it into a Fourier series: 
\begin{equation}
f(U_T,V_T)=\sum_k a_k(U_T)\exp(i k V_T)
\end{equation}
with
\begin{equation}
a_k(U_T)=\frac{1}{N_V}\sum_v f(U_T,v)\exp(-i k v)\label{eq:payoffDecomposition}
\end{equation}
where $v$ takes a value from the discretized grid for $V_t$, $k$ is from the
dual grid\footnote{The dual grid is defined as 
$\left\{ k\big| k v = 2\pi n 
\textnormal{ for some integer $n$ $\forall v$ in the grid}\right\} $}, and $N_V$ is 
the number of points in the grid.

For variables $Z_t$ and $X_t$, this step requires some clarification.
The Fourier series used here needs to be seen as an approximation of the original payoff function.
Inside the grid, it is a good approximation since it matches the value of the function
exactly for all points in the grid. 
However, it may not be such a good approximation outside of it.

When $\beta < 1$, locus $\{F_T = 0\}$ is of special concern. 
$F_T = 0$ implies $X_T = -\infty$, which is outside the grid, and it can be reached with non-zero
probability. To make sure that the Fourier series is still a good approximation for this case, we
slightly change the definition of $a_k(Z_T)$ when $Z_T = 0$:
\begin{equation}
a_k(0)=
\begin{cases}
0 &\textnormal{if } k \neq 0 \\
f(0,-\infty) &\textnormal{if } k = 0
\end{cases}
\end{equation}
As discussed in \cref{sec:scalingSymmetry}, we use variables $Z_t$ and $X_t$ only for payoffs that are
independent of $\alpha_T$ when $F_T = 0$. For such payoffs, 
$f(0,-\infty)$ is the value of the payoff function at $F_T = 0$.
Changing $a_k(0)$ can be understood as deforming the payoff function around $Z_T = 0$.
For example, the following deformation with very small $\epsilon > 0$ would produce the equivalent change in $a_k(0)$:
\begin{equation}
f(Z_T,X_T) \rightarrow \mathbbm{1}_{\{Z_T > \epsilon\}}f(Z_T,X_T) + \mathbbm{1}_{\{Z_T \le \epsilon\}}f(Z_T,X_T + \log\frac{Z_T}{\epsilon})
\end{equation}
The deformed payoff function will differ from the original payoff function when $Z_T$ is very small and $X_T$ is finite.
Since it implies $\alpha_T$ is very large and $F_T$ is finite, the probability of this happening is very small and it 
will not introduce much error in pricing.
Due to the singularity, $F_T = 0$ also implies $Z_T = 0$ and now, with the modified definition of $a_k(0)$, the Fourier
series matches the payoff function value when $F_T = 0$. 

When $X_T$ is large and outside the grid, the Fourier series will not be a good approximation.
For this, we need to make sure the grid is large enough so that the probability of reaching such $X_T$ is
very small.

The second set of variables, $W_t$ and $Y_t$, does not have similar issues.
$Y_t$ is a Normal process
and it cannot reach its boundaries at $+\infty$ and $-\infty$.
As long as the grid is sufficiently large, the Fourier series is a good approximation to
the original payoff function.
 
Using the same symmetry argument,
we conclude that the value of the option is given by
\begin{equation}
P_f(U_t,V_t,t,T)=\sum_k \exp(i k V_t) p_k(U_t,t)\label{eq:recombination}
\end{equation}
for some functions $p_k(U_t,t)$'s, where $k$ is taken from the
dual grid. $p_k(U_t,t)$'s are solutions of the same PDE's in \cref{eq:pdeImaginaryPower}
and can be obtained by solving them with terminal condition $p_k(U_T,T) = a_k(U_T)$.
With the symmetry argument, we have reduced the complexity
of the problem from solving an $1+2$ dimensional PDE to solving the $N_V$ uncoupled $1+1$
dimensional PDE's. Since they are uncoupled, they can be
solved parallelly, independent from each other. 

We can go further with
the symmetry argument. Once, $p_k(U_t,t)$'s are
computed, we can use them to price other options. Consider an option
whose payoff function is a convolutions of $f(U_T,V_T$) with
another function $g(V_T)$. For this payoff,
\begin{equation}
\begin{aligned}
(f \star g)(U_T,V_T) &= \sum_w f(U_T,V_T - w) g(w)\\
&= N_V \sum_k a_k(U_T) b_k \exp(i k V_T)
\end{aligned}
\end{equation}
with
\begin{equation}
b_k=\frac{1}{N_V}\sum_v g(v)\exp(-i k v)
\end{equation}
where $v$ and $w$ are taken from the grid and $k$ is from the dual
grid. Interchanging expectation and summation, one can show the option
value is given as follows
\begin{equation}
N_V\sum_k b_k \exp(i k V_t) p_k(U_t,t)
\end{equation}
With $p_k(U_t,t)$'s computed already, the option value can be obtained without
solving any more PDE's. By pricing one option, we can price a whole class of options 
whose payoff functions are related to the original payoff function by convolutions.

\subsection{Swaption Valuation}\label{sec:swaptionValuation}

The argument in \cref{sec:genericPayoff} can be applied to swaption valuation
and we can price swaptions with all strikes at once by solving the PDE's for one strike%
\footnote{One such choice is the at-the-money strike.}. To show this
explicitly, we consider the payer swaption with strike $K$:
\begin{equation}
C(F_t,\alpha_t,K,t,T)=\mathbb{E}\left[\left(F_T-K\right)^+ \big|\mathfrak{F}_t\right]
\end{equation}
Using the symmetry, one can show
\begin{equation}
C(F_t,\alpha_t,K,t,T) = \lambda C( \lambda^{-1} F_t, \lambda^{-(1-\beta)} \alpha_t, \lambda^{-1} K,t,T)
\end{equation}
for any $\lambda > 0$. We choose $\lambda = \frac{K}{K_0}$ for some fixed $K_0$ and
obtain
\begin{equation}
\begin{aligned}
C(F_t,\alpha_t,K,t,T) &= \frac{K}{K_0}C\!\left(\frac{K_0}{K} F_t, \left(\frac{K_0}{K}\right)^{1-\beta} \alpha_t, K_0,t,T\right)\\
&= \frac{K}{K_0}C_0\!\left(\frac{K_0}{K} F_t,\left(\frac{K_0}{K}\right)^{1-\beta} \alpha_t,t,T\right)
\end{aligned}\label{eq:swaptionSymmetry}
\end{equation}
where $C_0(F_t,\alpha_t,t,T)=\mathbb{E}\left[\left(F_T-K_0\right)^+ \big|\mathfrak{F}_t\right]$
is the value of the payer swaption with strike $K_0$. Therefore,
once we compute the value of this swaption, for example, following
the steps highlighted in \cref{sec:genericPayoff}, the swaption values
for all strikes can be obtained.

\section{Numerical Tests}\label{sec:numericalTests}
We numerically tested the swaption valuation method developed in the previous section.
We will call this method the ``PDE + Symmetry'' method.
We first solve the PDE's for the at-the-money swaption,
following the steps highlighted in \cref{sec:genericPayoff}. 
Then, we apply \cref{eq:swaptionSymmetry} to obtain swaption prices for different strikes.

The computational complexity of the method is following. To solve the PDE's, we discretize time and 
stochastic variables. We denote, by $N_t$, $N_U$, and $N_V$, the numbers of points used 
to discretize time, $U_t$, and $V_t$.
The decomposition of the payoff in \cref{eq:payoffDecomposition} and the recombination in \cref{eq:recombination}
can be broken into parallelizable independent $N_U$ computational 
units, each of which takes $\mathcal{O}(N_V\log N_V)$ operations. 
Usually, the number of operations in the these steps is negligible compared to
the number of operations in solving the $N_V$ PDE's. As noted before, the PDE's can be solved parallelly and 
each PDE takes $\mathcal{O}(N_t N_U)$ operations. 
All in all, the method takes $\mathcal{O}(N_t N_U N_V)$ operations to price swaptions with all strikes.

\subsection{Monte Carlo simulation}\label{sec:monteCarloSimulation}
To test our method, we compared pricing results with the Monte Carlo simulation. We used the Euler scheme to generate 
Monte Carlo paths and adjusted them for the absorbing boundary condition:
\begin{equation}
\begin{aligned}
\alpha_{t+\Delta t} &= \alpha_t \exp\left[ \nu \sqrt{\Delta t}Z_1 - \frac{1}{2}\nu^2\Delta t\right]\\
F^\prime_{t+\Delta t } &= F_t + \alpha_t F_t^\beta \sqrt{\Delta t}\left( \rho Z_1 + \sqrt{1-\rho^2} Z_2 \right)\\
F_{t+\Delta t } &= 
\begin{cases}
F^\prime_{t+\Delta t} & \textnormal{if } F^\prime_{t+\Delta t} > 0 \textnormal{ and }
U > \exp\left[-2 \frac{F_t F^\prime_{t+\Delta t}}{\alpha_t^2 F_t^{2\beta} \Delta t}\right]\\
0 & \textnormal{otherwise}
\end{cases}
\end{aligned}
\end{equation}
where $Z_1$ and $Z_2$ are standard normal random variables and $U$ is a uniform random variable. $Z_1$, $Z_2$ and $U$ are all
uncorrelated. Note that the above scheme consists of two steps. In the first step, we follow the standard Euler step to generate
values for $\alpha_t$ and $F_t^\prime$ at the next time step $t+\Delta t$. 
In the second step, we adjust $F^\prime_{t+\Delta t}$ for the absorbing boundary condition at $0$.
The uniform random variable $U$ is used to compensate the probability of hitting zero between $t$ and $t+\Delta t$ even if
$F^\prime_{t+\Delta t} > 0$. The approximate value of the probability 
\begin{equation}
\exp\left[-2 \frac{F_t F^\prime_{t+\Delta t}}{\alpha_t^2 F_t^{2\beta} \Delta t}\right]
\end{equation}
is obtained as follows\cite{glasserman:01}. Between time $t$ and $t+\Delta t$, $F_t$ is approximated with a normal process by keeping 
its drift and volatility constant to values given at time $t$. Then, we use the measure change to get rid of its drift and
use the reflection principle of the Brownian motion to compute the probability of hitting zero. 
For small value of $\beta$, this adjustment is not insignificant and sometimes changes swaption
values by as much as a few basis points. 

\subsection{Analytic Limit}
As a first test, we looked at a special limit of the SABR model where prices of swaptions are known analytically:
\begin{equation}
\textnormal{$\beta = 0$, $\nu = 0.0\%$, $\rho = 0.0\%$, $\alpha_0 = 1.0\%$, $F_0 = 5.0\%$, $T=5$}\label{eq:analyticParameters}
\end{equation}
Here, the parameter values are shown in their natural units. $\alpha_t$ is constant since $\nu=0.0\%$.
This, together with $\beta=0$, implies $F_t$ is a normal process with the absorbing boundary condition at $0$.
The analytic swaption price is obtained by applying the reflection principle of the Brownian motions:
\begin{equation}
\begin{split}
C(F_t,\alpha_t,K,t,T) &=\alpha_t\sqrt{T-t}\left(n(d_+)-n(d_-)\right)\\
&\qquad+F_t\left(N(d_+)+N(d_-)\right)-K\left(N(d_+)-N(d_-)\right)
\end{split}\label{eq:analyticFormula}
\end{equation}
where $d_+ = \frac{F_t-K}{\alpha_t\sqrt{T-t}}$ and $d_- = \frac{-F_t-K}{\alpha_t\sqrt{T-t}}$. 

\begin{figure}[htbp]
\centering
\includegraphics[width=1\columnwidth]{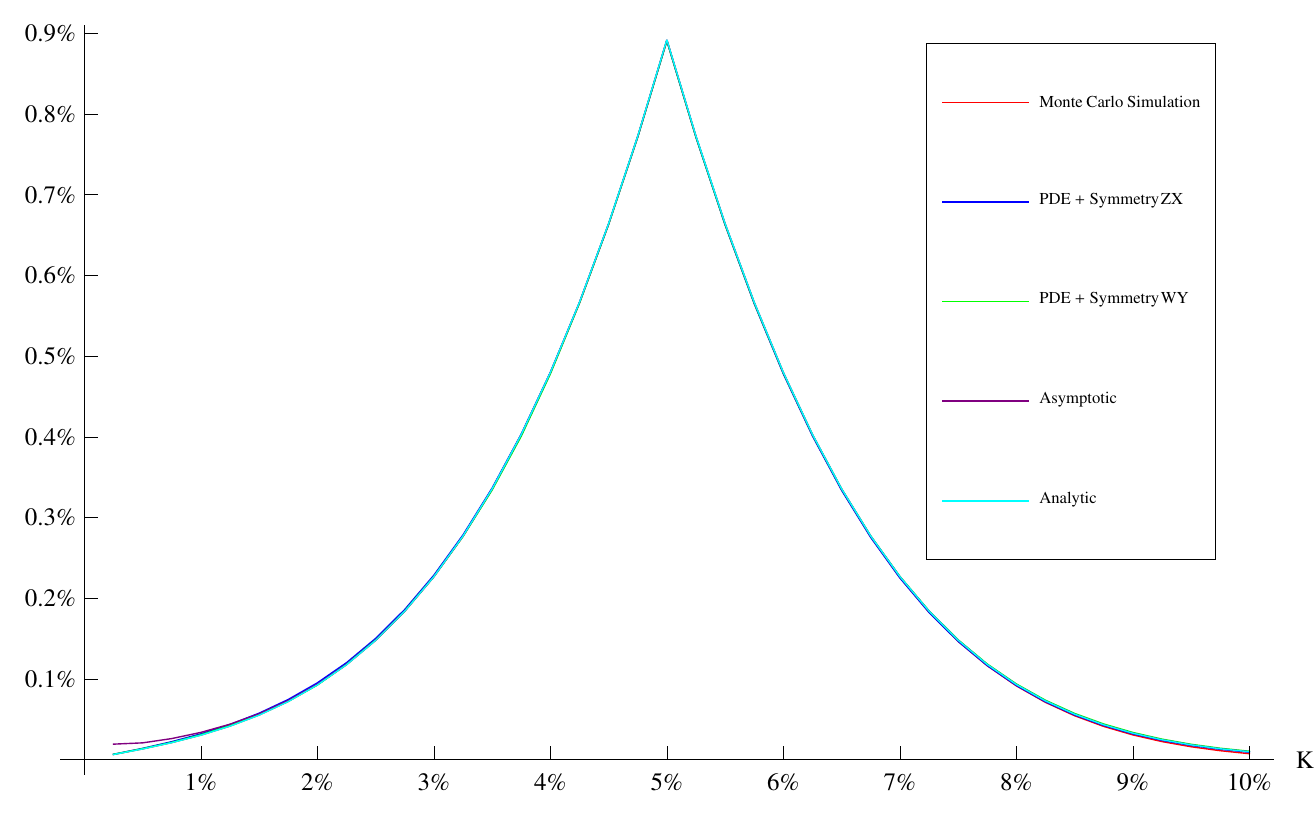}
\caption{\label{fig:analyticComparison}Swaption prices at the analytic limit. The time values of swaptions expressed in the unit of 
forward swap annuity were computed using different valuation methods and graphed as a function of $K$.
The SABR model parameters in \cref{eq:analyticParameters} were used.}
\end{figure}

The time values of swaptions in the unit of forward swap annuity: 
\begin{equation}
\mathbb{E}\left[\left(F_T-K\right)^+ \big|\mathfrak{F}_t\right]-\left(F_t-K\right)^+
\end{equation}
were computed using the following methods: 
were computed using the following methods: 
\begin{itemize}
\item Monte Carlo simulation: Monte Carlo method developed in \cref{sec:monteCarloSimulation}
\item PDE + Symmetry ZX: ``PDE + Symmetry'' method with the choice of variables $Z_t$ and $X_t$
\item PDE + Symmetry WY: ``PDE + Symmetry'' method with the choice of variables $W_t$ and $Y_t$
\item Asymptotic: The first order asymptotic series in \cite{obloj:08}
\item Analytic: analytic solution in \cref{eq:analyticFormula}
\end{itemize}
The results are shown 
in \cref{fig:analyticComparison} and \cref{tab:analyticComparison}.
The asymptotic series shows small difference from the other methods at very low strikes. 
All other methods produced virtually identical prices.
\begin{table}[htbp]
\centering
\caption{\label{tab:analyticComparison}Swaption prices at the analytic limit. The time values of swaptions expressed in the unit of 
forward swap annuity were computed using different valuation methods.
The SABR model parameters in \cref{eq:analyticParameters} were used.}
\resizebox{\columnwidth}{!}{%
\begin{tabular}{|c|c|c|c|c|c|}
\hline 
Strike & Monte Carlo simulation & PDE + Symmetry ZX & PDE + Symmetry WY & Asymptotic & Analytic\\
\hline
0.50\% & 0.01\% & 0.01\% & 0.01\%  & 0.02\%  &  0.01\% \\
1.00\% & 0.03\% & 0.03\% & 0.03\%  & 0.03\%  &  0.03\% \\
1.50\% & 0.06\% & 0.06\% & 0.06\%  & 0.06\%  &  0.06\% \\
2.00\% & 0.09\% & 0.10\% & 0.09\%  & 0.09\%  &  0.09\% \\
2.50\% & 0.15\% & 0.15\% & 0.15\%  & 0.15\%  &  0.15\% \\
3.00\% & 0.23\% & 0.23\% & 0.23\%  & 0.23\%  &  0.23\% \\
3.50\% & 0.34\% & 0.34\% & 0.33\%  & 0.34\%  &  0.34\% \\
4.00\% & 0.48\% & 0.48\% & 0.48\%  & 0.48\%  &  0.48\% \\
4.50\% & 0.66\% & 0.66\% & 0.66\%  & 0.66\%  &  0.66\% \\
5.00\% & 0.89\% & 0.89\% & 0.89\%  & 0.89\%  &  0.89\% \\
5.50\% & 0.66\% & 0.66\% & 0.66\%  & 0.66\%  &  0.66\% \\
6.00\% & 0.48\% & 0.48\% & 0.48\%  & 0.48\%  &  0.48\% \\
6.50\% & 0.33\% & 0.33\% & 0.34\%  & 0.34\%  &  0.34\% \\
7.00\% & 0.23\% & 0.23\% & 0.23\%  & 0.23\%  &  0.23\% \\
7.50\% & 0.15\% & 0.15\% & 0.15\%  & 0.15\%  &  0.15\% \\
8.00\% & 0.09\% & 0.09\% & 0.09\%  & 0.09\%  &  0.09\% \\
8.50\% & 0.05\% & 0.06\% & 0.06\%  & 0.06\%  &  0.06\% \\
9.00\% & 0.03\% & 0.03\% & 0.03\%  & 0.03\%  &  0.03\% \\
9.50\% & 0.02\% & 0.02\% & 0.02\%  & 0.02\%  &  0.02\% \\
10.00\% & 0.01\% & 0.01\% & 0.01\%  & 0.01\%  &  0.01\% \\
\hline
\end{tabular}%
}
\end{table}
\clearpage
\subsection{USD Swaption}
We compared the ``PDE + Symmetry'' method with the Monte Carlo simulation in actual
USD swaption valuation. 
We chose the USD swaption market on the following dates to represent various market conditions:
\begin{itemize}
\item October 9th, 2007: The S\&P 500 index reached its highest before the Great Recession.
\item September 15th, 2008: The Lehman Brothers filed for the bankruptcy protection.
\item March 9th, 2009: The S\&P 500 index reached its lowest during the Great Recession.
\item October 29th 2013: Recent market
\end{itemize}
We manually calibrated the model for various expiry-tenor combinations on these dates.
As noted in \cite{hagan:02}, $\beta$ and $\rho$ affect swaption prices in similar ways and it is difficult to 
determine both by fitting the market prices. Hence, we chose the value of $\beta$ arbitrarily and calibrated 
the rest of the model parameters by fitting market prices. Once the calibration was done, we compared swaptions prices computed using
our ``PDE + Symmetry'' method to the Monte Carlo simulation results to test the accuracy of our method.

\subsubsection{October 9th, 2007}
\begin{figure}[htbp]
\centering
\includegraphics[width=1\columnwidth]{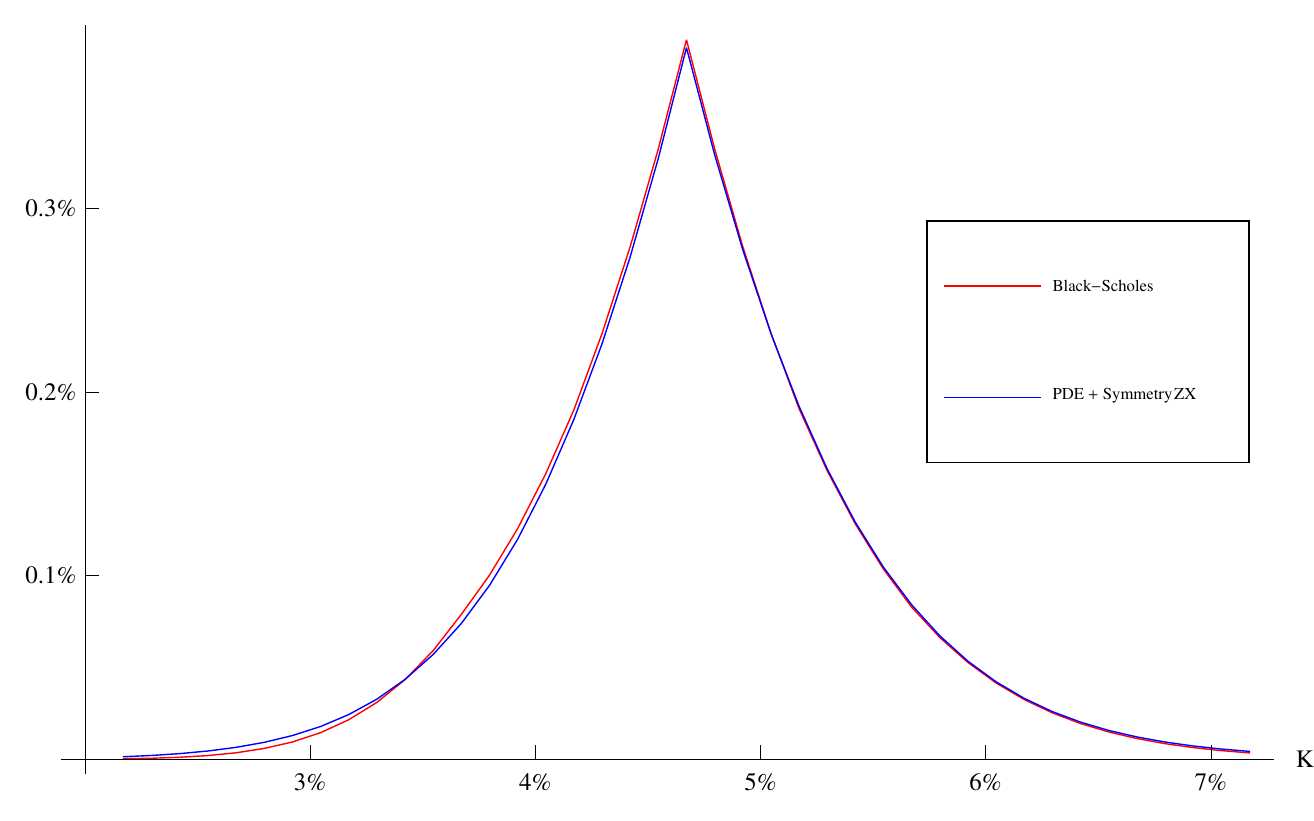}
\caption{\label{fig:USDCalibration.20071009.1Y1Y}Calibration results for USD 1Y1Y swaptions on October 9th, 2007. 
The time values of swaptions expressed in the unit of 
forward swap annuity were computed and graphed
as a function of $K$.
The red line was obtained by the Black-Scholes with linearly interpolated implied volatilities.
The blue line was computed using the ``PDE + Symmetry'' method with variables $Z_t$ and $X_t$. 
The SABR model parameters in \cref{eq:USDParameters.20071009.1Y1Y} were used.}  
\end{figure}

\Cref{tab:volatility.20071009.1Y1Y} shows the USD swaption
lognormal volatilities with 1 year expiry and 1 year tenor as of October 9th, 2007.
\begin{table}[htbp]
\centering
\caption{\label{tab:volatility.20071009.1Y1Y}USD Swaption lognormal volatilities with 1 year expiry and 1 year tenor as of October 9th, 2007.}
\resizebox{\columnwidth}{!}{%
\begin{tabular}{|*{10}{c|}}
\hline 
Strike(\%) & 2.67 & 3.67 & 4.17 & 4.42 & 4.67 & 4.92 & 5.17 & 5.67 & 6.67 \\
\hline 
Volatility(\%) & 24.70 & 23.70 & 22.20 & 21.63 & 21.06 & 20.50 & 19.97 & 19.39 & 18.69 \\
\hline
\end{tabular}%
}
\end{table}
The observed forward swap rate was 4.67\%. We chose the value of $\beta$ to be $0.90$ and manually
calibrated the model. For the calibration, the ``PDE + Symmetry'' method with variables $Z_t$ and $X_t$ was used.
We obtained the following values of parameters:
\begin{equation}
\textnormal{$\beta = 0.90$, $\nu = 30.0\%$, $\rho = -50.0\%$, $\alpha_0 = 15.50\%$, $F_0 = 4.67\%$, $T=1$}
\label{eq:USDParameters.20071009.1Y1Y}
\end{equation}
As before, the above parameters are in their natural units.
\Cref{fig:USDCalibration.20071009.1Y1Y} and \cref{tab:USDCalibration.20071009.1Y1Y} show the goodness of the calibration%
\footnote{The calibration may be improved by using a good optimizer, but for our purpose, it should be good enough.}%
.
\begin{table}[htbp]
\centering
\caption{\label{tab:USDCalibration.20071009.1Y1Y}Calibration results for USD 1Y1Y swaptions on October 9th, 2007. 
The time values of swaptions expressed in the unit of 
forward swap annuity were computed.
For Black-Scholes, the linearly interpolated implied volatilities were used.
For SABR, the ``PDE + Symmetry'' method with variables $Z_t$ and $X_t$ was used.
The SABR model parameters in \cref{eq:USDParameters.20071009.1Y1Y} were used.}
\begin{tabular}{|c|c|c|}
\hline 
Strike & Black-Scholes & SABR \\
\hline
2.67\% & 0.00\% & 0.01\%  \\
3.67\% & 0.08\% & 0.07\%  \\
4.17\% & 0.19\% & 0.19\%  \\
4.42\% & 0.28\% & 0.27\%  \\
4.67\% & 0.39\% & 0.39\%  \\
4.92\% & 0.28\% & 0.28\%  \\
5.17\% & 0.19\% & 0.19\%  \\
5.67\% & 0.08\% & 0.08\%  \\
6.67\% & 0.01\% & 0.01\%  \\
\hline
\end{tabular}
\end{table} 

\begin{figure}[htbp]
\centering
\includegraphics[width=1\columnwidth]{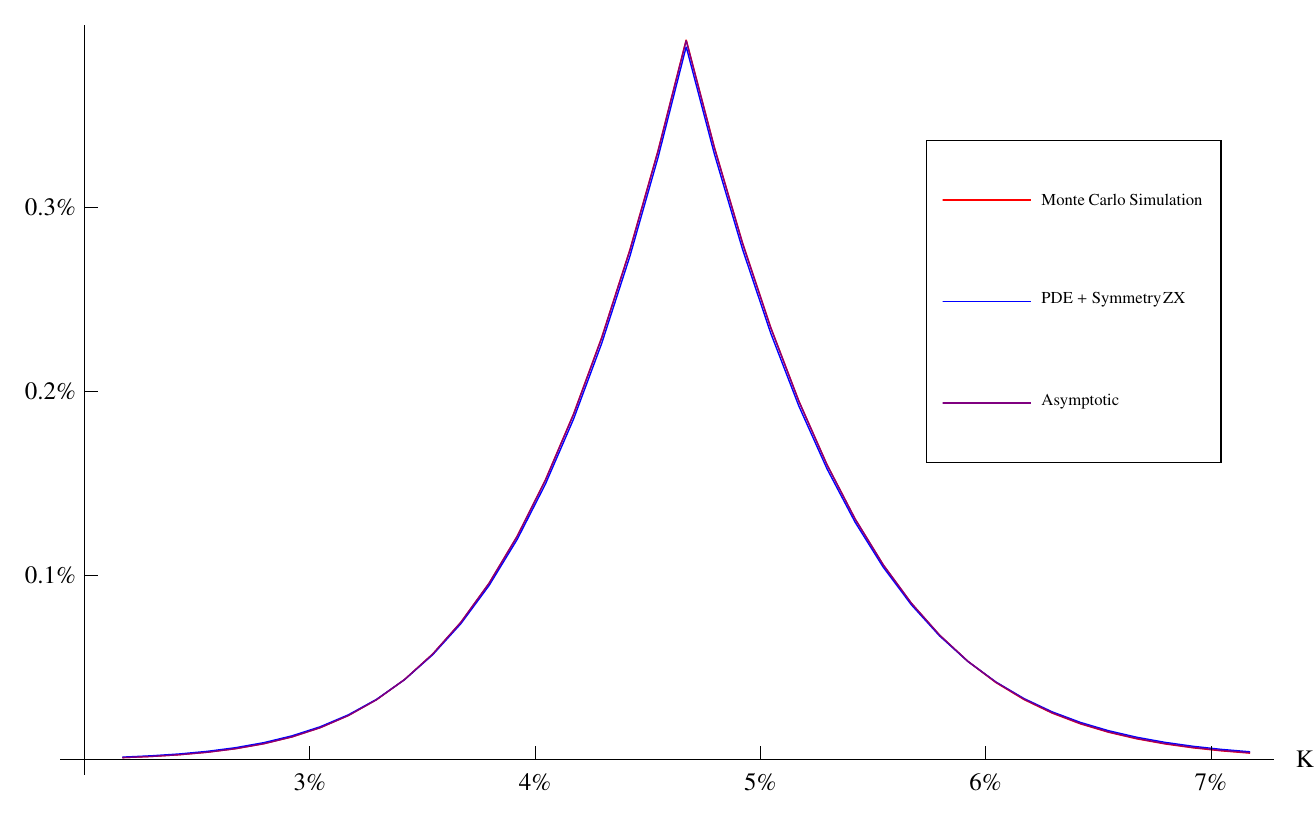}
\caption{\label{fig:comparison.20071009.1Y1Y}Pricing method comparison for USD 1Y1Y swaptions on October 9th, 2007. 
The time values of swaptions expressed in the unit of 
forward swap annuity were computed using various pricing methods.
The SABR model parameters in \cref{eq:USDParameters.20071009.1Y1Y} were used.}
\end{figure}

With the calibration done, we priced swaptions with various strikes using the Monte Carlo simulation, 
the ``PDE + Symmetry'' method with variables $Z_t$ and $X_t$, and the asymptotic formula.
The value of $\beta$ was too high for
$W_t$ and $Y_t$ variables, causing large numerical errors in swaption prices.
The results are shown in \cref{fig:comparison.20071009.1Y1Y} and \cref{tab:comparison.20071009.1Y1Y}.
The results from all three methods show very little difference. 
\begin{table}[htbp]
\centering
\caption{\label{tab:comparison.20071009.1Y1Y}Pricing method comparison for USD 1Y1Y swaptions on October 9th, 2007.
The time values of swaptions expressed in the unit of 
forward swap annuity were computed using various pricing methods. The SABR model parameters in \cref{eq:USDParameters.20071009.1Y1Y} were used.}
\begin{tabular}{|c|c|c|c|}
\hline 
Strike & Monte Carlo simulation  & PDE + Symmetry ZX & Asymptotic \\
\hline
2.17\% & 0.00\% & 0.00\%  & 0.00\% \\
2.67\% & 0.01\% & 0.01\%  & 0.01\% \\
3.17\% & 0.02\% & 0.02\%  & 0.02\% \\
3.67\% & 0.07\% & 0.07\%  & 0.07\% \\
4.17\% & 0.19\% & 0.19\%  & 0.19\% \\
4.42\% & 0.28\% & 0.27\%  & 0.28\% \\
4.67\% & 0.39\% & 0.39\%  & 0.39\% \\
4.92\% & 0.28\% & 0.28\%  & 0.28\% \\
5.17\% & 0.19\% & 0.19\%  & 0.19\% \\
5.67\% & 0.08\% & 0.08\%  & 0.08\% \\
6.17\% & 0.03\% & 0.03\%  & 0.03\% \\
6.67\% & 0.01\% & 0.01\%  & 0.01\% \\
7.17\% & 0.00\% & 0.00\%  & 0.00\% \\
\hline
\end{tabular}
\end{table}

\begin{figure}[htbp]
\centering
\includegraphics[width=1\columnwidth]{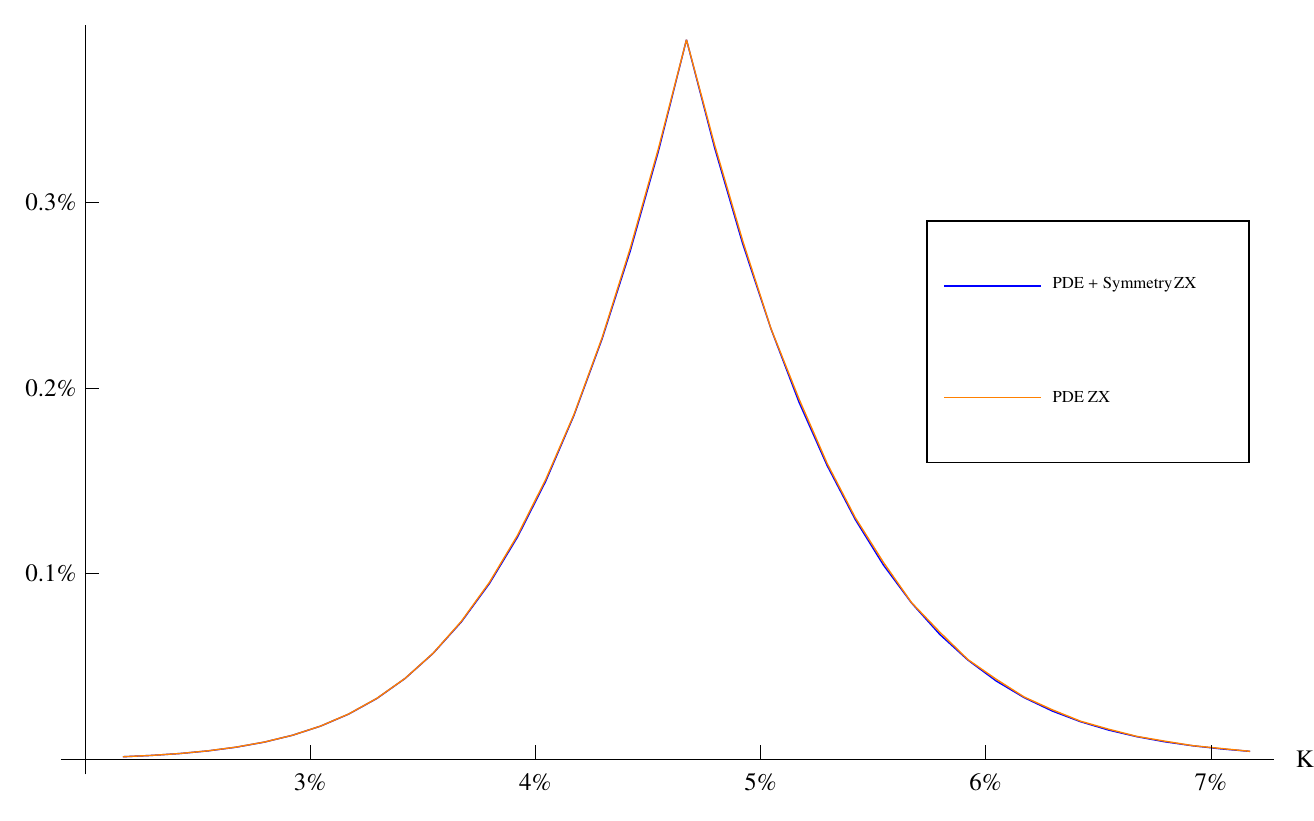}
\caption{\label{fig:pdeVSpdeSym.20071009.1Y1Y}Pricing method comparison for USD 1Y1Y swaptions on October 9th, 2007.
The time values of swaptions expressed in the unit of 
forward swap annuity were computed in the ``PDE + Symmetry'' method with variables $Z_t$ and $X_t$. 
For the blue line, the symmetry argument in \cref{sec:swaptionValuation} was
used to compute swaption prices for different strikes at once.
For the orange line, swaption prices were computed separately for each strike.
The SABR model parameters in \cref{eq:USDParameters.20071009.1Y1Y} were used.}
\end{figure}

To further test how well the symmetry is respected in the solutions of the PDE's, we computed the swaption prices 
by solving the PDE's for each strike separately and compared them with the results of the ``PDE + Symmetry'' method.
Note that numerically computed solutions do not necessarily show the symmetry since the discretization of 
the variables breaks it.    
\Cref{fig:pdeVSpdeSym.20071009.1Y1Y} and \cref{tab:pdeVSpdeSym.20071009.1Y1Y} show the comparison result.
These two methods produced identical results with no noticeable differences.
\begin{table}[htbp]
\centering
\caption{\label{tab:pdeVSpdeSym.20071009.1Y1Y}Pricing method comparison for USD 1Y1Y swaptions on October 9th, 2007.
The time values of swaptions expressed in the unit of 
forward swap annuity were computed in the ``PDE + Symmetry'' method with variables $Z_t$ and $X_t$.  
In the PDE ZX column, swaption prices were computed separately for each strike. 
The SABR model parameters in \cref{eq:USDParameters.20071009.1Y1Y} were used.}
\begin{tabular}{|c|c|c|}
\hline 
Strike & PDE + Symmetry ZX & PDE ZX\\
\hline
2.17\% & 0.00\%  & 0.00\% \\
2.67\% & 0.01\%  & 0.01\% \\
3.17\% & 0.02\%  & 0.02\% \\
3.67\% & 0.07\%  & 0.07\% \\
4.17\% & 0.19\%  & 0.19\% \\
4.42\% & 0.27\%  & 0.27\% \\
4.67\% & 0.39\%  & 0.39\% \\
4.92\% & 0.28\%  & 0.28\% \\
5.17\% & 0.19\%  & 0.19\% \\
5.67\% & 0.08\%  & 0.08\% \\
6.17\% & 0.03\%  & 0.03\% \\
6.67\% & 0.01\%  & 0.01\% \\
7.17\% & 0.00\%  & 0.00\% \\
\hline
\end{tabular}
\end{table}
\clearpage

\begin{figure}[htbp]
\centering
\includegraphics[width=1\columnwidth]{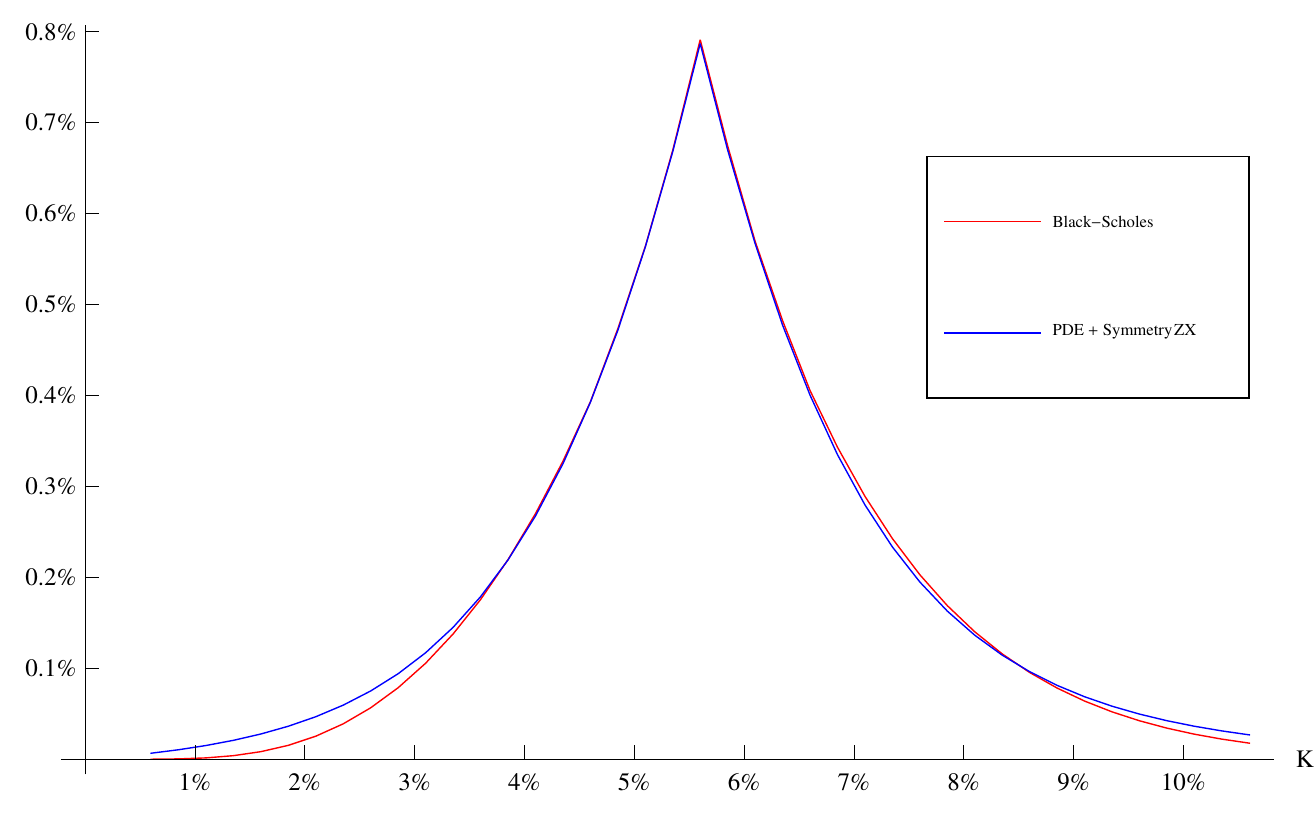}
\caption{\label{fig:USDCalibration.20071009.5Y5Y}Calibration results for USD 5Y5Y swaptions on October 9th, 2007.
The time values of swaptions expressed in the unit of 
forward swap annuity were computed and graphed
as a function of $K$. 
The red line was obtained by the Black-Scholes with linearly interpolated implied volatilities.
The blue line was computed using the ``PDE + Symmetry'' method with variables $Z_t$ and $X_t$. 
The SABR model parameters in \cref{eq:USDParameters.20071009.5Y5Y} were used.}  
\end{figure}

\begin{figure}[htbp]
\centering
\includegraphics[width=1\columnwidth]{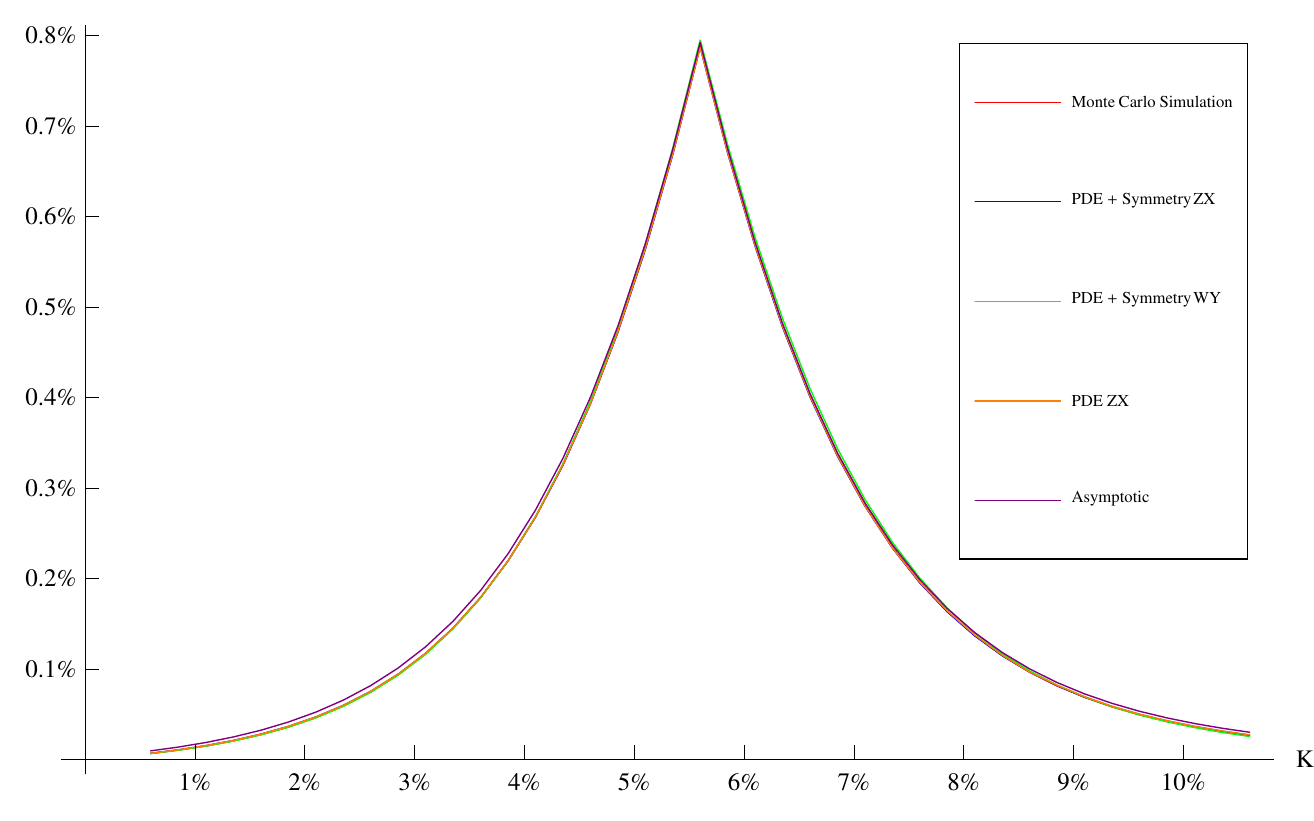}
\caption{\label{fig:comparison.20071009.5Y5Y}Pricing method comparison for USD 5Y5Y swaptions on October 9th, 2007. 
The time values of swaptions expressed in the unit of 
forward swap annuity were computed using various pricing methods.
The SABR model parameters in \cref{eq:USDParameters.20071009.5Y5Y} were used.}
\end{figure}

We repeated the same tests for 5 year expiry and 5 year tenor, 10 year expiry and 10 year tenor, and 20 year expiry and 20 year tenor combinations.
Let's start with the 5Y5Y combination. The 5Y5Y forward swap rate was 
$5.60\%$ and the market volatilities are shown in \cref{tab:volatility.20071009.5Y5Y}.
\begin{table}[htbp]
\centering
\caption{\label{tab:volatility.20071009.5Y5Y}USD Swaption lognormal volatilities with 5 year expiry and 5 year tenor as of October 9th, 2007.}
\resizebox{\columnwidth}{!}{%
\begin{tabular}{|*{10}{c|}}
\hline 
Strike(\%) & 3.60 & 4.60 & 5.10 & 5.35 & 5.60 & 5.85 & 6.10 & 6.60 & 7.60 \\
\hline 
Volatility(\%) & 20.29 & 17.64 & 16.63 & 16.23 & 15.92 & 15.61 & 15.31 & 14.97 & 14.71 \\
\hline
\end{tabular}%
}
\end{table}
We chose $\beta = 0.40$ and calibrated the rest of the parameters:
\begin{equation}
\textnormal{$\beta = 0.40$, $\nu = 30.0\%$, $\rho = -20.0\%$, $\alpha_0 = 2.74\%$, $F_0 = 5.60\%$, $T=5$}
\label{eq:USDParameters.20071009.5Y5Y}
\end{equation}
\Cref{fig:USDCalibration.20071009.5Y5Y} and \cref{tab:USDCalibration.20071009.5Y5Y} show the goodness of the calibration.
\begin{table}[htbp]
\centering
\caption{\label{tab:USDCalibration.20071009.5Y5Y}Calibration results for USD 5Y5Y swaptions on October 9th, 2007. 
The time values of swaptions expressed in the unit of 
forward swap annuity were computed.
For Black-Scholes, the linearly interpolated implied volatilities were used.
For SABR, the ``PDE + Symmetry'' method with variables $Z_t$ and $X_t$ was used.
The SABR model parameters in \cref{eq:USDParameters.20071009.5Y5Y} were used.}
\begin{tabular}{|c|c|c|}
\hline 
Strike & Black-Scholes & SABR \\
\hline
3.60\% & 0.18\% & 0.18\%  \\
4.60\% & 0.39\% & 0.39\%  \\
5.10\% & 0.56\% & 0.56\%  \\
5.35\% & 0.67\% & 0.67\%  \\
5.60\% & 0.79\% & 0.79\%  \\
5.85\% & 0.67\% & 0.67\%  \\
6.10\% & 0.57\% & 0.57\%  \\
6.60\% & 0.41\% & 0.40\%  \\
7.60\% & 0.20\% & 0.19\%  \\
\hline
\end{tabular}
\end{table} 
With the calibrated parameters in \cref{eq:USDParameters.20071009.5Y5Y}, we priced swaptions in the various methods and compared
the results in \cref{fig:comparison.20071009.5Y5Y} and \cref{tab:comparison.20071009.5Y5Y}.
\begin{table}[htbp]
\centering
\caption{\label{tab:comparison.20071009.5Y5Y}Pricing method comparison for USD 5Y5Y swaptions on October 9th, 2007. 
The time values of swaptions expressed in the unit of 
forward swap annuity were computed using various pricing methods.
The SABR model parameters in \cref{eq:USDParameters.20071009.5Y5Y} were used.}
\resizebox{\columnwidth}{!}{%
\begin{tabular}{|*{6}{c|}}
\hline 
Strike & Monte Carlo simulation  & PDE + Symmetry ZX & PDE + Symmetry WY & PDE ZX & Asymptotic\\
\hline 
0.60\% & 0.01\% & 0.01\%  & 0.01\% & 0.01\% & 0.01\% \\
1.10\% & 0.01\% & 0.02\%  & 0.01\% & 0.02\% & 0.02\% \\
1.60\% & 0.03\% & 0.03\%  & 0.03\% & 0.03\% & 0.03\% \\
2.10\% & 0.05\% & 0.05\%  & 0.05\% & 0.05\% & 0.05\% \\
2.60\% & 0.07\% & 0.08\%  & 0.07\% & 0.08\% & 0.08\% \\
3.10\% & 0.12\% & 0.12\%  & 0.12\% & 0.12\% & 0.12\% \\
3.60\% & 0.18\% & 0.18\%  & 0.18\% & 0.18\% & 0.19\% \\
4.10\% & 0.27\% & 0.27\%  & 0.27\% & 0.27\% & 0.28\% \\
4.60\% & 0.39\% & 0.39\%  & 0.40\% & 0.39\% & 0.40\% \\
5.10\% & 0.56\% & 0.56\%  & 0.57\% & 0.56\% & 0.57\% \\
5.35\% & 0.67\% & 0.67\%  & 0.68\% & 0.67\% & 0.67\% \\
5.60\% & 0.79\% & 0.79\%  & 0.80\% & 0.79\% & 0.79\% \\
5.85\% & 0.67\% & 0.67\%  & 0.68\% & 0.67\% & 0.67\% \\
6.10\% & 0.57\% & 0.57\%  & 0.58\% & 0.57\% & 0.57\% \\
6.60\% & 0.40\% & 0.40\%  & 0.41\% & 0.40\% & 0.40\% \\
7.10\% & 0.28\% & 0.28\%  & 0.29\% & 0.28\% & 0.28\% \\
7.60\% & 0.20\% & 0.19\%  & 0.20\% & 0.20\% & 0.20\% \\
8.10\% & 0.14\% & 0.14\%  & 0.14\% & 0.14\% & 0.14\% \\
8.60\% & 0.10\% & 0.10\%  & 0.10\% & 0.10\% & 0.10\% \\
9.10\% & 0.07\% & 0.07\%  & 0.07\% & 0.07\% & 0.07\% \\
9.60\% & 0.05\% & 0.05\%  & 0.05\% & 0.05\% & 0.05\% \\
10.10\% & 0.04\% & 0.04\%  & 0.04\% & 0.04\% & 0.04\% \\
10.60\% & 0.03\% & 0.03\%  & 0.03\% & 0.03\% & 0.03\% \\
\hline
\end{tabular}%
}
\end{table}
This time, the value of $\beta$ was small enough to use variables $W_t$ and $Y_t$.
All methods produced the results that were very close to each other. 
It is a little difficult to see, but upon closer inspection at \cref{fig:comparison.20071009.5Y5Y},
it was noted that
the asymptotic formula produced prices slightly different from the other methods.
\clearpage

\begin{figure}[htbp]
\centering
\includegraphics[width=1\columnwidth]{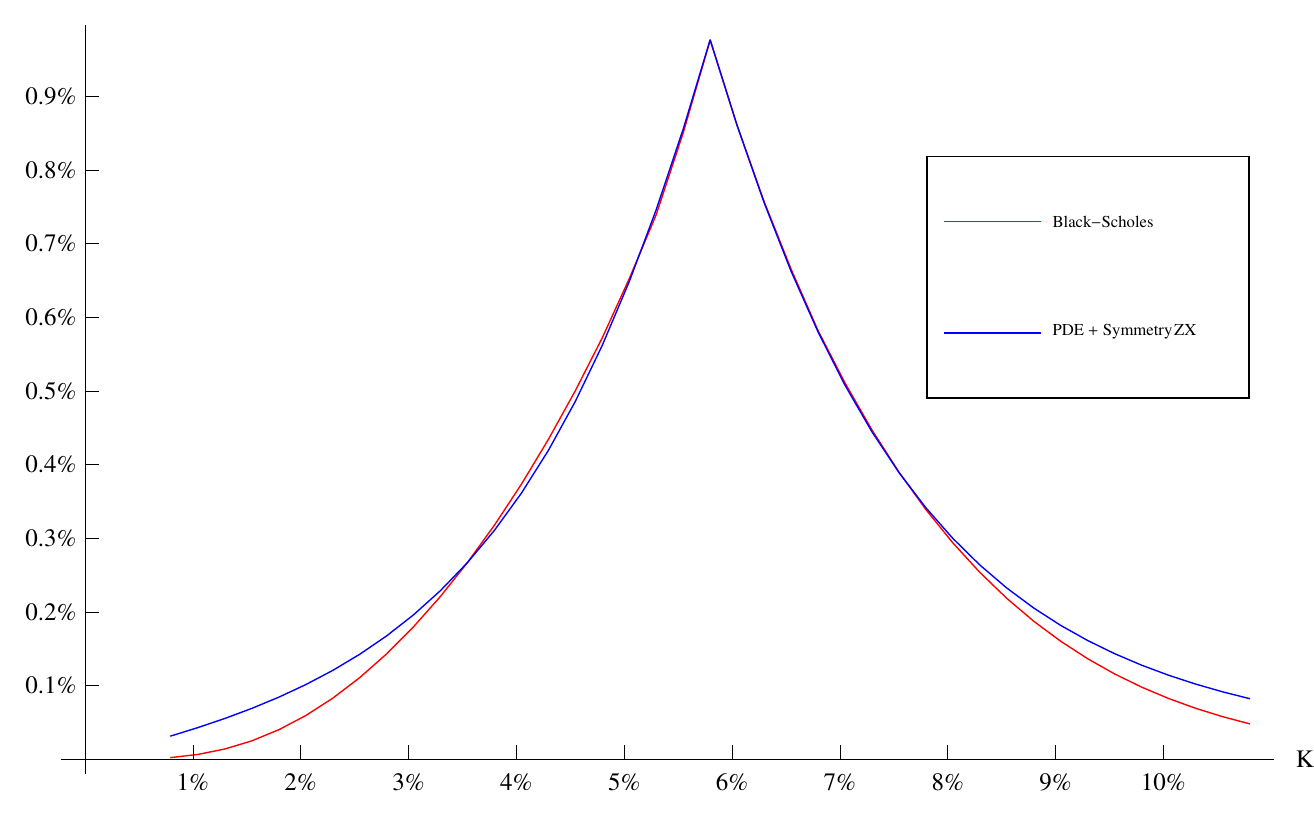}
\caption{\label{fig:USDCalibration.20071009.10Y10Y}Calibration results for USD 10Y10Y swaptions on October 9th, 2007.
The time values of swaptions expressed in the unit of 
forward swap annuity were computed and graphed
as a function of $K$. 
The red line was obtained by the Black-Scholes with linearly interpolated implied volatilities.
The blue line was computed using the ``PDE + Symmetry'' method with variables $Z_t$ and $X_t$. 
The SABR model parameters in \cref{eq:USDParameters.20071009.10Y10Y} were used.}  
\end{figure}

\begin{figure}[htbp]
\centering
\includegraphics[width=1\columnwidth]{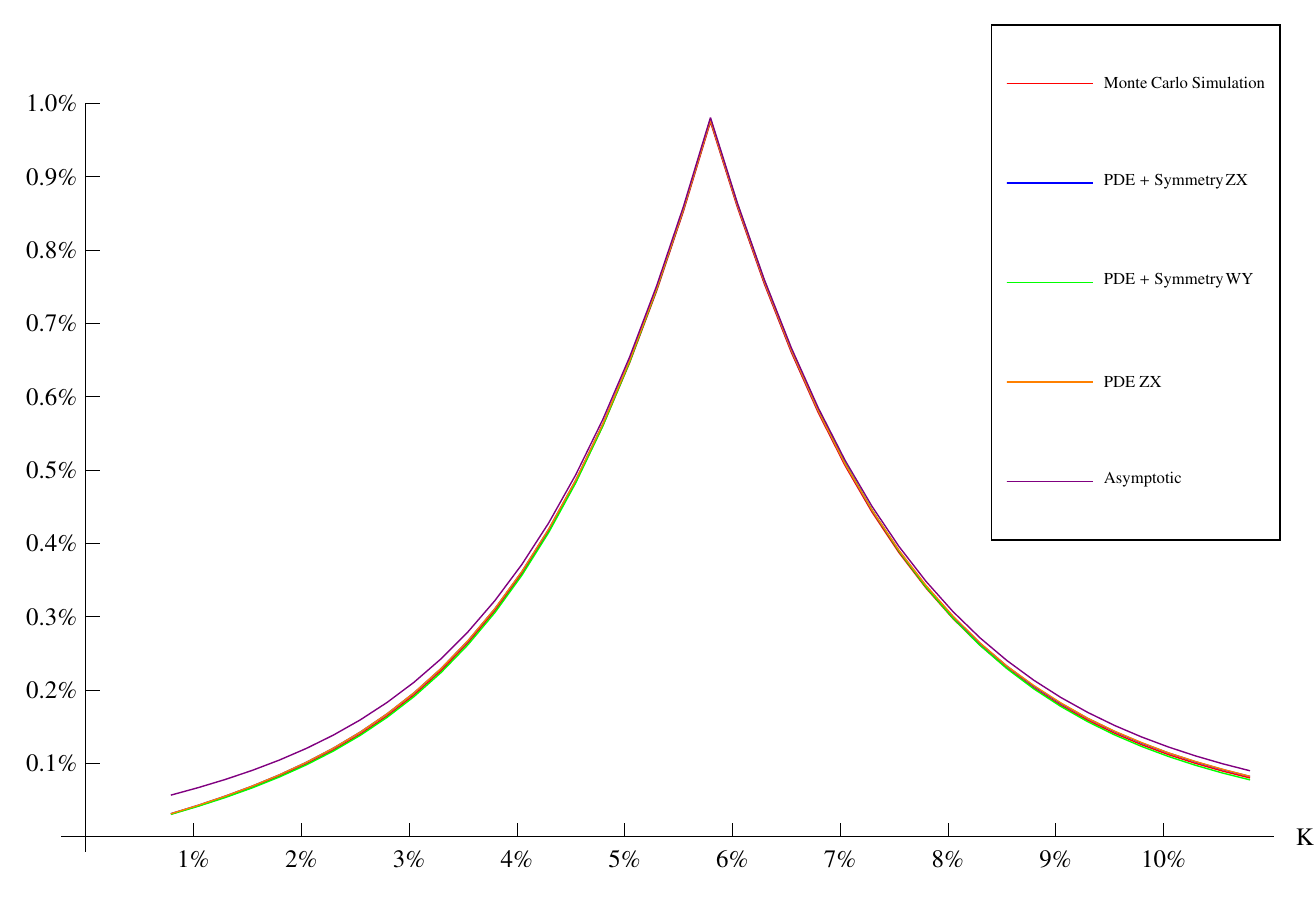}
\caption{\label{fig:comparison.20071009.10Y10Y}Pricing method comparison for USD 10Y10Y swaptions on October 9th, 2007. 
The time values of swaptions expressed in the unit of 
forward swap annuity were computed using various pricing methods.
The SABR model parameters in \cref{eq:USDParameters.20071009.10Y10Y} were used.}
\end{figure}

\begin{table}[htbp]
\centering
\caption{\label{tab:volatility.20071009.10Y10Y}USD Swaption lognormal volatilities with 10 year expiry and 10 year tenor as of October 9th, 2007.}
\resizebox{\columnwidth}{!}{%
\begin{tabular}{|*{10}{c|}}
\hline 
Strike(\%) & 3.80 & 4.80 & 5.30 & 5.55 & 5.80 & 6.05 & 6.30 & 6.80 & 7.80 \\
\hline 
Volatility(\%) & 17.34 & 15.05 & 13.97 & 13.67 & 13.45 & 13.22 & 13.04 & 12.75 & 12.41 \\
\hline
\end{tabular}%
}
\end{table}
\Cref{tab:volatility.20071009.10Y10Y} shows the USD 10Y10Y swaption volatilities.
The forward swap rate was $5.80\%$ and $\beta = 0.10$ was chosen. 
The calibrated parameters are:
\begin{equation}
\textnormal{$\beta = 0.10$, $\nu = 25.0\%$, $\rho = 0.0\%$, $\alpha_0 = 0.98\%$, $F_0 = 5.80\%$, $T=10$}
\label{eq:USDParameters.20071009.10Y10Y}
\end{equation}
\Cref{fig:USDCalibration.20071009.10Y10Y} and \cref{tab:USDCalibration.20071009.10Y10Y} show the goodness of the calibration and
\cref{fig:comparison.20071009.10Y10Y} and \cref{tab:comparison.20071009.10Y10Y} show how different methods priced the swaptions.
All methods other than the asymptotic formula showed almost identical results. 
The difference between the asymptotic formula and the other methods is noticeable, more so at low strikes.
\begin{table}[htbp]
\centering
\caption{\label{tab:USDCalibration.20071009.10Y10Y}Calibration results for USD 10Y10Y swaptions on October 9th, 2007.
The time values of swaptions expressed in the unit of 
forward swap annuity were computed.
For Black-Scholes, the linearly interpolated implied volatilities were used.
For SABR, the ``PDE + Symmetry'' method with variables $Z_t$ and $X_t$ was used.
The SABR model parameters in \cref{eq:USDParameters.20071009.10Y10Y} were used.}
\begin{tabular}{|c|c|c|}
\hline 
Strike & Black-Scholes & SABR \\
\hline
3.80\% & 0.32\% & 0.31\%  \\
4.80\% & 0.57\% & 0.56\%  \\
5.30\% & 0.74\% & 0.75\%  \\
5.55\% & 0.85\% & 0.86\%  \\
5.80\% & 0.98\% & 0.98\%  \\
6.05\% & 0.86\% & 0.86\%  \\
6.30\% & 0.76\% & 0.76\%  \\
6.80\% & 0.58\% & 0.58\%  \\
7.80\% & 0.34\% & 0.34\%  \\
\hline
\end{tabular}
\end{table} 
\begin{table}[htbp]
\centering
\caption{\label{tab:comparison.20071009.10Y10Y}Pricing method comparison for USD 10Y10Y swaptions on October 9th, 2007. 
The time values of swaptions expressed in the unit of 
forward swap annuity were computed using various pricing methods.
The SABR model parameters in \cref{eq:USDParameters.20071009.10Y10Y} were used.}
\resizebox{\columnwidth}{!}{%
\begin{tabular}{|*{6}{c|}}
\hline 
Strike & Monte Carlo simulation  & PDE + Symmetry ZX & PDE + Symmetry WY & PDE ZX & Asymptotic\\
\hline 
0.80\% & 0.03\% & 0.03\%  & 0.03\% & 0.03\% & 0.06\% \\
1.30\% & 0.05\% & 0.06\%  & 0.05\% & 0.06\% & 0.08\% \\
1.80\% & 0.08\% & 0.08\%  & 0.08\% & 0.08\% & 0.10\% \\
2.30\% & 0.12\% & 0.12\%  & 0.12\% & 0.12\% & 0.14\% \\
2.80\% & 0.16\% & 0.17\%  & 0.16\% & 0.17\% & 0.18\% \\
3.30\% & 0.23\% & 0.23\%  & 0.22\% & 0.23\% & 0.24\% \\
3.80\% & 0.31\% & 0.31\%  & 0.31\% & 0.31\% & 0.32\% \\
4.30\% & 0.42\% & 0.42\%  & 0.42\% & 0.42\% & 0.43\% \\
4.80\% & 0.56\% & 0.56\%  & 0.56\% & 0.56\% & 0.57\% \\
5.30\% & 0.74\% & 0.75\%  & 0.75\% & 0.75\% & 0.75\% \\
5.55\% & 0.85\% & 0.86\%  & 0.86\% & 0.86\% & 0.86\% \\
5.80\% & 0.97\% & 0.98\%  & 0.98\% & 0.98\% & 0.98\% \\
6.05\% & 0.86\% & 0.86\%  & 0.86\% & 0.86\% & 0.86\% \\
6.30\% & 0.75\% & 0.76\%  & 0.76\% & 0.76\% & 0.76\% \\
6.80\% & 0.58\% & 0.58\%  & 0.58\% & 0.58\% & 0.58\% \\
7.30\% & 0.44\% & 0.44\%  & 0.45\% & 0.45\% & 0.45\% \\
7.80\% & 0.34\% & 0.34\%  & 0.34\% & 0.34\% & 0.35\% \\
8.30\% & 0.26\% & 0.26\%  & 0.26\% & 0.26\% & 0.27\% \\
8.80\% & 0.20\% & 0.21\%  & 0.20\% & 0.21\% & 0.21\% \\
9.30\% & 0.16\% & 0.16\%  & 0.16\% & 0.16\% & 0.17\% \\
9.80\% & 0.13\% & 0.13\%  & 0.12\% & 0.13\% & 0.14\% \\
10.30\% & 0.10\% & 0.10\%  & 0.10\% & 0.10\% & 0.11\% \\
10.80\% & 0.08\% & 0.08\%  & 0.08\% & 0.08\% & 0.09\% \\
\hline
\end{tabular}%
}
\end{table}
\clearpage

\begin{figure}[htbp]
\centering
\includegraphics[width=1\columnwidth]{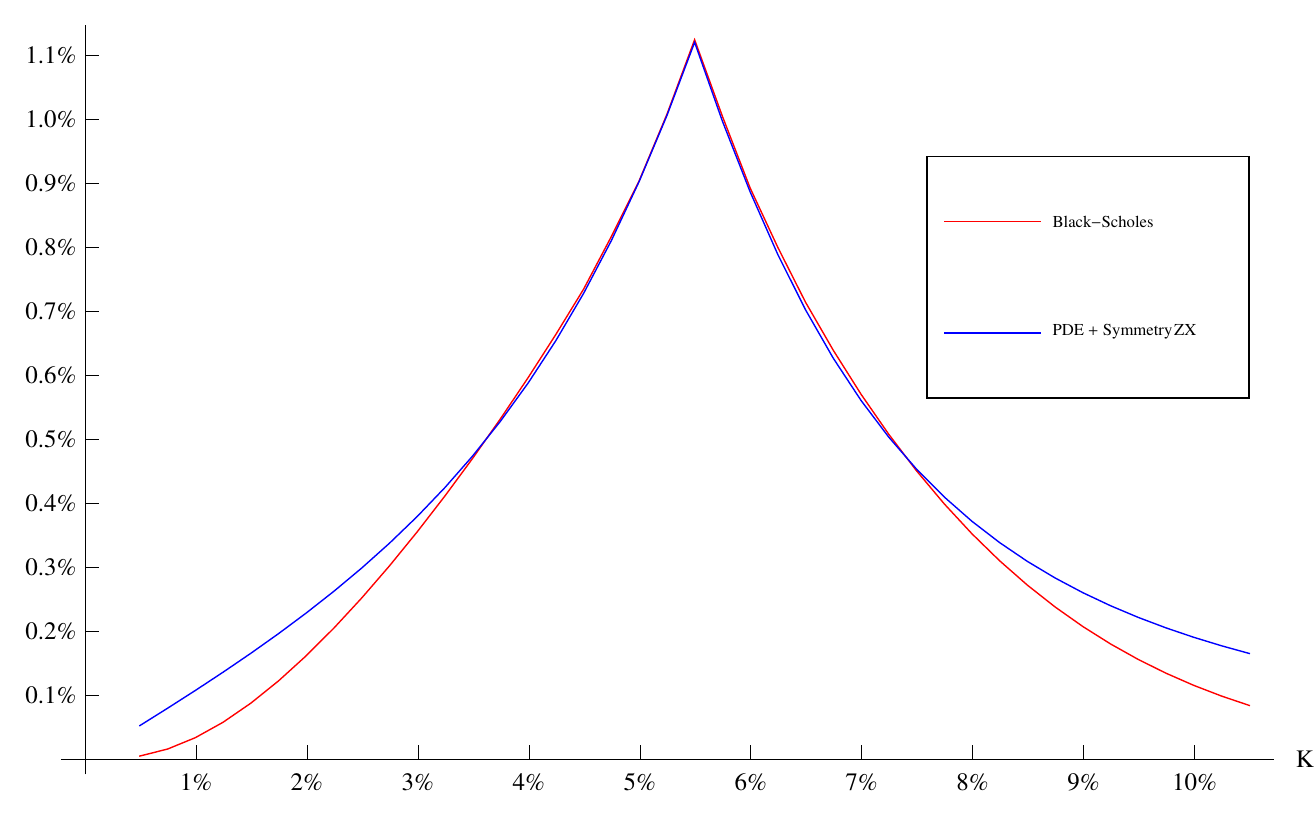}
\caption{\label{fig:USDCalibration.20071009.20Y20Y}Calibration results for USD 20Y20Y swaptions on October 9th, 2007.
The time values of swaptions expressed in the unit of 
forward swap annuity were computed and graphed
as a function of $K$. 
The red line was obtained by the Black-Scholes with linearly interpolated implied volatilities.
The blue line was computed using the ``PDE + Symmetry'' method with variables $Z_t$ and $X_t$. 
The SABR model parameters in \cref{eq:USDParameters.20071009.20Y20Y} were used.}  
\end{figure}

\begin{figure}[htbp]
\centering
\includegraphics[width=1\columnwidth]{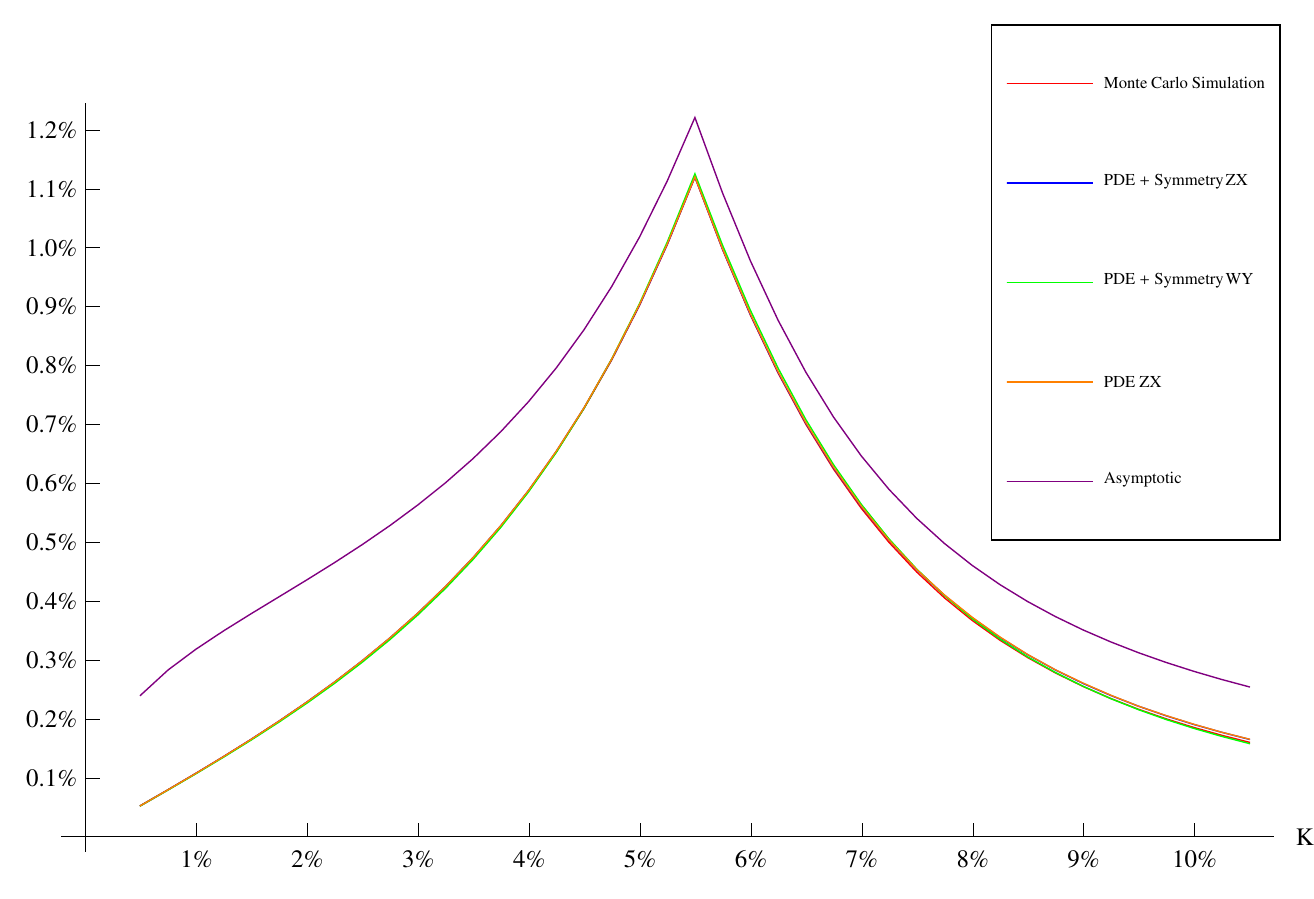}
\caption{\label{fig:comparison.20071009.20Y20Y}Pricing method comparison for USD 20Y20Y swaptions on October 9th, 2007. 
The time values of swaptions expressed in the unit of 
forward swap annuity were computed using various pricing methods.
The SABR model parameters in \cref{eq:USDParameters.20071009.20Y20Y} were used.}
\end{figure}

Finally, we look at the USD 20Y20Y swaptions. 
The lognormal volatilities are shown in \cref{tab:volatility.20071009.20Y20Y}.
\begin{table}[htbp]
\centering
\caption{\label{tab:volatility.20071009.20Y20Y}USD Swaption lognormal volatilities with 20 year expiry and 20 year tenor as of October 9th, 2007.}
\resizebox{\columnwidth}{!}{%
\begin{tabular}{|*{10}{c|}}
\hline 
Strike(\%) & 3.49 & 4.49 & 4.99 & 5.24 & 5.49 & 5.74 & 5.99 & 6.49 & 7.49 \\
\hline 
Volatility(\%) & 15.77 & 13.34 & 12.31 & 11.92 & 11.60 & 11.35 & 11.10 & 10.84 & 10.48 \\
\hline
\end{tabular}%
}
\end{table}
The forward swap rate was $5.49\%$ and $\beta = 0.0$ was chosen. 
The calibrated parameters are:
\begin{equation}
\textnormal{$\beta = 0.0$, $\nu = 28.0\%$, $\rho = -15.0\%$, $\alpha_0 = 0.60\%$, $F_0 = 5.49\%$, $T=20$}
\label{eq:USDParameters.20071009.20Y20Y}
\end{equation}
The calibration results are shown in \cref{fig:USDCalibration.20071009.20Y20Y} and \cref{tab:USDCalibration.20071009.20Y20Y}.
\Cref{fig:comparison.20071009.20Y20Y} and \cref{tab:comparison.20071009.20Y20Y} show swaption pricing results.
Now, with longer expiry, the asymptotic series shows fairly large difference from the other methods' pricing results.
The difference is larger at lower strikes.
All other methods produced virtually identical prices. 
\begin{table}[htbp]
\centering
\caption{\label{tab:USDCalibration.20071009.20Y20Y}Calibration results for USD 20Y20Y swaptions on October 9th, 2007.
The time values of swaptions expressed in the unit of 
forward swap annuity were computed.
For Black-Scholes, the linearly interpolated implied volatilities were used.
For SABR, the ``PDE + Symmetry'' method with variables $Z_t$ and $X_t$ was used.
The SABR model parameters in \cref{eq:USDParameters.20071009.20Y20Y} were used.}
\begin{tabular}{|c|c|c|}
\hline 
Strike & Black-Scholes & SABR \\
\hline
3.49\% & 0.47\% & 0.47\%  \\
4.49\% & 0.73\% & 0.73\%  \\
4.99\% & 0.90\% & 0.90\%  \\
5.24\% & 1.01\% & 1.01\%  \\
5.49\% & 1.12\% & 1.12\%  \\
5.74\% & 1.00\% & 1.00\%  \\
5.99\% & 0.89\% & 0.89\%  \\
6.49\% & 0.71\% & 0.70\%  \\
7.49\% & 0.45\% & 0.45\%  \\
\hline
\end{tabular}
\end{table} 
\begin{table}[htbp]
\centering
\caption{\label{tab:comparison.20071009.20Y20Y}Pricing method comparison for USD 20Y20Y swaptions on October 9th, 2007. 
The time values of swaptions expressed in the unit of 
forward swap annuity were computed using various pricing methods.
The SABR model parameters in \cref{eq:USDParameters.20071009.20Y20Y} were used.}
\resizebox{\columnwidth}{!}{%
\begin{tabular}{|*{6}{c|}}
\hline 
Strike & Monte Carlo simulation  & PDE + Symmetry ZX & PDE + Symmetry WY & PDE ZX & Asymptotic\\
\hline 
0.49\% & 0.05\% & 0.05\%  & 0.05\% & 0.05\% & 0.24\% \\
0.99\% & 0.11\% & 0.11\%  & 0.11\% & 0.11\% & 0.32\% \\
1.49\% & 0.17\% & 0.17\%  & 0.16\% & 0.17\% & 0.38\% \\
1.99\% & 0.23\% & 0.23\%  & 0.23\% & 0.23\% & 0.44\% \\
2.49\% & 0.30\% & 0.30\%  & 0.30\% & 0.30\% & 0.50\% \\
2.99\% & 0.38\% & 0.38\%  & 0.38\% & 0.38\% & 0.56\% \\
3.49\% & 0.47\% & 0.47\%  & 0.47\% & 0.47\% & 0.64\% \\
3.99\% & 0.59\% & 0.59\%  & 0.59\% & 0.59\% & 0.74\% \\
4.49\% & 0.73\% & 0.73\%  & 0.73\% & 0.73\% & 0.86\% \\
4.99\% & 0.90\% & 0.90\%  & 0.90\% & 0.90\% & 1.02\% \\
5.24\% & 1.01\% & 1.01\%  & 1.01\% & 1.01\% & 1.11\% \\
5.49\% & 1.12\% & 1.12\%  & 1.13\% & 1.12\% & 1.22\% \\
5.74\% & 1.00\% & 1.00\%  & 1.00\% & 1.00\% & 1.09\% \\
5.99\% & 0.89\% & 0.89\%  & 0.89\% & 0.89\% & 0.98\% \\
6.49\% & 0.70\% & 0.70\%  & 0.71\% & 0.70\% & 0.79\% \\
6.99\% & 0.56\% & 0.56\%  & 0.56\% & 0.56\% & 0.65\% \\
7.49\% & 0.45\% & 0.45\%  & 0.45\% & 0.45\% & 0.54\% \\
7.99\% & 0.37\% & 0.37\%  & 0.37\% & 0.37\% & 0.46\% \\
8.49\% & 0.30\% & 0.31\%  & 0.31\% & 0.31\% & 0.40\% \\
8.99\% & 0.26\% & 0.26\%  & 0.26\% & 0.26\% & 0.35\% \\
9.49\% & 0.22\% & 0.22\%  & 0.22\% & 0.22\% & 0.31\% \\
9.99\% & 0.19\% & 0.19\%  & 0.18\% & 0.19\% & 0.28\% \\
10.49\% & 0.16\% & 0.17\%  & 0.16\% & 0.17\% & 0.25\% \\
\hline
\end{tabular}%
}
\end{table}

The same tests were performed for the USD swaption market on different dates. 
Since the testing results were qualitatively similar to the previous one, we present the results without much explanations.
\clearpage

\subsubsection{September 15th, 2008}
We make some comments on the tests done for the USD swaption market on September 15th, 2008.
First, the calibration results for the 20Y20Y swaptions were not quite good. 
We could not find the model parameters that match volatilities at all strikes equally good. 
The implied volatility($22.51\%$) at the lowest strike($2.55\%$) seems to be too high for the SABR model to match. 
Hence, we calibrated the model trying to match market prices of swaptions at other strikes better.

Also, the ``PDE + Symmetry'' method with variable $W_t$ and $Y_t$ produced large errors for the 1Y1Y and 20Y20Y swaptions, and
was not included in the pricing method comparisons.
This method seems to struggle with the high value(0.9) of $\beta$ in the 1Y1Y case and with the large value(50\%)
of $\nu$ in the 20Y20Y case.
The ``PDE + Symmetry'' method with variable $Z_t$ and $X_t$ performed well for the both cases. 
\begin{table}[htbp]
\centering
\caption{\label{tab:volatility.20080915}USD Swaption lognormal volatilities and the forward swap rates as of September 15th, 2008.}
\resizebox{\columnwidth}{!}{%
\begin{tabular}{|*{11}{c|}}
\hline 
Expiry/Tenor & -200bp & -100bp & -50bp & -25bp & 0bp & +25bp & +50bp & +100bp & +200bp & ATM\\
\hline 
1Y1Y & 61.09\% & 54.18\% & 50.70\% & 49.76\% & 48.82\% & 47.84\% & 46.03\% & 44.37\% & 41.97\% & 2.99\% \\
5Y5Y & 27.00\% & 23.54\% & 21.95\% & 21.55\% & 21.08\% & 20.60\% & 20.12\% & 19.40\% & 18.30\% & 4.67\% \\
10Y10Y & 22.56\% & 18.49\% & 16.67\% & 16.39\% & 16.11\% & 15.83\% & 15.57\% & 15.08\% & 14.52\% & 4.82\% \\
20Y20Y & 22.51\% & 16.71\% & 14.88\% & 14.13\% & 13.73\% & 13.43\% & 13.12\% & 12.76\% & 12.36\% & 4.55\% \\
\hline
\end{tabular}%
}
\end{table}
\begin{table}[htbp]
\centering
\caption{\label{tab:SABRparameters.20080915}The SABR parameters calibrated to the USD Swaption market on September 15th, 2008.}
\begin{tabular}{|*{7}{c|}}
\hline 
Expiry/Tenor & $\beta$ & $\nu$ & $\rho$ & $\alpha_0$ & $F_0$ & $T$\\
\hline 
1Y1Y & 0.90 & 30.0\% & -75.0\% & 35.40\% & 2.99\% & 1 \\
5Y5Y & 0.40 & 25.0\% & -20.0\% & 3.29\% & 4.67\% & 5 \\
10Y10Y & 0.10 & 30.0\% & -10.0\% & 1.00\% & 4.82\% & 10 \\
20Y20Y & 0.00 & 50.0\% & -25.0\% & 0.72\% & 4.55\% & 20 \\
\hline
\end{tabular}%
\end{table}
\begin{figure}[htbp]
\centering
\includegraphics[width=1\columnwidth]{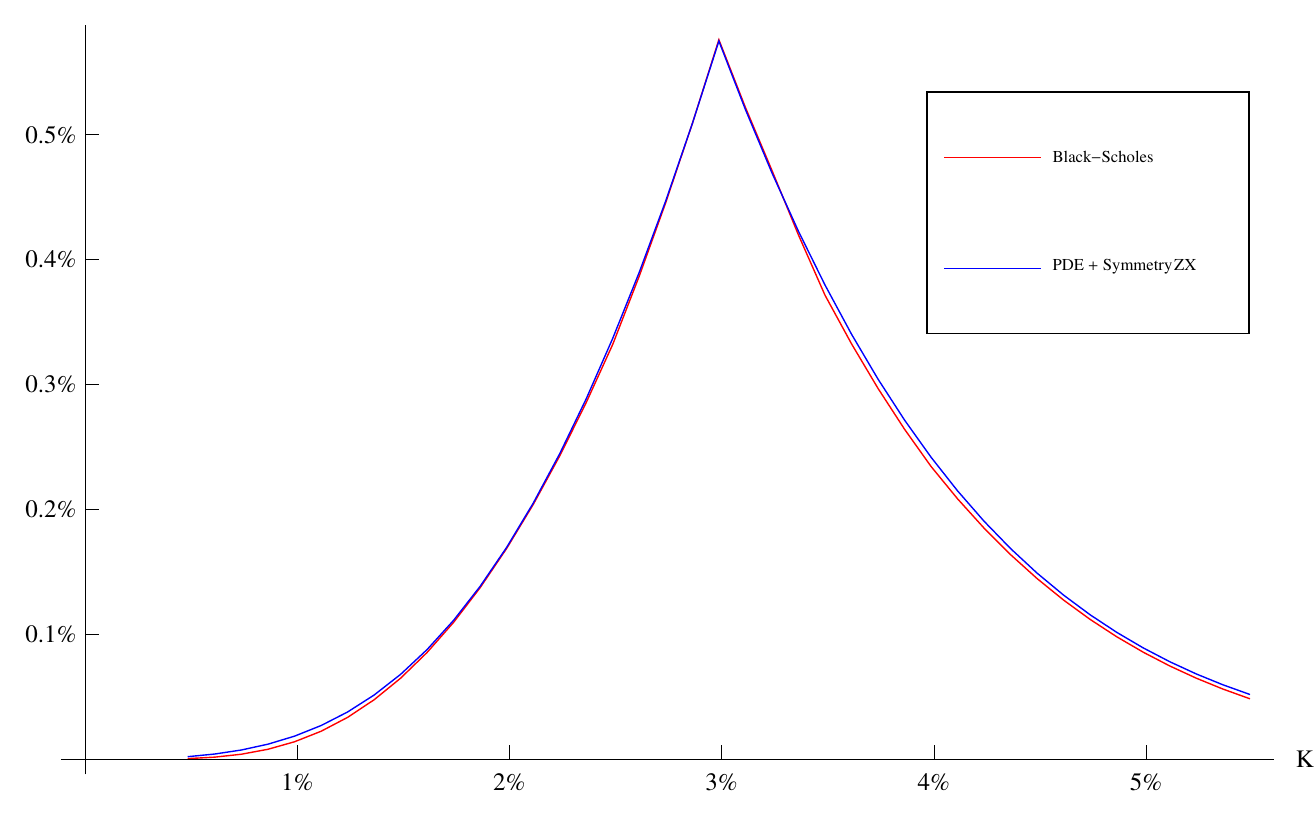}
\caption{\label{fig:USDCalibration.20080915.1Y1Y}Calibration results for USD 1Y1Y swaptions on September 15th, 2008.
The time values of swaptions expressed in the unit of 
forward swap annuity were computed and graphed
as a function of $K$. 
The red line was obtained by the Black-Scholes with linearly interpolated implied volatilities.
The blue line was computed using the ``PDE + Symmetry'' method with variables $Z_t$ and $X_t$. 
The SABR model parameters in \cref{tab:SABRparameters.20080915} were used.}  
\end{figure}
\begin{table}[htbp]
\centering
\caption{\label{tab:USDCalibration.20080915.1Y1Y}Calibration results for USD 1Y1Y swaptions on September 15th, 2008.
The time values of swaptions expressed in the unit of 
forward swap annuity were computed.
For Black-Scholes, the linearly interpolated implied volatilities were used.
For SABR, the ``PDE + Symmetry'' method with variables $Z_t$ and $X_t$ was used.
The SABR model parameters in \cref{tab:SABRparameters.20080915} were used.}
\begin{tabular}{|c|c|c|}
\hline 
Strike & Black-Scholes & SABR \\
\hline
0.99\% & 0.01\% & 0.02\%  \\
1.99\% & 0.17\% & 0.17\%  \\
2.49\% & 0.33\% & 0.34\%  \\
2.74\% & 0.45\% & 0.45\%  \\
2.99\% & 0.58\% & 0.57\%  \\
3.24\% & 0.47\% & 0.47\%  \\
3.49\% & 0.37\% & 0.38\%  \\
3.99\% & 0.23\% & 0.24\%  \\
4.99\% & 0.09\% & 0.09\%  \\
\hline
\end{tabular}
\end{table} 
\begin{figure}[htbp]
\centering
\includegraphics[width=1\columnwidth]{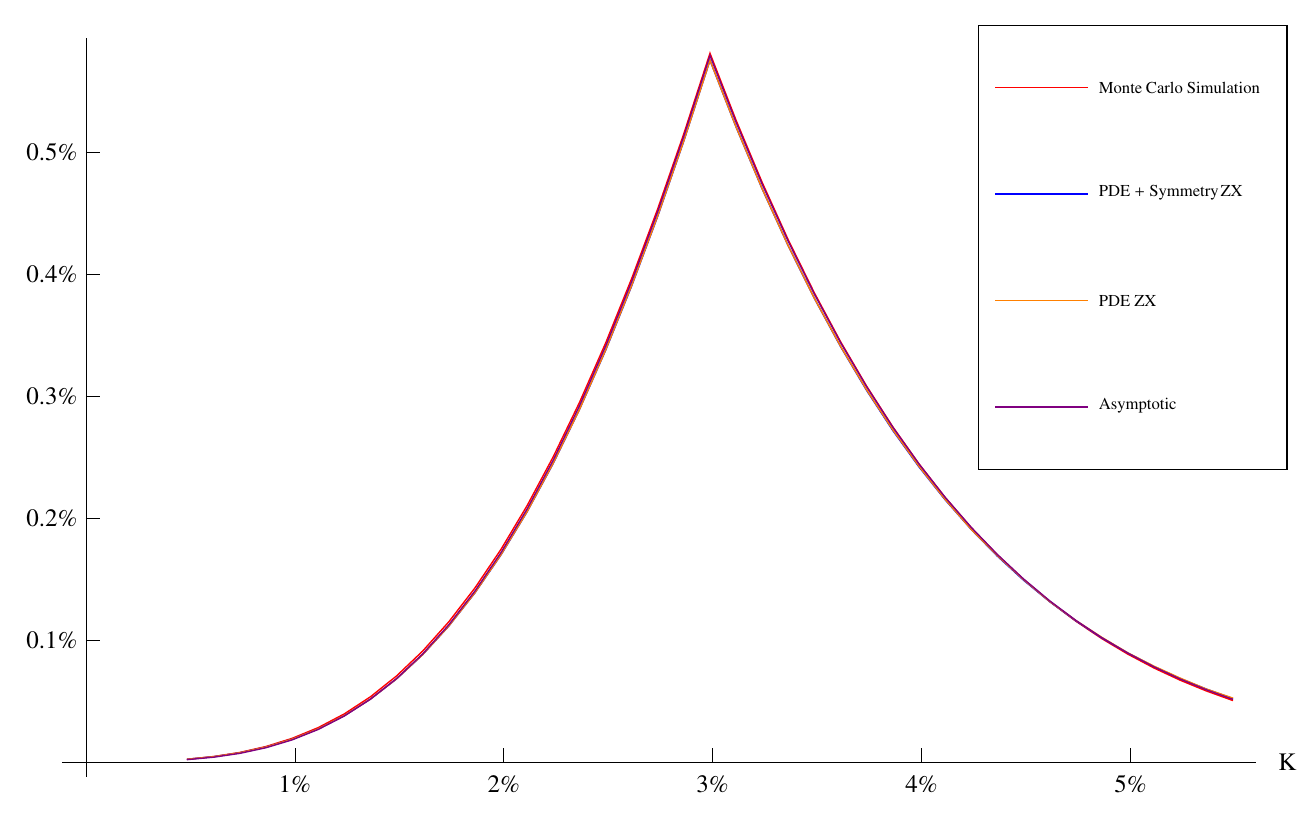}
\caption{\label{fig:comparison.20080915.1Y1Y}Pricing method comparison for USD 1Y1Y swaptions on September 15th, 2008. 
The time values of swaptions expressed in the unit of 
forward swap annuity were computed using various pricing methods.
The SABR model parameters in \cref{tab:SABRparameters.20080915} were used.}
\end{figure}
\begin{table}[htbp]
\centering
\caption{\label{tab:comparison.20080915.1Y1Y}Pricing method comparison for USD 1Y1Y swaptions on September 15th, 2008. 
The time values of swaptions expressed in the unit of 
forward swap annuity were computed using various pricing methods.
The SABR model parameters in \cref{tab:SABRparameters.20080915} were used.}
\resizebox{\columnwidth}{!}{%
\begin{tabular}{|*{5}{c|}}
\hline 
Strike & Monte Carlo simulation  & PDE + Symmetry ZX & PDE ZX & Asymptotic\\
\hline 
0.49\% & 0.00\% & 0.00\%  & 0.00\% & 0.00\% \\
0.99\% & 0.02\% & 0.02\%  & 0.02\% & 0.02\% \\
1.49\% & 0.07\% & 0.07\%  & 0.07\% & 0.07\% \\
1.99\% & 0.17\% & 0.17\%  & 0.17\% & 0.17\% \\
2.49\% & 0.34\% & 0.34\%  & 0.34\% & 0.34\% \\
2.74\% & 0.45\% & 0.45\%  & 0.45\% & 0.45\% \\
2.99\% & 0.58\% & 0.57\%  & 0.57\% & 0.58\% \\
3.24\% & 0.47\% & 0.47\%  & 0.47\% & 0.47\% \\
3.49\% & 0.38\% & 0.38\%  & 0.38\% & 0.38\% \\
3.99\% & 0.24\% & 0.24\%  & 0.24\% & 0.24\% \\
4.49\% & 0.15\% & 0.15\%  & 0.15\% & 0.15\% \\
4.99\% & 0.09\% & 0.09\%  & 0.09\% & 0.09\% \\
5.49\% & 0.05\% & 0.05\%  & 0.05\% & 0.05\% \\
5.99\% & 0.03\% & 0.03\%  & 0.03\% & 0.03\% \\
6.49\% & 0.01\% & 0.02\%  & 0.02\% & 0.02\% \\
6.99\% & 0.01\% & 0.01\%  & 0.01\% & 0.01\% \\
7.49\% & 0.00\% & 0.01\%  & 0.01\% & 0.00\% \\
7.99\% & 0.00\% & 0.00\%  & 0.00\% & 0.00\% \\
8.49\% & 0.00\% & 0.00\%  & 0.00\% & 0.00\% \\
8.99\% & 0.00\% & 0.00\%  & 0.00\% & 0.00\% \\
9.49\% & 0.00\% & 0.00\%  & 0.00\% & 0.00\% \\
9.99\% & 0.00\% & 0.00\%  & 0.00\% & 0.00\% \\
10.49\% & 0.00\% & 0.00\%  & 0.00\% & 0.00\% \\
\hline
\end{tabular}%
}
\end{table}
\clearpage
\begin{figure}[htbp]
\centering
\includegraphics[width=1\columnwidth]{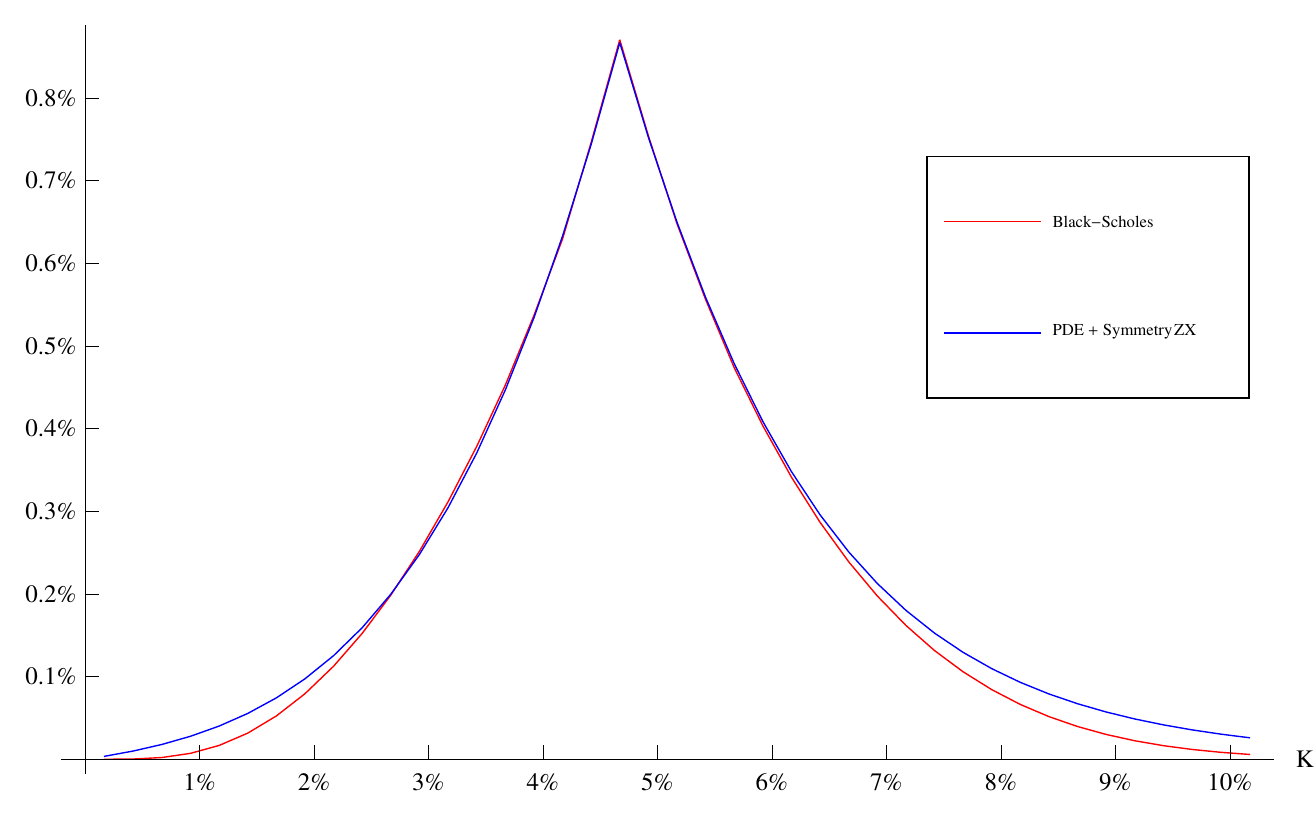}
\caption{\label{fig:USDCalibration.20080915.5Y5Y}Calibration results for USD 5Y5Y swaptions on September 15th, 2008.
The time values of swaptions expressed in the unit of 
forward swap annuity were computed and graphed
as a function of $K$. 
The red line was obtained by the Black-Scholes with linearly interpolated implied volatilities.
The blue line was computed using the ``PDE + Symmetry'' method with variables $Z_t$ and $X_t$. 
The SABR model parameters in \cref{tab:SABRparameters.20080915} were used.}  
\end{figure}
\begin{table}[htbp]
\centering
\caption{\label{tab:USDCalibration.20080915.5Y5Y}Calibration results for USD 5Y5Y swaptions on September 15th, 2008.
The time values of swaptions expressed in the unit of 
forward swap annuity were computed.
For Black-Scholes, the linearly interpolated implied volatilities were used.
For SABR, the ``PDE + Symmetry'' method with variables $Z_t$ and $X_t$ was used.
The SABR model parameters in \cref{tab:SABRparameters.20080915} were used.}
\begin{tabular}{|c|c|c|}
\hline 
Strike & Black-Scholes & SABR \\
\hline
2.67\% & 0.20\% & 0.20\%  \\
3.67\% & 0.45\% & 0.45\%  \\
4.17\% & 0.63\% & 0.63\%  \\
4.42\% & 0.75\% & 0.74\%  \\
4.67\% & 0.87\% & 0.87\%  \\
4.92\% & 0.75\% & 0.75\%  \\
5.17\% & 0.65\% & 0.65\%  \\
5.67\% & 0.47\% & 0.48\%  \\
6.67\% & 0.24\% & 0.25\%  \\
\hline
\end{tabular}
\end{table} 
\begin{figure}[htbp]
\centering
\includegraphics[width=1\columnwidth]{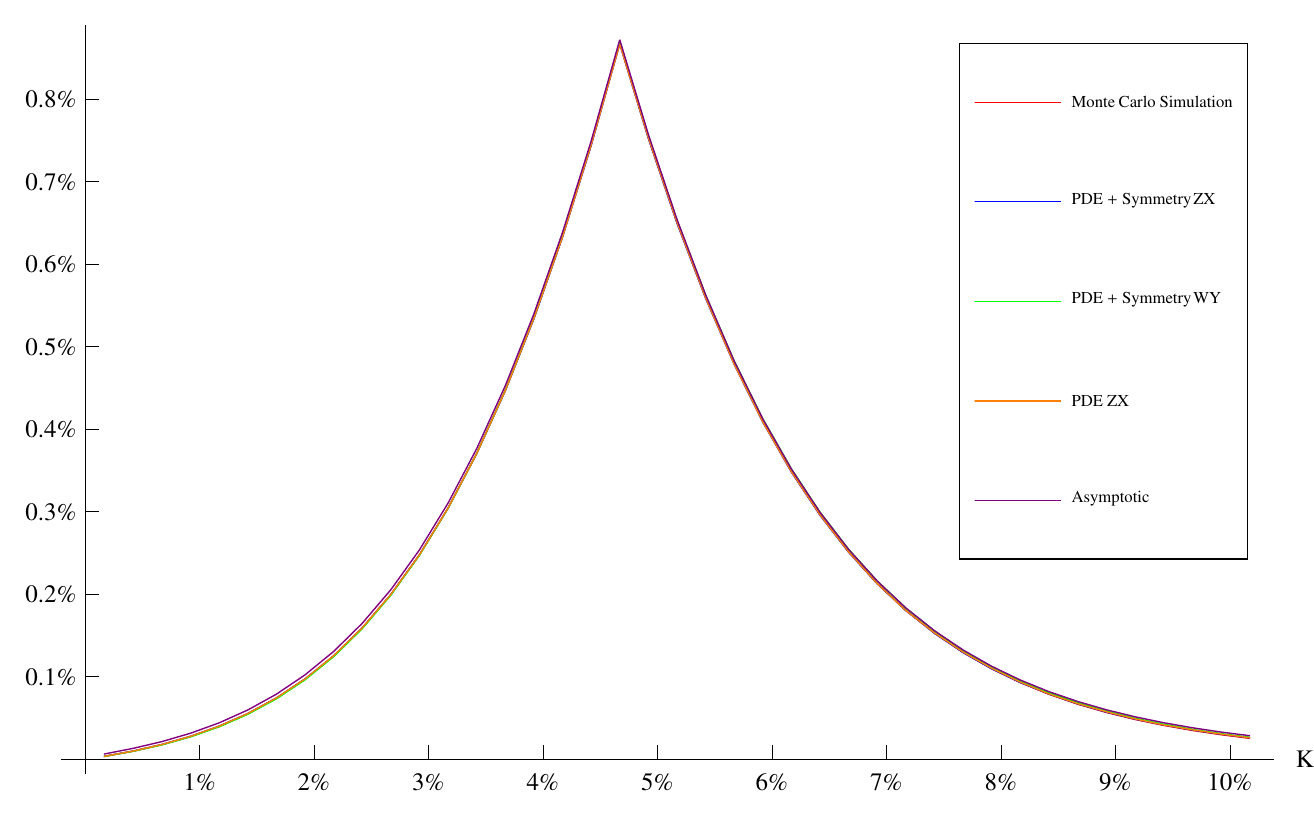}
\caption{\label{fig:comparison.20080915.5Y5Y}Pricing method comparison for USD 5Y5Y swaptions on September 15th, 2008. 
The time values of swaptions expressed in the unit of 
forward swap annuity were computed using various pricing methods.
The SABR model parameters in \cref{tab:SABRparameters.20080915} were used.}
\end{figure}
\begin{table}[htbp]
\centering
\caption{\label{tab:comparison.20080915.5Y5Y}Pricing method comparison for USD 5Y5Y swaptions on September 15th, 2008. 
The time values of swaptions expressed in the unit of 
forward swap annuity were computed using various pricing methods.
The SABR model parameters in \cref{tab:SABRparameters.20080915} were used.}
\resizebox{\columnwidth}{!}{%
\begin{tabular}{|*{6}{c|}}
\hline 
Strike & Monte Carlo simulation  & PDE + Symmetry ZX & PDE + Symmetry WY & PDE ZX & Asymptotic\\
\hline 
0.17\% & 0.00\% & 0.00\%  & 0.00\% & 0.00\% & 0.01\% \\
0.67\% & 0.02\% & 0.02\%  & 0.02\% & 0.02\% & 0.02\% \\
1.17\% & 0.04\% & 0.04\%  & 0.04\% & 0.04\% & 0.04\% \\
1.67\% & 0.07\% & 0.07\%  & 0.07\% & 0.07\% & 0.08\% \\
2.17\% & 0.12\% & 0.13\%  & 0.12\% & 0.13\% & 0.13\% \\
2.67\% & 0.20\% & 0.20\%  & 0.20\% & 0.20\% & 0.21\% \\
3.17\% & 0.30\% & 0.30\%  & 0.30\% & 0.30\% & 0.31\% \\
3.67\% & 0.45\% & 0.45\%  & 0.45\% & 0.45\% & 0.45\% \\
4.17\% & 0.63\% & 0.63\%  & 0.63\% & 0.63\% & 0.64\% \\
4.42\% & 0.74\% & 0.74\%  & 0.75\% & 0.74\% & 0.75\% \\
4.67\% & 0.87\% & 0.87\%  & 0.87\% & 0.87\% & 0.87\% \\
4.92\% & 0.75\% & 0.75\%  & 0.76\% & 0.75\% & 0.76\% \\
5.17\% & 0.65\% & 0.65\%  & 0.65\% & 0.65\% & 0.65\% \\
5.67\% & 0.48\% & 0.48\%  & 0.48\% & 0.48\% & 0.48\% \\
6.17\% & 0.35\% & 0.35\%  & 0.35\% & 0.35\% & 0.35\% \\
6.67\% & 0.25\% & 0.25\%  & 0.25\% & 0.25\% & 0.25\% \\
7.17\% & 0.18\% & 0.18\%  & 0.18\% & 0.18\% & 0.18\% \\
7.67\% & 0.13\% & 0.13\%  & 0.13\% & 0.13\% & 0.13\% \\
8.17\% & 0.09\% & 0.09\%  & 0.09\% & 0.09\% & 0.10\% \\
8.67\% & 0.07\% & 0.07\%  & 0.07\% & 0.07\% & 0.07\% \\
9.17\% & 0.05\% & 0.05\%  & 0.05\% & 0.05\% & 0.05\% \\
9.67\% & 0.04\% & 0.04\%  & 0.04\% & 0.04\% & 0.04\% \\
10.17\% & 0.03\% & 0.03\%  & 0.03\% & 0.03\% & 0.03\% \\
\hline
\end{tabular}%
}
\end{table}
\clearpage
\begin{figure}[htbp]
\centering
\includegraphics[width=1\columnwidth]{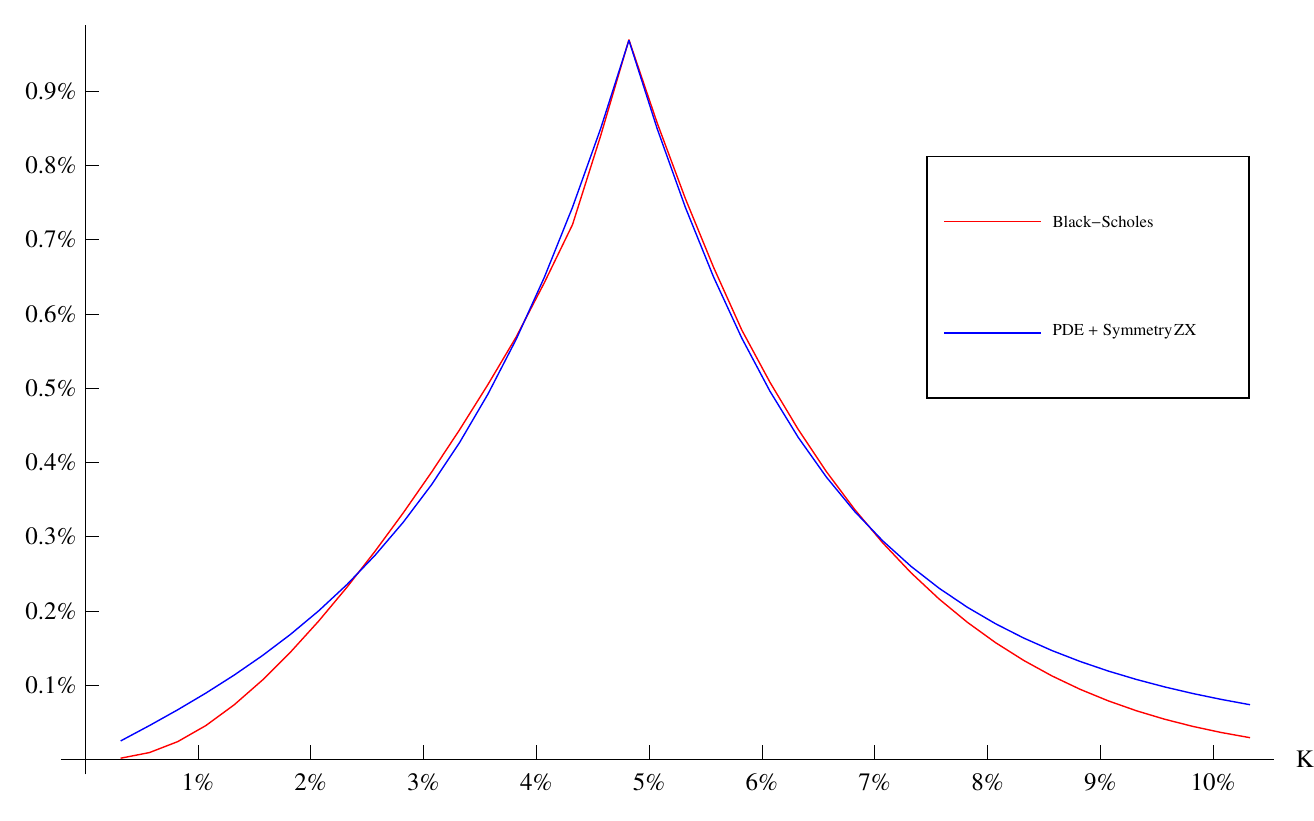}
\caption{\label{fig:USDCalibration.20080915.10Y10Y}Calibration results for USD 10Y10Y swaptions on September 15th, 2008.
The time values of swaptions expressed in the unit of 
forward swap annuity were computed and graphed
as a function of $K$. 
The red line was obtained by the Black-Scholes with linearly interpolated implied volatilities.
The blue line was computed using the ``PDE + Symmetry'' method with variables $Z_t$ and $X_t$. 
The SABR model parameters in \cref{tab:SABRparameters.20080915} were used.}  
\end{figure}
\begin{table}[htbp]
\centering
\caption{\label{tab:USDCalibration.20080915.10Y10Y}Calibration results for USD 10Y10Y swaptions on September 15th, 2008.
The time values of swaptions expressed in the unit of 
forward swap annuity were computed.
For Black-Scholes, the linearly interpolated implied volatilities were used.
For SABR, the ``PDE + Symmetry'' method with variables $Z_t$ and $X_t$ was used.
The SABR model parameters in \cref{tab:SABRparameters.20080915} were used.}
\begin{tabular}{|c|c|c|}
\hline 
Strike & Black-Scholes & SABR \\
\hline
2.82\% & 0.33\% & 0.32\%  \\
3.82\% & 0.57\% & 0.57\%  \\
4.32\% & 0.72\% & 0.74\%  \\
4.57\% & 0.84\% & 0.85\%  \\
4.82\% & 0.97\% & 0.97\%  \\
5.07\% & 0.86\% & 0.85\%  \\
5.32\% & 0.76\% & 0.74\%  \\
5.82\% & 0.58\% & 0.57\%  \\
6.82\% & 0.34\% & 0.33\%  \\
\hline
\end{tabular}
\end{table} 
\begin{figure}[htbp]
\centering
\includegraphics[width=1\columnwidth]{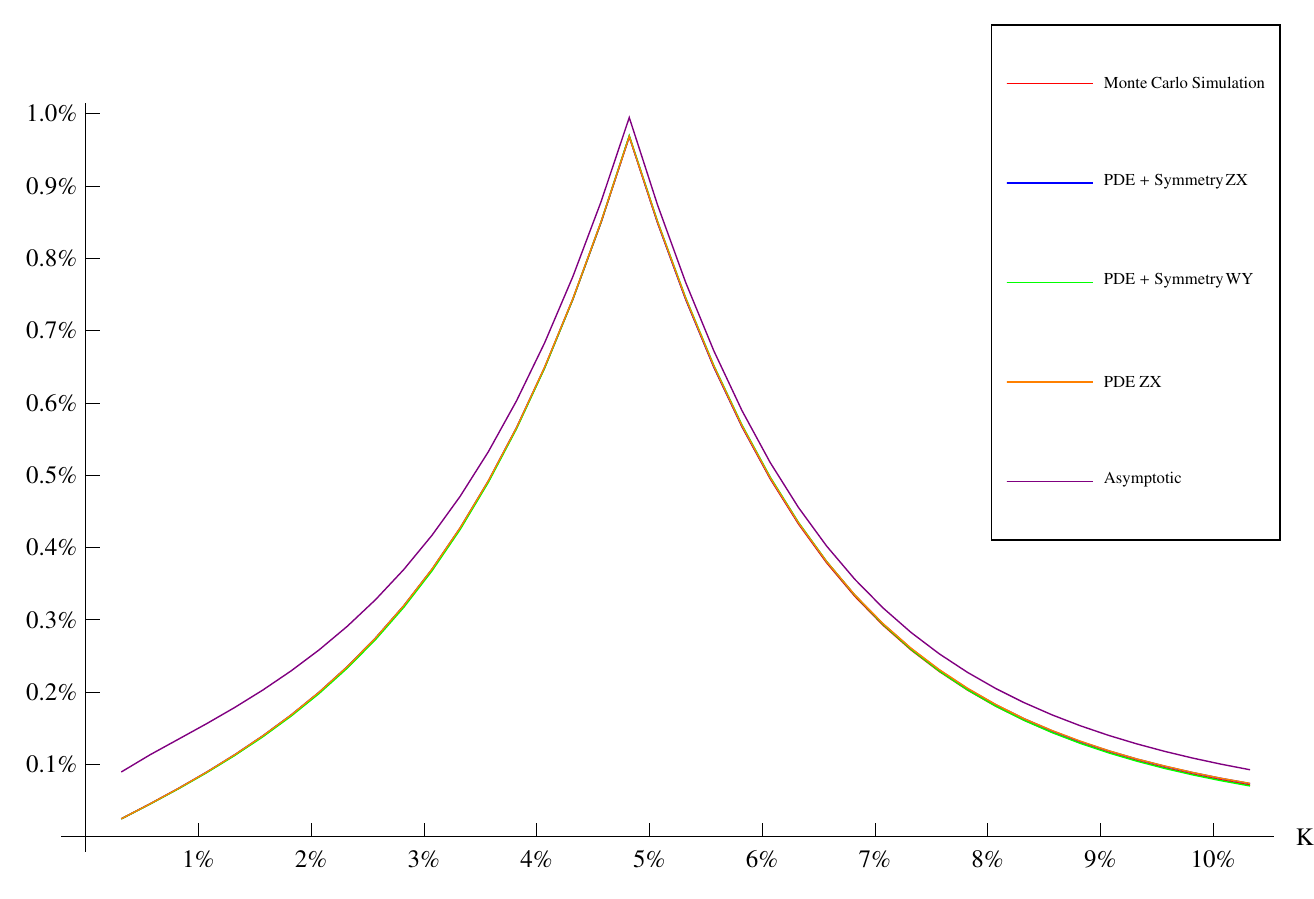}
\caption{\label{fig:comparison.20080915.10Y10Y}Pricing method comparison for USD 10Y10Y swaptions on September 15th, 2008. 
The time values of swaptions expressed in the unit of 
forward swap annuity were computed using various pricing methods.
The SABR model parameters in \cref{tab:SABRparameters.20080915} were used.}
\end{figure}
\begin{table}[htbp]
\centering
\caption{\label{tab:comparison.20080915.10Y10Y}Pricing method comparison for USD 10Y10Y swaptions on September 15th, 2008. 
The time values of swaptions expressed in the unit of 
forward swap annuity were computed using various pricing methods.
The SABR model parameters in \cref{tab:SABRparameters.20080915} were used.}
\resizebox{\columnwidth}{!}{%
\begin{tabular}{|*{6}{c|}}
\hline 
Strike & Monte Carlo simulation  & PDE + Symmetry ZX & PDE + Symmetry WY & PDE ZX & Asymptotic\\
\hline 
0.32\% & 0.03\% & 0.03\%  & 0.03\% & 0.03\% & 0.09\% \\
0.82\% & 0.07\% & 0.07\%  & 0.07\% & 0.07\% & 0.13\% \\
1.32\% & 0.11\% & 0.11\%  & 0.11\% & 0.11\% & 0.18\% \\
1.82\% & 0.17\% & 0.17\%  & 0.17\% & 0.17\% & 0.23\% \\
2.32\% & 0.23\% & 0.24\%  & 0.23\% & 0.24\% & 0.29\% \\
2.82\% & 0.32\% & 0.32\%  & 0.32\% & 0.32\% & 0.37\% \\
3.32\% & 0.43\% & 0.43\%  & 0.42\% & 0.43\% & 0.47\% \\
3.82\% & 0.57\% & 0.57\%  & 0.56\% & 0.57\% & 0.60\% \\
4.32\% & 0.74\% & 0.74\%  & 0.74\% & 0.74\% & 0.77\% \\
4.57\% & 0.85\% & 0.85\%  & 0.85\% & 0.85\% & 0.88\% \\
4.82\% & 0.97\% & 0.97\%  & 0.97\% & 0.97\% & 0.99\% \\
5.07\% & 0.85\% & 0.85\%  & 0.85\% & 0.85\% & 0.87\% \\
5.57\% & 0.65\% & 0.65\%  & 0.65\% & 0.65\% & 0.67\% \\
6.07\% & 0.49\% & 0.50\%  & 0.50\% & 0.50\% & 0.52\% \\
6.57\% & 0.38\% & 0.38\%  & 0.38\% & 0.38\% & 0.40\% \\
7.07\% & 0.29\% & 0.29\%  & 0.29\% & 0.29\% & 0.32\% \\
7.57\% & 0.23\% & 0.23\%  & 0.23\% & 0.23\% & 0.25\% \\
8.07\% & 0.18\% & 0.18\%  & 0.18\% & 0.18\% & 0.20\% \\
8.57\% & 0.14\% & 0.15\%  & 0.14\% & 0.15\% & 0.17\% \\
9.07\% & 0.12\% & 0.12\%  & 0.12\% & 0.12\% & 0.14\% \\
9.57\% & 0.10\% & 0.10\%  & 0.09\% & 0.10\% & 0.12\% \\
10.07\% & 0.08\% & 0.08\%  & 0.08\% & 0.08\% & 0.10\% \\
10.57\% & 0.07\% & 0.07\%  & 0.06\% & 0.07\% & 0.09\% \\
\hline
\end{tabular}%
}
\end{table}
\clearpage
\begin{figure}[htbp]
\centering
\includegraphics[width=1\columnwidth]{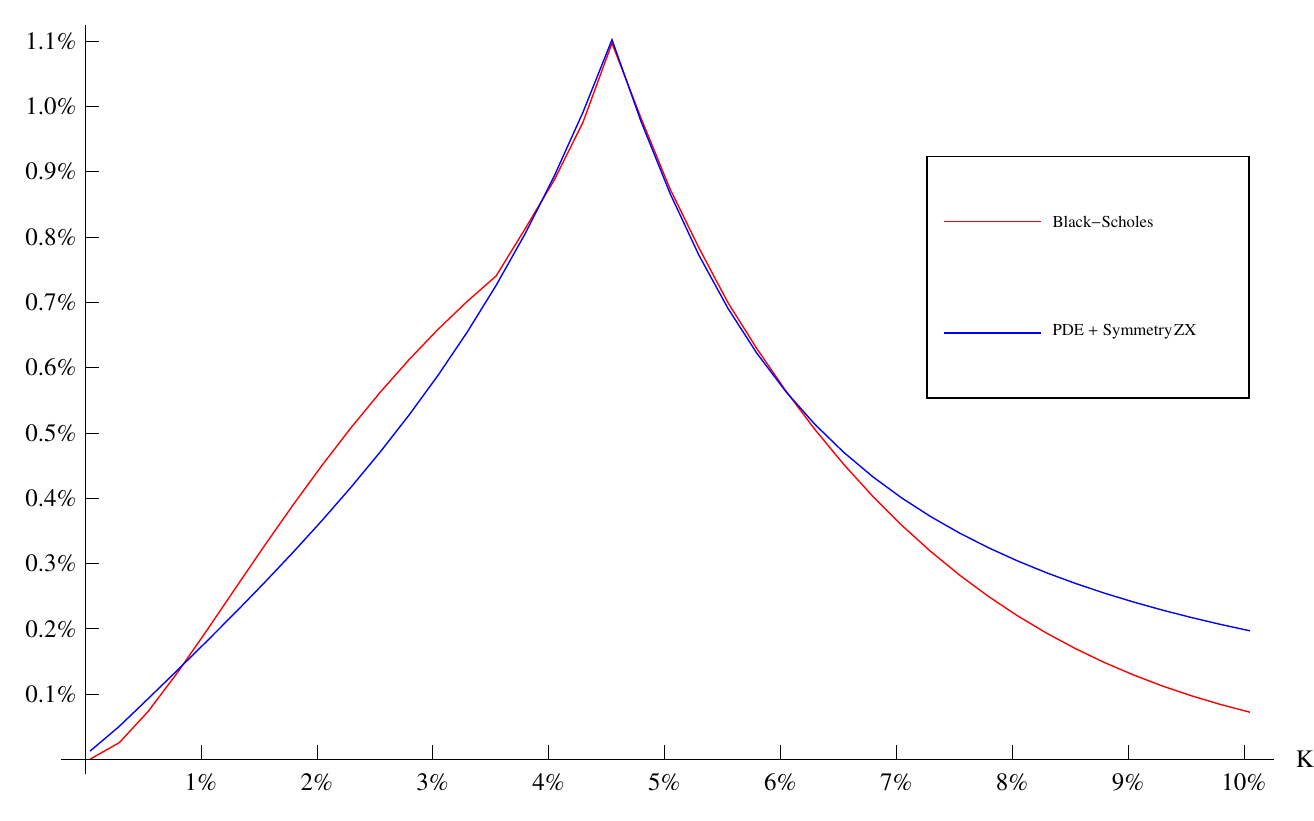}
\caption{\label{fig:USDCalibration.20080915.20Y20Y}Calibration results for USD 20Y20Y swaptions on September 15th, 2008.
The time values of swaptions expressed in the unit of 
forward swap annuity were computed and graphed
as a function of $K$. 
The red line was obtained by the Black-Scholes with linearly interpolated implied volatilities.
The blue line was computed using the ``PDE + Symmetry'' method with variables $Z_t$ and $X_t$. 
The SABR model parameters in \cref{tab:SABRparameters.20080915} were used.}  
\end{figure}
\begin{table}[htbp]
\centering
\caption{\label{tab:USDCalibration.20080915.20Y20Y}Calibration results for USD 20Y20Y swaptions on September 15th, 2008.
The time values of swaptions expressed in the unit of 
forward swap annuity were computed.
For Black-Scholes, the linearly interpolated implied volatilities were used.
For SABR, the ``PDE + Symmetry'' method with variables $Z_t$ and $X_t$ was used.
The SABR model parameters in \cref{tab:SABRparameters.20080915} were used.}
\begin{tabular}{|c|c|c|}
\hline 
Strike & Black-Scholes & SABR \\
\hline
2.55\% & 0.56\% & 0.47\%  \\
3.55\% & 0.74\% & 0.73\%  \\
4.05\% & 0.89\% & 0.89\%  \\
4.30\% & 0.98\% & 0.99\%  \\
4.55\% & 1.10\% & 1.10\%  \\
4.80\% & 0.98\% & 0.98\%  \\
5.05\% & 0.87\% & 0.87\%  \\
5.55\% & 0.70\% & 0.69\%  \\
6.55\% & 0.45\% & 0.47\%  \\
\hline
\end{tabular}
\end{table} 
\begin{figure}[htbp]
\centering
\includegraphics[width=1\columnwidth]{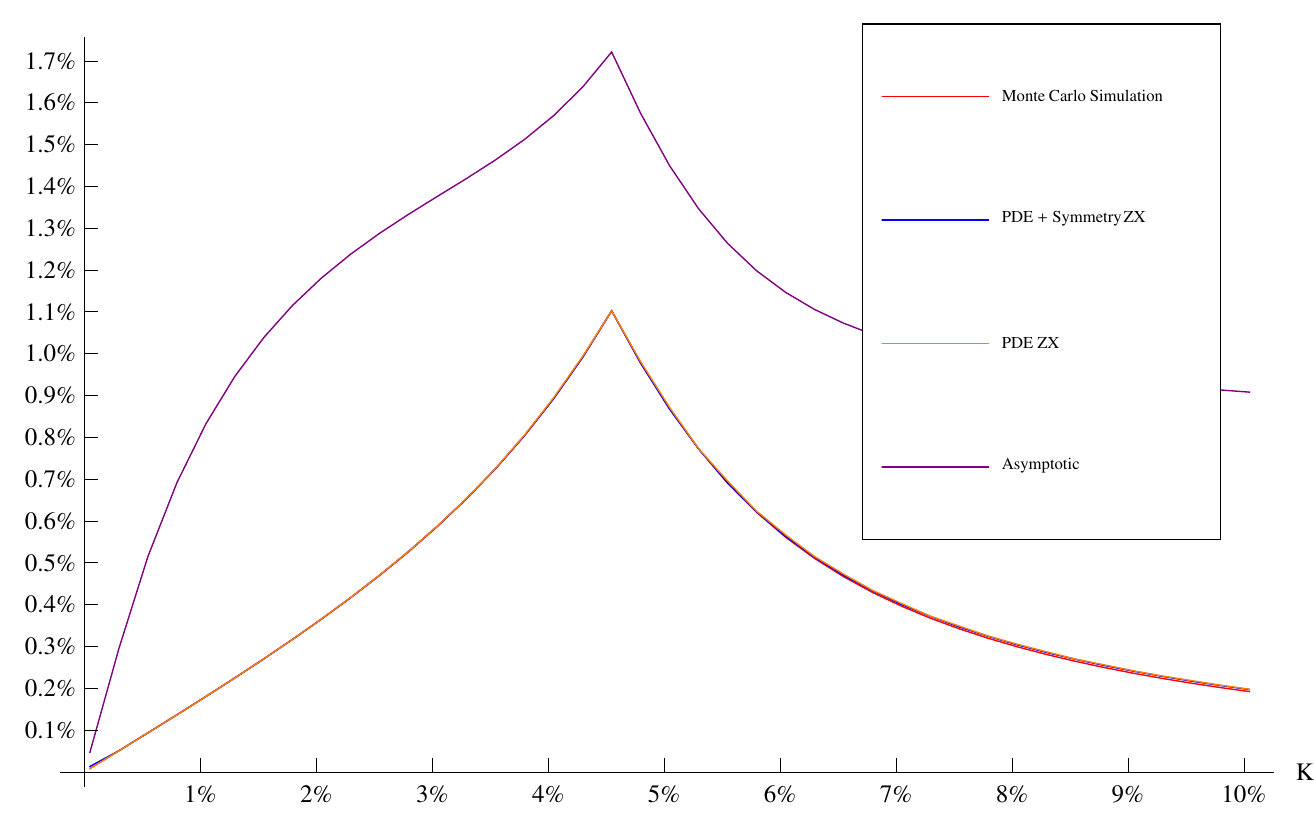}
\caption{\label{fig:comparison.20080915.20Y20Y}Pricing method comparison for USD 20Y20Y swaptions on September 15th, 2008. 
The time values of swaptions expressed in the unit of 
forward swap annuity were computed using various pricing methods.
The SABR model parameters in \cref{tab:SABRparameters.20080915} were used.}
\end{figure}
\begin{table}[htbp]
\centering
\caption{\label{tab:comparison.20080915.20Y20Y}Pricing method comparison for USD 20Y20Y swaptions on September 15th, 2008. 
The time values of swaptions expressed in the unit of 
forward swap annuity were computed using various pricing methods.
The SABR model parameters in \cref{tab:SABRparameters.20080915} were used.}
\resizebox{\columnwidth}{!}{%
\begin{tabular}{|*{5}{c|}}
\hline 
Strike & Monte Carlo simulation  & PDE + Symmetry ZX & PDE ZX & Asymptotic\\
\hline 
0.05\% & 0.01\% & 0.01\%  & 0.01\% & 0.05\% \\
0.55\% & 0.09\% & 0.09\%  & 0.09\% & 0.51\% \\
1.05\% & 0.18\% & 0.18\%  & 0.18\% & 0.83\% \\
1.55\% & 0.27\% & 0.27\%  & 0.27\% & 1.04\% \\
2.05\% & 0.37\% & 0.37\%  & 0.37\% & 1.18\% \\
2.55\% & 0.47\% & 0.47\%  & 0.47\% & 1.29\% \\
3.05\% & 0.59\% & 0.59\%  & 0.59\% & 1.38\% \\
3.55\% & 0.73\% & 0.73\%  & 0.73\% & 1.46\% \\
4.05\% & 0.89\% & 0.89\%  & 0.89\% & 1.57\% \\
4.30\% & 0.99\% & 0.99\%  & 0.99\% & 1.64\% \\
4.55\% & 1.10\% & 1.10\%  & 1.10\% & 1.72\% \\
4.80\% & 0.98\% & 0.98\%  & 0.98\% & 1.57\% \\
5.30\% & 0.77\% & 0.77\%  & 0.77\% & 1.35\% \\
5.80\% & 0.62\% & 0.62\%  & 0.62\% & 1.20\% \\
6.30\% & 0.51\% & 0.51\%  & 0.51\% & 1.11\% \\
6.80\% & 0.43\% & 0.43\%  & 0.43\% & 1.05\% \\
7.30\% & 0.37\% & 0.37\%  & 0.37\% & 1.01\% \\
7.80\% & 0.32\% & 0.32\%  & 0.32\% & 0.98\% \\
8.30\% & 0.28\% & 0.29\%  & 0.29\% & 0.96\% \\
8.80\% & 0.25\% & 0.25\%  & 0.26\% & 0.94\% \\
9.30\% & 0.22\% & 0.23\%  & 0.23\% & 0.92\% \\
9.80\% & 0.20\% & 0.21\%  & 0.21\% & 0.91\% \\
10.30\% & 0.18\% & 0.19\%  & 0.19\% & 0.90\% \\
\hline
\end{tabular}%
}
\end{table}
\clearpage
\subsubsection{March 9th, 2009}
As before, the ``PDE + Symmetry'' method with variable $W_t$ and $Y_t$ did not perform well for the 1Y1Y swaptions due to 
the high value(0.9) of $\beta$ and was not included in the pricing method comparison.
\begin{table}[htbp]
\centering
\caption{\label{tab:volatility.20090309}USD Swaption lognormal volatilities and the forward swap rates as of March 9th, 2009.}
\resizebox{\columnwidth}{!}{%
\begin{tabular}{|*{11}{c|}}
\hline 
Expiry/Tenor & -200bp & -100bp & -50bp & -25bp & 0bp & +25bp & +50bp & +100bp & +200bp & ATM\\
\hline 
1Y1Y &  & 58.43\% & 54.53\% & 52.59\% & 50.88\% & 49.23\% & 48.11\% & 46.26\% & 43.35\% & 2.00\% \\
5Y5Y & 37.69\% & 33.09\% & 31.42\% & 30.76\% & 30.18\% & 29.67\% & 29.24\% & 28.51\% & 27.38\% & 3.73\% \\
10Y10Y & 28.88\% & 25.59\% & 24.23\% & 23.82\% & 23.44\% & 23.08\% & 22.86\% & 22.47\% & 22.03\% & 3.51\% \\
20Y20Y & 25.30\% & 21.25\% & 20.16\% & 19.61\% & 19.12\% & 18.96\% & 18.77\% & 18.50\% & 18.36\% & 3.19\% \\
\hline
\end{tabular}%
}
\end{table}
\begin{table}[htbp]
\centering
\caption{\label{tab:SABRparameters.20090309}The SABR parameters calibrated to the USD Swaption market on March 9th, 2009.}
\begin{tabular}{|*{7}{c|}}
\hline 
Expiry/Tenor & $\beta$ & $\nu$ & $\rho$ & $\alpha_0$ & $F_0$ & $T$\\
\hline 
1Y1Y & 0.90 & 30.0\% & -75.0\% & 35.44\% & 2.00\% & 1 \\
5Y5Y & 0.40 & 15.0\% & 15.0\% & 4.15\% & 3.73\% & 5 \\
10Y10Y & 0.10 & 21.0\% & 60.0\% & 1.10\% & 3.51\% & 10 \\
20Y20Y & 0.00 & 30.0\% & 40.0\% & 0.60\% & 3.19\% & 20 \\
\hline
\end{tabular}%
\end{table}
\begin{figure}[htbp]
\centering
\includegraphics[width=1\columnwidth]{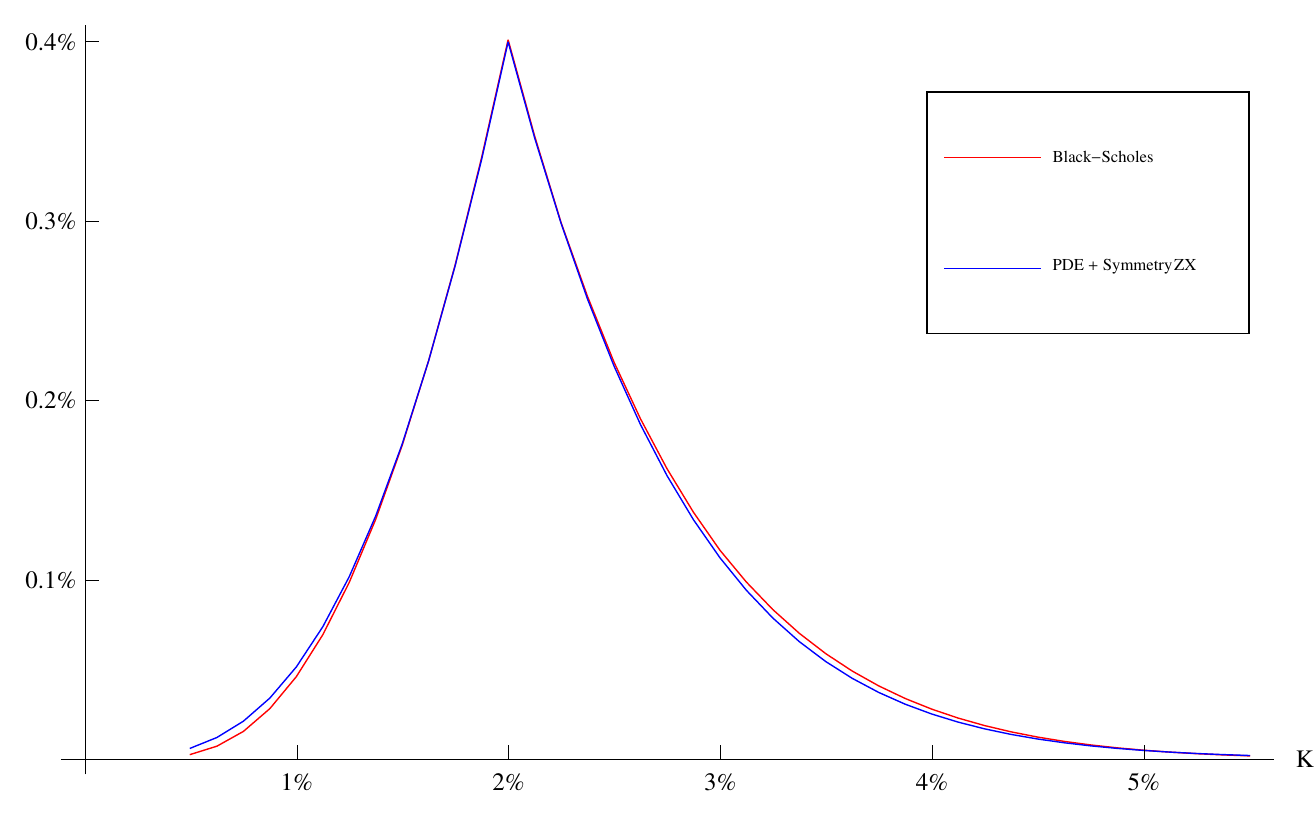}
\caption{\label{fig:USDCalibration.20090309.1Y1Y}Calibration results for USD 1Y1Y swaptions on March 9th, 2009.
The time values of swaptions expressed in the unit of 
forward swap annuity were computed and graphed
as a function of $K$. 
The red line was obtained by the Black-Scholes with linearly interpolated implied volatilities.
The blue line was computed using the ``PDE + Symmetry'' method with variables $Z_t$ and $X_t$. 
The SABR model parameters in \cref{tab:SABRparameters.20090309} were used.}  
\end{figure}
\begin{table}[htbp]
\centering
\caption{\label{tab:USDCalibration.20090309.1Y1Y}Calibration results for USD 1Y1Y swaptions on March 9th, 2009.
The time values of swaptions expressed in the unit of 
forward swap annuity were computed.
For Black-Scholes, the linearly interpolated implied volatilities were used.
For SABR, the ``PDE + Symmetry'' method with variables $Z_t$ and $X_t$ was used.
The SABR model parameters in \cref{tab:SABRparameters.20090309} were used.}
\begin{tabular}{|c|c|c|}
\hline 
Strike & Black-Scholes & SABR \\
\hline
1.00\% & 0.05\% & 0.05\%  \\
1.50\% & 0.18\% & 0.18\%  \\
1.75\% & 0.28\% & 0.28\%  \\
2.00\% & 0.40\% & 0.40\%  \\
2.25\% & 0.30\% & 0.30\%  \\
2.50\% & 0.22\% & 0.22\%  \\
3.00\% & 0.12\% & 0.11\%  \\
4.00\% & 0.03\% & 0.03\%  \\
\hline
\end{tabular}
\end{table} 
\begin{figure}[htbp]
\centering
\includegraphics[width=1\columnwidth]{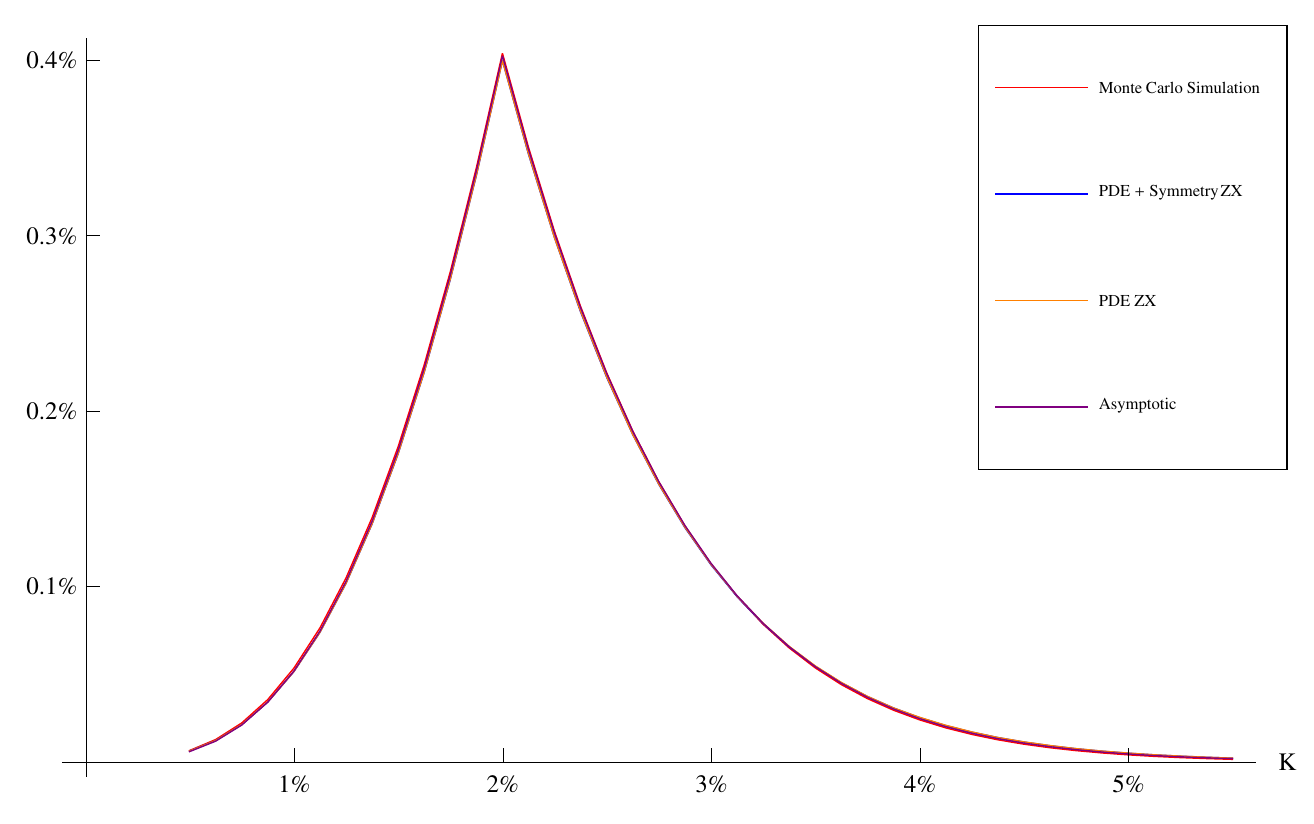}
\caption{\label{fig:comparison.20090309.1Y1Y}Pricing method comparison for USD 1Y1Y swaptions on March 9th, 2009. 
The time values of swaptions expressed in the unit of 
forward swap annuity were computed using various pricing methods.
The SABR model parameters in \cref{tab:SABRparameters.20090309} were used.}
\end{figure}
\begin{table}[htbp]
\centering
\caption{\label{tab:comparison.20090309.1Y1Y}Pricing method comparison for USD 1Y1Y swaptions on March 9th, 2009. 
The time values of swaptions expressed in the unit of 
forward swap annuity were computed using various pricing methods.
The SABR model parameters in \cref{tab:SABRparameters.20090309} were used.}
\resizebox{\columnwidth}{!}{%
\begin{tabular}{|*{5}{c|}}
\hline 
Strike & Monte Carlo simulation  & PDE + Symmetry ZX & PDE ZX & Asymptotic\\
\hline 
0.50\% & 0.01\% & 0.01\%  & 0.01\% & 0.01\% \\
1.00\% & 0.05\% & 0.05\%  & 0.05\% & 0.05\% \\
1.50\% & 0.18\% & 0.18\%  & 0.18\% & 0.18\% \\
1.75\% & 0.28\% & 0.28\%  & 0.28\% & 0.28\% \\
2.00\% & 0.40\% & 0.40\%  & 0.40\% & 0.40\% \\
2.25\% & 0.30\% & 0.30\%  & 0.30\% & 0.30\% \\
2.50\% & 0.22\% & 0.22\%  & 0.22\% & 0.22\% \\
3.00\% & 0.11\% & 0.11\%  & 0.11\% & 0.11\% \\
3.50\% & 0.05\% & 0.05\%  & 0.05\% & 0.05\% \\
4.00\% & 0.02\% & 0.03\%  & 0.03\% & 0.02\% \\
4.50\% & 0.01\% & 0.01\%  & 0.01\% & 0.01\% \\
5.00\% & 0.00\% & 0.00\%  & 0.00\% & 0.00\% \\
\hline
\end{tabular}%
}
\end{table}
\clearpage
\begin{figure}[htbp]
\centering
\includegraphics[width=1\columnwidth]{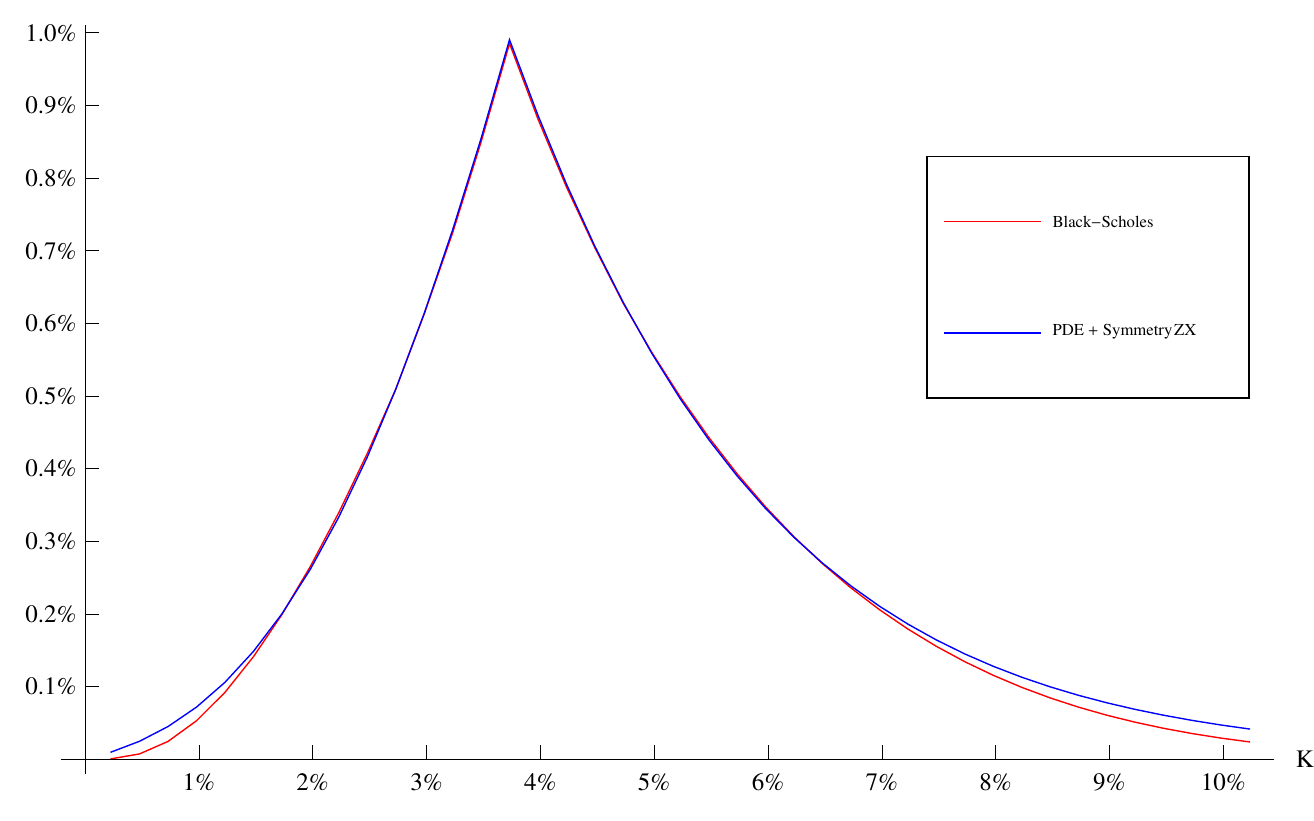}
\caption{\label{fig:USDCalibration.20090309.5Y5Y}Calibration results for USD 5Y5Y swaptions on March 9th, 2009.
The time values of swaptions expressed in the unit of 
forward swap annuity were computed and graphed
as a function of $K$. 
The red line was obtained by the Black-Scholes with linearly interpolated implied volatilities.
The blue line was computed using the ``PDE + Symmetry'' method with variables $Z_t$ and $X_t$. 
The SABR model parameters in \cref{tab:SABRparameters.20090309} were used.}  
\end{figure}
\begin{table}[htbp]
\centering
\caption{\label{tab:USDCalibration.20090309.5Y5Y}Calibration results for USD 5Y5Y swaptions on March 9th, 2009.
The time values of swaptions expressed in the unit of 
forward swap annuity were computed.
For Black-Scholes, the linearly interpolated implied volatilities were used.
For SABR, the ``PDE + Symmetry'' method with variables $Z_t$ and $X_t$ was used.
The SABR model parameters in \cref{tab:SABRparameters.20090309} were used.}
\begin{tabular}{|c|c|c|}
\hline 
Strike & Black-Scholes & SABR \\
\hline
1.73\% & 0.20\% & 0.20\%  \\
2.73\% & 0.51\% & 0.51\%  \\
3.23\% & 0.72\% & 0.73\%  \\
3.48\% & 0.85\% & 0.85\%  \\
3.73\% & 0.99\% & 0.99\%  \\
3.98\% & 0.88\% & 0.89\%  \\
4.23\% & 0.79\% & 0.79\%  \\
4.73\% & 0.63\% & 0.63\%  \\
5.73\% & 0.39\% & 0.39\%  \\
\hline
\end{tabular}
\end{table} 
\begin{figure}[htbp]
\centering
\includegraphics[width=1\columnwidth]{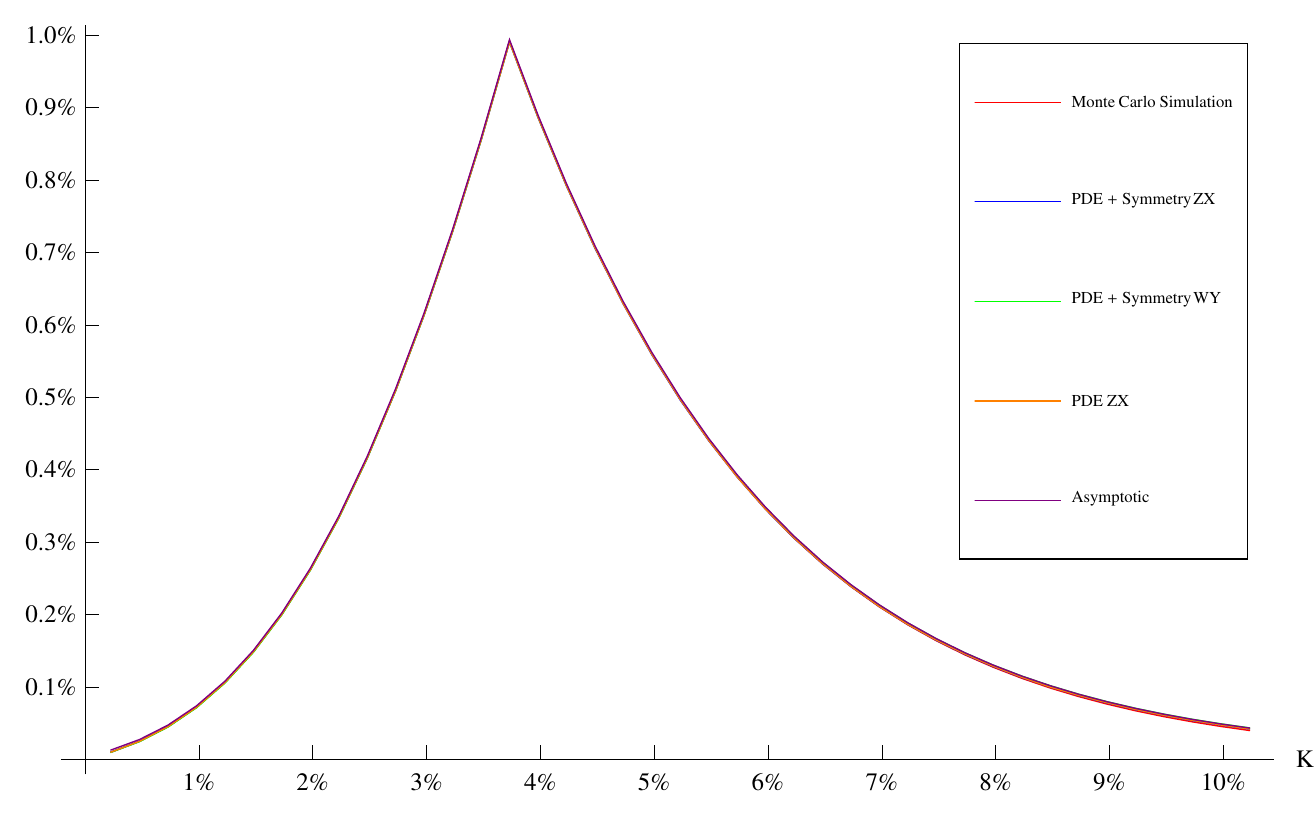}
\caption{\label{fig:comparison.20090309.5Y5Y}Pricing method comparison for USD 5Y5Y swaptions on March 9th, 2009. 
The time values of swaptions expressed in the unit of 
forward swap annuity were computed using various pricing methods.
The SABR model parameters in \cref{tab:SABRparameters.20090309} were used.}
\end{figure}
\begin{table}[htbp]
\centering
\caption{\label{tab:comparison.20090309.5Y5Y}Pricing method comparison for USD 5Y5Y swaptions on March 9th, 2009. 
The time values of swaptions expressed in the unit of 
forward swap annuity were computed using various pricing methods.
The SABR model parameters in \cref{tab:SABRparameters.20090309} were used.}
\resizebox{\columnwidth}{!}{%
\begin{tabular}{|*{6}{c|}}
\hline 
Strike & Monte Carlo simulation  & PDE + Symmetry ZX & PDE + Symmetry WY & PDE ZX & Asymptotic\\
\hline 
0.23\% & 0.01\% & 0.01\%  & 0.01\% & 0.01\% & 0.01\% \\
0.73\% & 0.05\% & 0.05\%  & 0.04\% & 0.05\% & 0.05\% \\
1.23\% & 0.11\% & 0.11\%  & 0.11\% & 0.11\% & 0.11\% \\
1.73\% & 0.20\% & 0.20\%  & 0.20\% & 0.20\% & 0.20\% \\
2.23\% & 0.33\% & 0.33\%  & 0.33\% & 0.33\% & 0.34\% \\
2.73\% & 0.51\% & 0.51\%  & 0.51\% & 0.51\% & 0.51\% \\
3.23\% & 0.73\% & 0.73\%  & 0.73\% & 0.73\% & 0.73\% \\
3.48\% & 0.86\% & 0.85\%  & 0.85\% & 0.85\% & 0.86\% \\
3.73\% & 0.99\% & 0.99\%  & 0.99\% & 0.99\% & 0.99\% \\
3.98\% & 0.89\% & 0.89\%  & 0.89\% & 0.89\% & 0.89\% \\
4.23\% & 0.79\% & 0.79\%  & 0.79\% & 0.79\% & 0.80\% \\
4.73\% & 0.63\% & 0.63\%  & 0.63\% & 0.63\% & 0.63\% \\
5.23\% & 0.50\% & 0.50\%  & 0.50\% & 0.50\% & 0.50\% \\
5.73\% & 0.39\% & 0.39\%  & 0.39\% & 0.39\% & 0.39\% \\
6.23\% & 0.31\% & 0.31\%  & 0.31\% & 0.31\% & 0.31\% \\
6.73\% & 0.24\% & 0.24\%  & 0.24\% & 0.24\% & 0.24\% \\
7.23\% & 0.19\% & 0.19\%  & 0.19\% & 0.19\% & 0.19\% \\
7.73\% & 0.14\% & 0.14\%  & 0.15\% & 0.15\% & 0.15\% \\
8.23\% & 0.11\% & 0.11\%  & 0.11\% & 0.11\% & 0.11\% \\
8.73\% & 0.09\% & 0.09\%  & 0.09\% & 0.09\% & 0.09\% \\
9.23\% & 0.07\% & 0.07\%  & 0.07\% & 0.07\% & 0.07\% \\
9.73\% & 0.05\% & 0.05\%  & 0.06\% & 0.05\% & 0.06\% \\
10.23\% & 0.04\% & 0.04\%  & 0.04\% & 0.04\% & 0.04\% \\
\hline
\end{tabular}%
}
\end{table}
\clearpage
\begin{figure}[htbp]
\centering
\includegraphics[width=1\columnwidth]{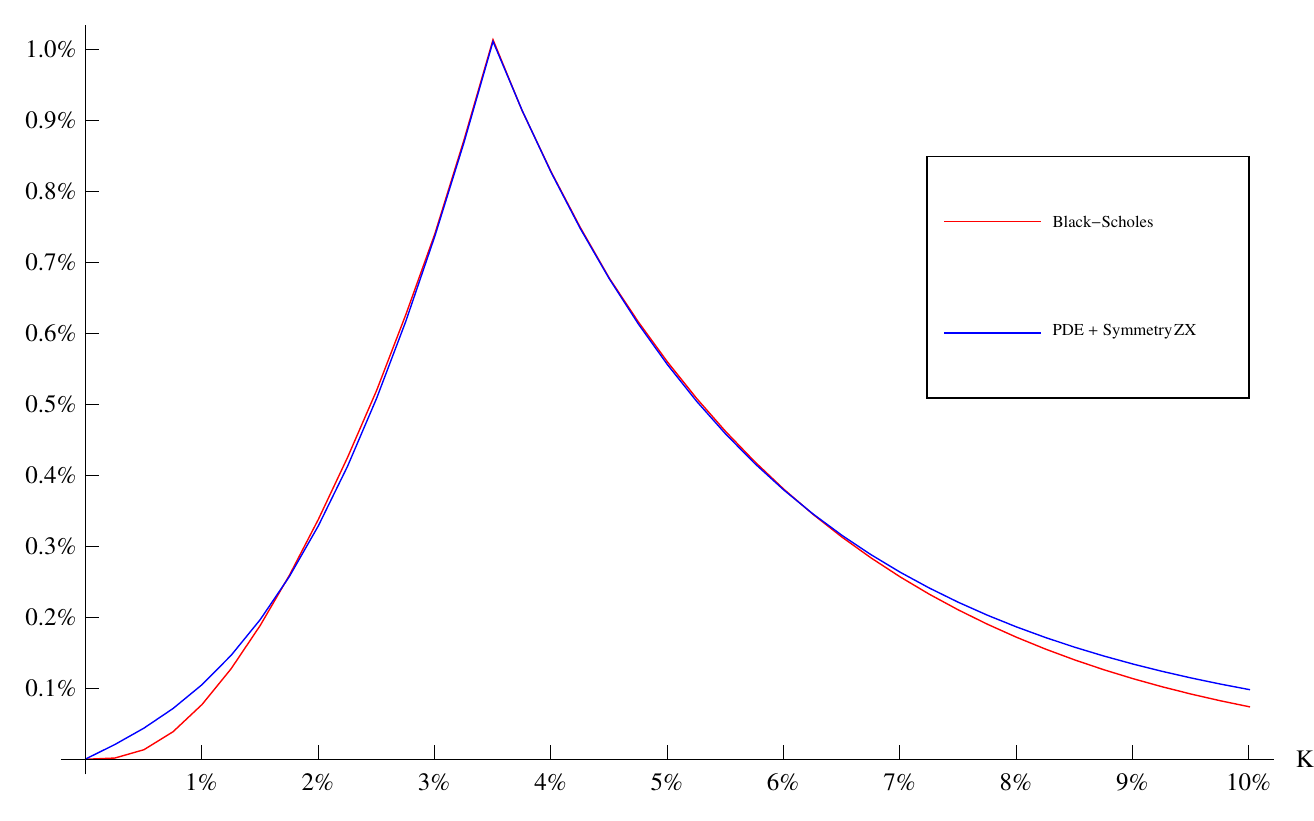}
\caption{\label{fig:USDCalibration.20090309.10Y10Y}Calibration results for USD 10Y10Y swaptions on March 9th, 2009.
The time values of swaptions expressed in the unit of 
forward swap annuity were computed and graphed
as a function of $K$. 
The red line was obtained by the Black-Scholes with linearly interpolated implied volatilities.
The blue line was computed using the ``PDE + Symmetry'' method with variables $Z_t$ and $X_t$. 
The SABR model parameters in \cref{tab:SABRparameters.20090309} were used.}  
\end{figure}
\begin{table}[htbp]
\centering
\caption{\label{tab:USDCalibration.20090309.10Y10Y}Calibration results for USD 10Y10Y swaptions on March 9th, 2009.
The time values of swaptions expressed in the unit of 
forward swap annuity were computed.
For Black-Scholes, the linearly interpolated implied volatilities were used.
For SABR, the ``PDE + Symmetry'' method with variables $Z_t$ and $X_t$ was used.
The SABR model parameters in \cref{tab:SABRparameters.20090309} were used.}
\begin{tabular}{|c|c|c|}
\hline 
Strike & Black-Scholes & SABR \\
\hline
1.51\% & 0.19\% & 0.20\%  \\
2.51\% & 0.52\% & 0.51\%  \\
3.01\% & 0.74\% & 0.74\%  \\
3.26\% & 0.87\% & 0.87\%  \\
3.51\% & 1.01\% & 1.01\%  \\
3.76\% & 0.91\% & 0.91\%  \\
4.01\% & 0.83\% & 0.83\%  \\
4.51\% & 0.68\% & 0.68\%  \\
5.51\% & 0.46\% & 0.46\%  \\
\hline
\end{tabular}
\end{table} 
\begin{figure}[htbp]
\centering
\includegraphics[width=1\columnwidth]{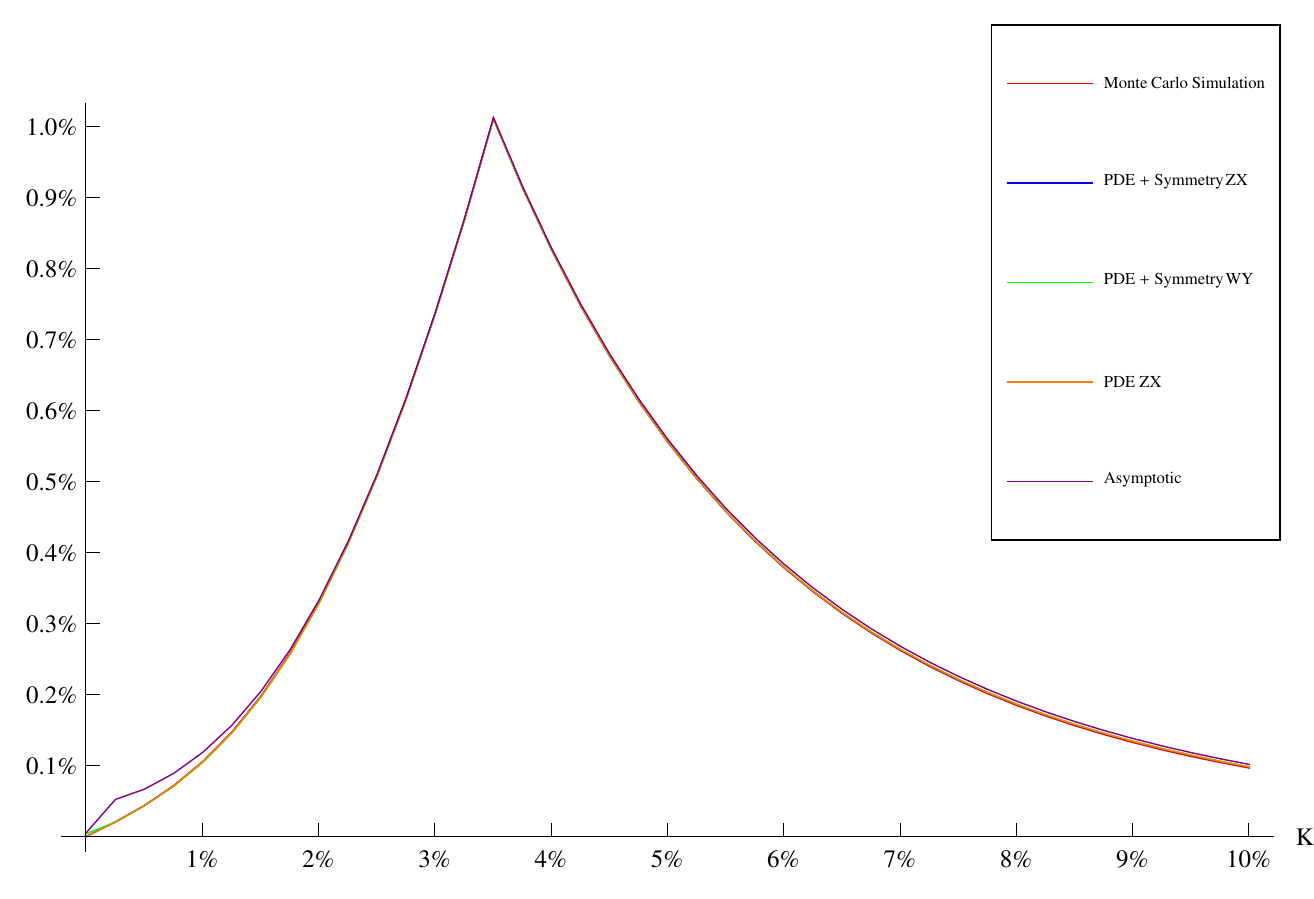}
\caption{\label{fig:comparison.20090309.10Y10Y}Pricing method comparison for USD 10Y10Y swaptions on March 9th, 2009. 
The time values of swaptions expressed in the unit of 
forward swap annuity were computed using various pricing methods.
The SABR model parameters in \cref{tab:SABRparameters.20090309} were used.}
\end{figure}
\begin{table}[htbp]
\centering
\caption{\label{tab:comparison.20090309.10Y10Y}Pricing method comparison for USD 10Y10Y swaptions on March 9th, 2009. 
The time values of swaptions expressed in the unit of 
forward swap annuity were computed using various pricing methods.
The SABR model parameters in \cref{tab:SABRparameters.20090309} were used.}
\resizebox{\columnwidth}{!}{%
\begin{tabular}{|*{6}{c|}}
\hline 
Strike & Monte Carlo simulation  & PDE + Symmetry ZX & PDE + Symmetry WY & PDE ZX & Asymptotic\\
\hline 
0.01\% & 0.00\% & 0.00\%  & 0.00\% & 0.00\% & 0.01\% \\
0.51\% & 0.04\% & 0.04\%  & 0.04\% & 0.04\% & 0.07\% \\
1.01\% & 0.11\% & 0.11\%  & 0.10\% & 0.11\% & 0.12\% \\
1.51\% & 0.20\% & 0.20\%  & 0.20\% & 0.20\% & 0.20\% \\
2.01\% & 0.33\% & 0.33\%  & 0.33\% & 0.33\% & 0.33\% \\
2.51\% & 0.51\% & 0.51\%  & 0.51\% & 0.51\% & 0.51\% \\
3.01\% & 0.74\% & 0.74\%  & 0.74\% & 0.74\% & 0.74\% \\
3.26\% & 0.87\% & 0.87\%  & 0.87\% & 0.87\% & 0.87\% \\
3.51\% & 1.01\% & 1.01\%  & 1.01\% & 1.01\% & 1.01\% \\
3.76\% & 0.91\% & 0.91\%  & 0.91\% & 0.91\% & 0.92\% \\
4.01\% & 0.83\% & 0.83\%  & 0.83\% & 0.83\% & 0.83\% \\
4.51\% & 0.68\% & 0.68\%  & 0.68\% & 0.68\% & 0.68\% \\
5.01\% & 0.56\% & 0.56\%  & 0.56\% & 0.56\% & 0.56\% \\
5.51\% & 0.46\% & 0.46\%  & 0.46\% & 0.46\% & 0.46\% \\
6.01\% & 0.38\% & 0.38\%  & 0.38\% & 0.38\% & 0.38\% \\
6.51\% & 0.31\% & 0.32\%  & 0.32\% & 0.32\% & 0.32\% \\
7.01\% & 0.26\% & 0.26\%  & 0.26\% & 0.26\% & 0.27\% \\
7.51\% & 0.22\% & 0.22\%  & 0.22\% & 0.22\% & 0.23\% \\
8.01\% & 0.18\% & 0.19\%  & 0.19\% & 0.19\% & 0.19\% \\
8.51\% & 0.16\% & 0.16\%  & 0.16\% & 0.16\% & 0.16\% \\
9.01\% & 0.13\% & 0.13\%  & 0.13\% & 0.13\% & 0.14\% \\
9.51\% & 0.11\% & 0.11\%  & 0.11\% & 0.11\% & 0.12\% \\
10.01\% & 0.10\% & 0.10\%  & 0.10\% & 0.10\% & 0.10\% \\
\hline
\end{tabular}%
}
\end{table}
\clearpage
\begin{figure}[htbp]
\centering
\includegraphics[width=1\columnwidth]{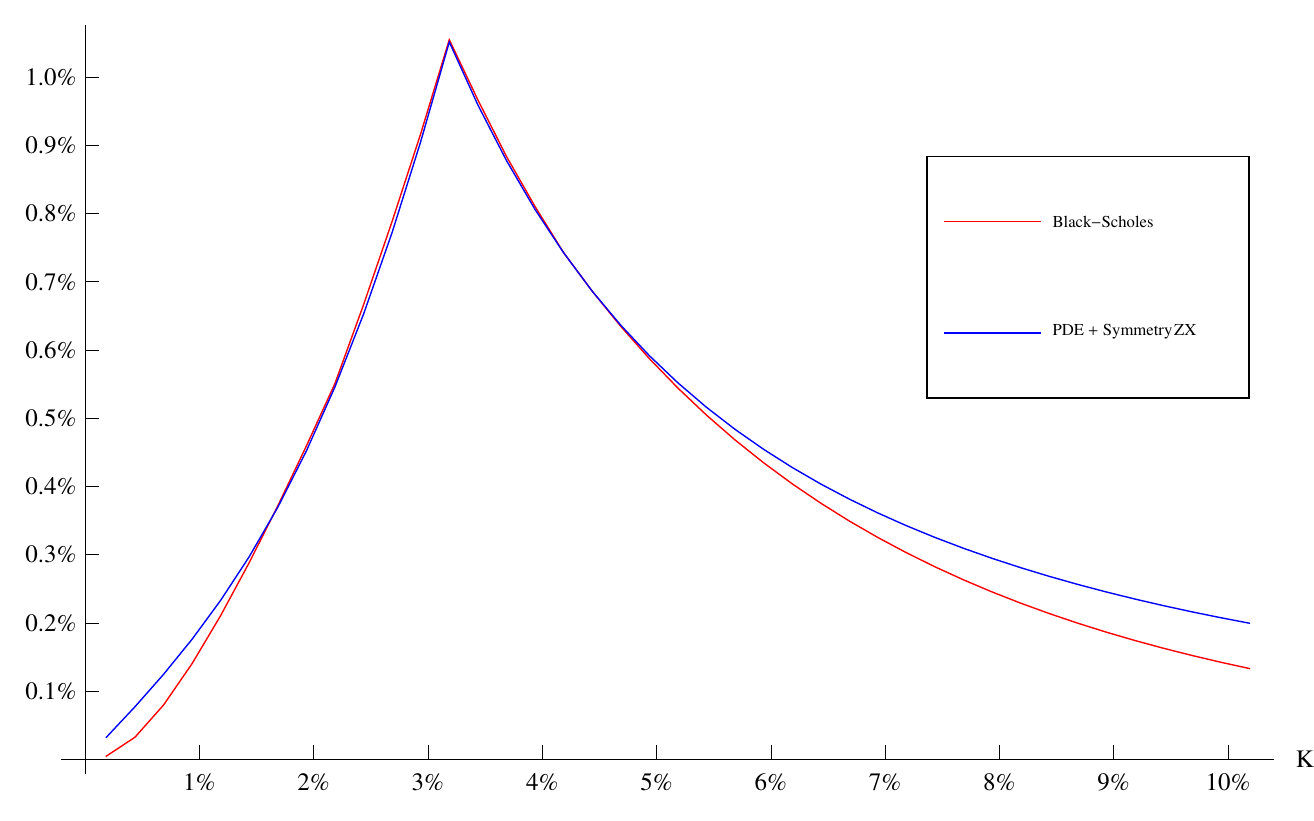}
\caption{\label{fig:USDCalibration.20090309.20Y20Y}Calibration results for USD 20Y20Y swaptions on March 9th, 2009.
The time values of swaptions expressed in the unit of 
forward swap annuity were computed and graphed
as a function of $K$. 
The red line was obtained by the Black-Scholes with linearly interpolated implied volatilities.
The blue line was computed using the ``PDE + Symmetry'' method with variables $Z_t$ and $X_t$. 
The SABR model parameters in \cref{tab:SABRparameters.20090309} were used.}  
\end{figure}
\begin{table}[htbp]
\centering
\caption{\label{tab:USDCalibration.20090309.20Y20Y}Calibration results for USD 20Y20Y swaptions on March 9th, 2009.
The time values of swaptions expressed in the unit of 
forward swap annuity were computed.
For Black-Scholes, the linearly interpolated implied volatilities were used.
For SABR, the ``PDE + Symmetry'' method with variables $Z_t$ and $X_t$ was used.
The SABR model parameters in \cref{tab:SABRparameters.20090309} were used.}
\begin{tabular}{|c|c|c|}
\hline 
Strike & Black-Scholes & SABR \\
\hline
1.19\% & 0.21\% & 0.23\%  \\
2.19\% & 0.55\% & 0.55\%  \\
2.69\% & 0.79\% & 0.77\%  \\
2.94\% & 0.92\% & 0.91\%  \\
3.19\% & 1.05\% & 1.05\%  \\
3.44\% & 0.97\% & 0.96\%  \\
3.69\% & 0.88\% & 0.88\%  \\
4.19\% & 0.74\% & 0.74\%  \\
5.19\% & 0.54\% & 0.55\%  \\
\hline
\end{tabular}
\end{table} 
\begin{figure}[htbp]
\centering
\includegraphics[width=1\columnwidth]{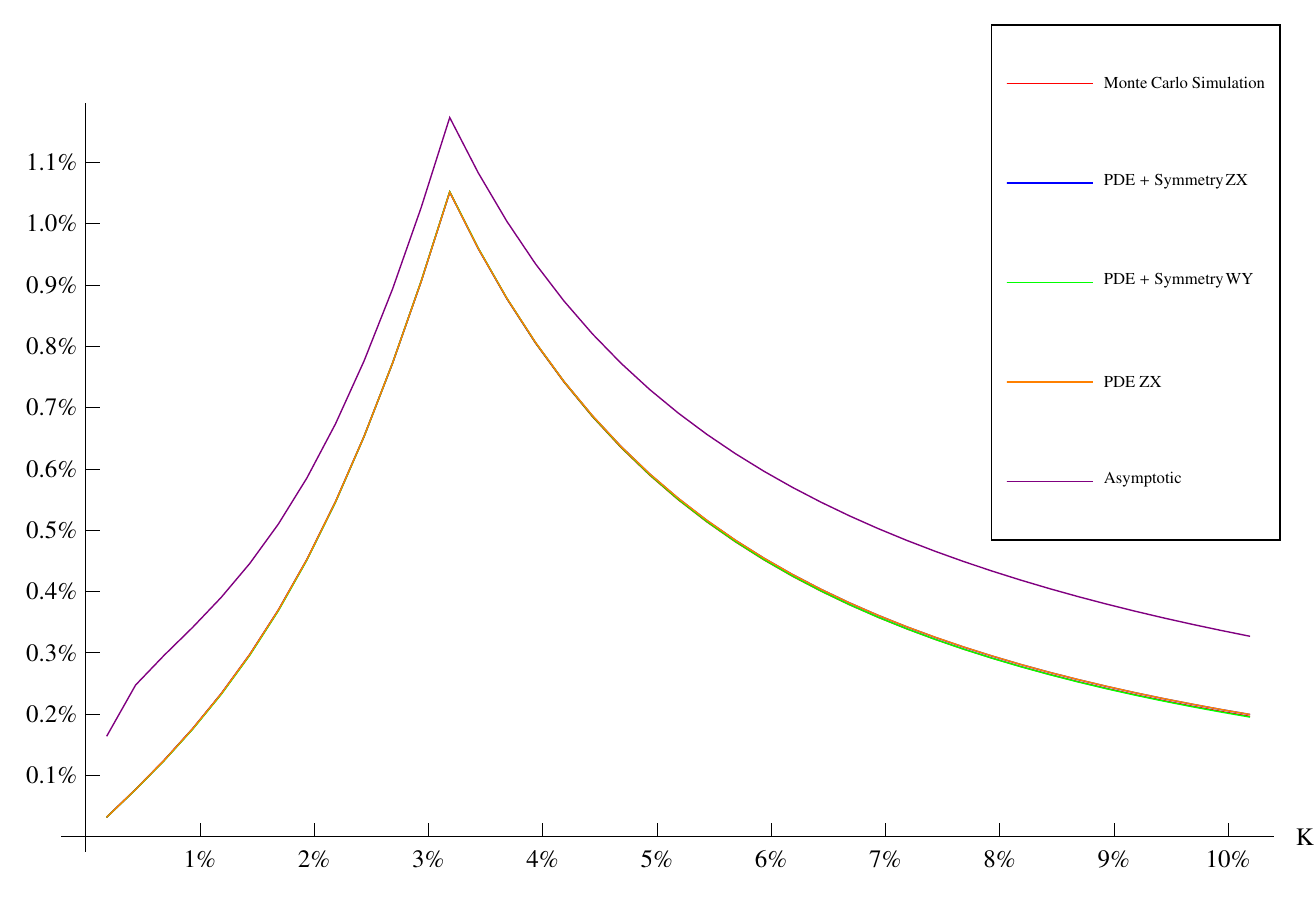}
\caption{\label{fig:comparison.20090309.20Y20Y}Pricing method comparison for USD 20Y20Y swaptions on March 9th, 2009. 
The time values of swaptions expressed in the unit of 
forward swap annuity were computed using various pricing methods.
The SABR model parameters in \cref{tab:SABRparameters.20090309} were used.}
\end{figure}
\begin{table}[htbp]
\centering
\caption{\label{tab:comparison.20090309.20Y20Y}Pricing method comparison for USD 20Y20Y swaptions on March 9th, 2009. 
The time values of swaptions expressed in the unit of 
forward swap annuity were computed using various pricing methods.
The SABR model parameters in \cref{tab:SABRparameters.20090309} were used.}
\resizebox{\columnwidth}{!}{%
\begin{tabular}{|*{6}{c|}}
\hline 
Strike & Monte Carlo simulation  & PDE + Symmetry ZX & PDE + Symmetry WY & PDE ZX & Asymptotic\\
\hline 
0.19\% & 0.03\% & 0.03\%  & 0.03\% & 0.03\% & 0.16\% \\
0.69\% & 0.12\% & 0.12\%  & 0.12\% & 0.12\% & 0.30\% \\
1.19\% & 0.23\% & 0.23\%  & 0.23\% & 0.23\% & 0.39\% \\
1.69\% & 0.37\% & 0.37\%  & 0.37\% & 0.37\% & 0.51\% \\
2.19\% & 0.55\% & 0.55\%  & 0.55\% & 0.55\% & 0.67\% \\
2.69\% & 0.77\% & 0.77\%  & 0.77\% & 0.77\% & 0.89\% \\
2.94\% & 0.91\% & 0.91\%  & 0.91\% & 0.91\% & 1.03\% \\
3.19\% & 1.05\% & 1.05\%  & 1.05\% & 1.05\% & 1.17\% \\
3.44\% & 0.96\% & 0.96\%  & 0.96\% & 0.96\% & 1.08\% \\
3.69\% & 0.88\% & 0.88\%  & 0.88\% & 0.88\% & 1.00\% \\
4.19\% & 0.74\% & 0.74\%  & 0.74\% & 0.74\% & 0.87\% \\
4.69\% & 0.63\% & 0.64\%  & 0.64\% & 0.64\% & 0.77\% \\
5.19\% & 0.55\% & 0.55\%  & 0.55\% & 0.55\% & 0.69\% \\
5.69\% & 0.48\% & 0.48\%  & 0.48\% & 0.48\% & 0.62\% \\
6.19\% & 0.42\% & 0.43\%  & 0.43\% & 0.43\% & 0.57\% \\
6.69\% & 0.38\% & 0.38\%  & 0.38\% & 0.38\% & 0.52\% \\
7.19\% & 0.34\% & 0.34\%  & 0.34\% & 0.34\% & 0.48\% \\
7.69\% & 0.31\% & 0.31\%  & 0.31\% & 0.31\% & 0.45\% \\
8.19\% & 0.28\% & 0.28\%  & 0.28\% & 0.28\% & 0.42\% \\
8.69\% & 0.25\% & 0.26\%  & 0.25\% & 0.26\% & 0.39\% \\
9.19\% & 0.23\% & 0.23\%  & 0.23\% & 0.23\% & 0.37\% \\
9.69\% & 0.21\% & 0.22\%  & 0.21\% & 0.22\% & 0.35\% \\
10.19\% & 0.20\% & 0.20\%  & 0.20\% & 0.20\% & 0.33\% \\
\hline
\end{tabular}%
}
\end{table}
\clearpage
\subsubsection{October 29th, 2013}
Again, the ``PDE + Symmetry'' method with variable $W_t$ and $Y_t$ did not perform well for the 1Y1Y swaptions due to 
the high value(0.9) of $\beta$ and the method was not included in the pricing method comparison.
\begin{table}[htbp]
\centering
\caption{\label{tab:volatility.20131029}USD Swaption lognormal volatilities and the forward swap rates as of October 29th, 2013.}
\resizebox{\columnwidth}{!}{%
\begin{tabular}{|*{11}{c|}}
\hline 
Expiry/Tenor & -200bp & -100bp & -50bp & -25bp & 0bp & +25bp & +50bp & +100bp & +200bp & ATM\\
\hline 
1Y1Y &  &  & 65.43\% & 65.43\% & 65.43\% & 65.42\% & 65.41\% & 65.39\% & 65.38\% & .56\% \\
5Y5Y & 27.31\% & 25.55\% & 24.64\% & 24.33\% & 24.03\% & 23.78\% & 23.54\% & 23.20\% & 22.69\% & 4.07\% \\
10Y10Y & 20.01\% & 18.35\% & 17.80\% & 17.66\% & 17.55\% & 17.44\% & 17.33\% & 17.34\% & 17.51\% & 4.50\% \\
20Y20Y & 13.80\% & 13.10\% & 12.74\% & 12.68\% & 12.62\% & 12.55\% & 12.49\% & 12.43\% & 12.58\% & 4.01\% \\
\hline
\end{tabular}%
}
\end{table}
\begin{table}[htbp]
\centering
\caption{\label{tab:SABRparameters.20131029}The SABR parameters calibrated to the USD Swaption market on October 29th, 2013.}
\begin{tabular}{|*{7}{c|}}
\hline 
Expiry/Tenor & $\beta$ & $\nu$ & $\rho$ & $\alpha_0$ & $F_0$ & $T$\\
\hline 
1Y1Y & 0.90 & 30.0\% & 5.0\% & 38.98\% & 0.56\% & 1 \\
5Y5Y & 0.40 & 10.0\% & 60.0\% & 3.48\% & 4.07\% & 5 \\
10Y10Y & 0.10 & 23.0\% & 55.5\% & 1.03\% & 4.50\% & 10 \\
20Y20Y & 0.00 & 13.0\% & 85.0\% & 0.50\% & 4.01\% & 20 \\
\hline
\end{tabular}%
\end{table}
\begin{figure}[htbp]
\centering
\includegraphics[width=1\columnwidth]{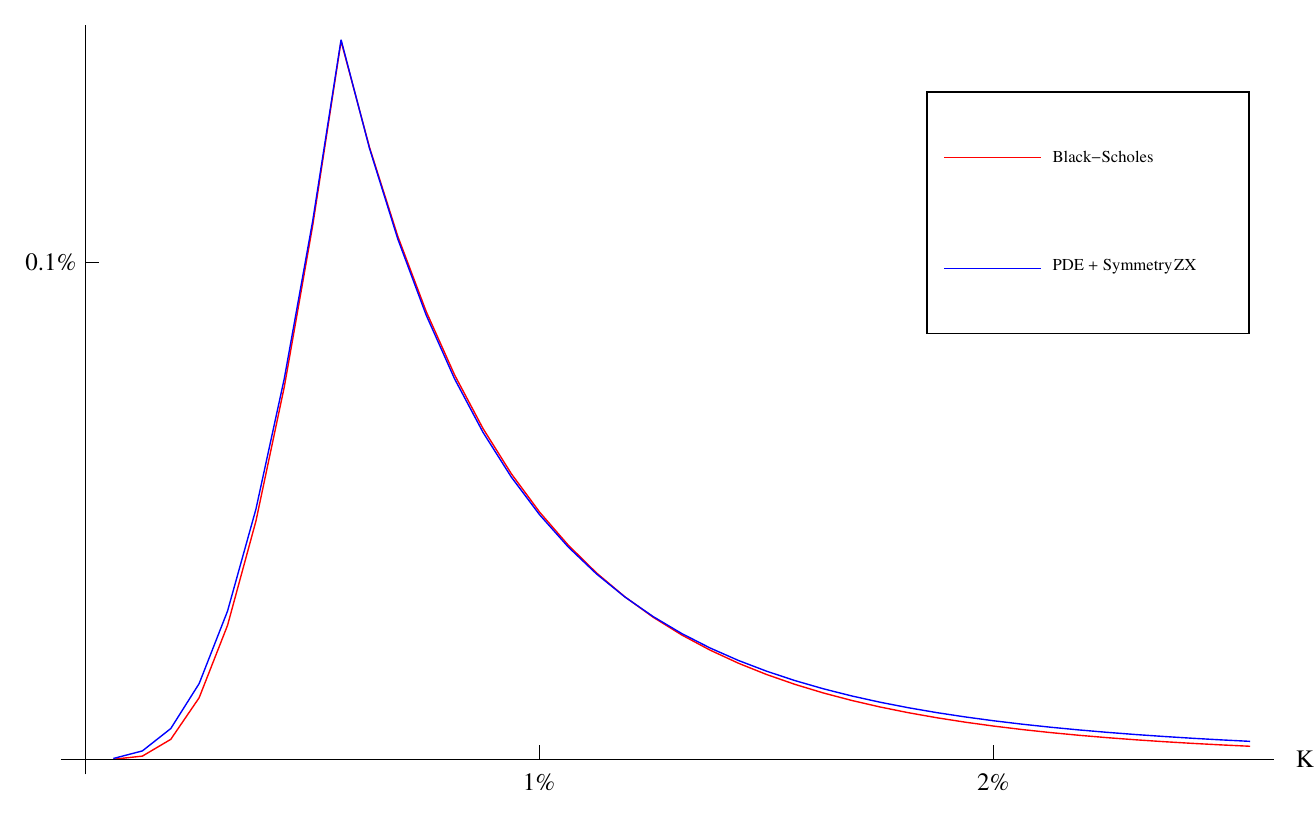}
\caption{\label{fig:USDCalibration.20131029.1Y1Y}Calibration results for USD 1Y1Y swaptions on October 29th, 2013.
The time values of swaptions expressed in the unit of 
forward swap annuity were computed and graphed
as a function of $K$. 
The red line was obtained by the Black-Scholes with linearly interpolated implied volatilities.
The blue line was computed using the ``PDE + Symmetry'' method with variables $Z_t$ and $X_t$. 
The SABR model parameters in \cref{tab:SABRparameters.20131029} were used.}  
\end{figure}
\begin{table}[htbp]
\centering
\caption{\label{tab:USDCalibration.20131029.1Y1Y}Calibration results for USD 1Y1Y swaptions on October 29th, 2013.
The time values of swaptions expressed in the unit of 
forward swap annuity were computed.
For Black-Scholes, the linearly interpolated implied volatilities were used.
For SABR, the ``PDE + Symmetry'' method with variables $Z_t$ and $X_t$ was used.
The SABR model parameters in \cref{tab:SABRparameters.20131029} were used.}
\begin{tabular}{|c|c|c|}
\hline 
Strike & Black-Scholes & SABR \\
\hline
0.06\% & 0.00\% & 0.00\%  \\
0.31\% & 0.03\% & 0.03\%  \\
0.56\% & 0.14\% & 0.14\%  \\
0.81\% & 0.08\% & 0.08\%  \\
1.06\% & 0.04\% & 0.04\%  \\
1.56\% & 0.02\% & 0.02\%  \\
2.56\% & 0.00\% & 0.00\%  \\
\hline
\end{tabular}
\end{table} 
\begin{figure}[htbp]
\centering
\includegraphics[width=1\columnwidth]{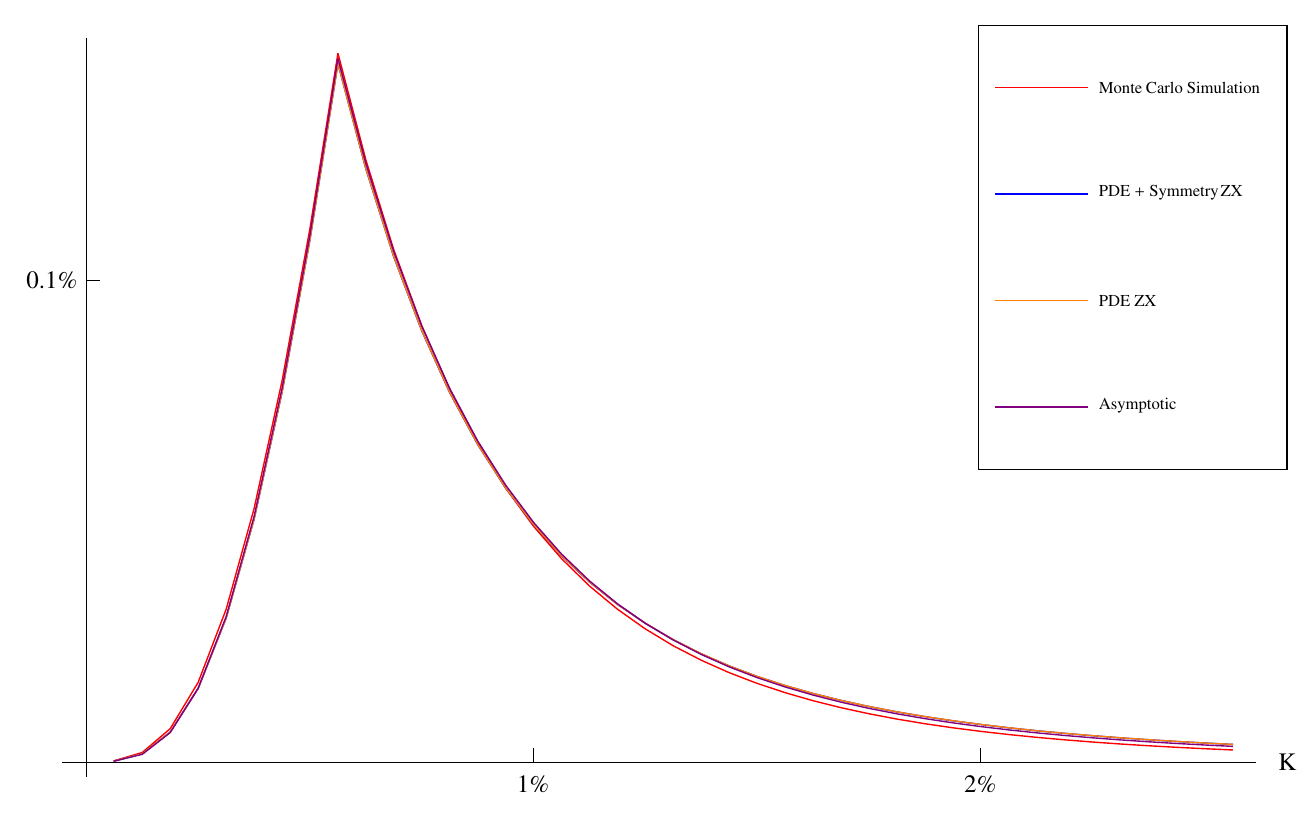}
\caption{\label{fig:comparison.20131029.1Y1Y}Pricing method comparison for USD 1Y1Y swaptions on October 29th, 2013. 
The time values of swaptions expressed in the unit of 
forward swap annuity were computed using various pricing methods.
The SABR model parameters in \cref{tab:SABRparameters.20131029} were used.}
\end{figure}
\begin{table}[htbp]
\centering
\caption{\label{tab:comparison.20131029.1Y1Y}Pricing method comparison for USD 1Y1Y swaptions on October 29th, 2013. 
The time values of swaptions expressed in the unit of 
forward swap annuity were computed using various pricing methods.
The SABR model parameters in \cref{tab:SABRparameters.20131029} were used.}
\resizebox{\columnwidth}{!}{%
\begin{tabular}{|*{5}{c|}}
\hline 
Strike & Monte Carlo simulation  & PDE + Symmetry ZX & PDE ZX & Asymptotic\\
\hline 
0.06\% & 0.00\% & 0.00\%  & 0.00\% & 0.00\% \\
0.31\% & 0.03\% & 0.03\%  & 0.03\% & 0.03\% \\
0.56\% & 0.15\% & 0.14\%  & 0.14\% & 0.15\% \\
0.81\% & 0.08\% & 0.08\%  & 0.08\% & 0.08\% \\
1.06\% & 0.04\% & 0.04\%  & 0.04\% & 0.04\% \\
1.56\% & 0.01\% & 0.02\%  & 0.02\% & 0.02\% \\
2.06\% & 0.01\% & 0.01\%  & 0.01\% & 0.01\% \\
2.56\% & 0.00\% & 0.00\%  & 0.00\% & 0.00\% \\
\hline
\end{tabular}%
}
\end{table}
\clearpage
\begin{figure}[htbp]
\centering
\includegraphics[width=1\columnwidth]{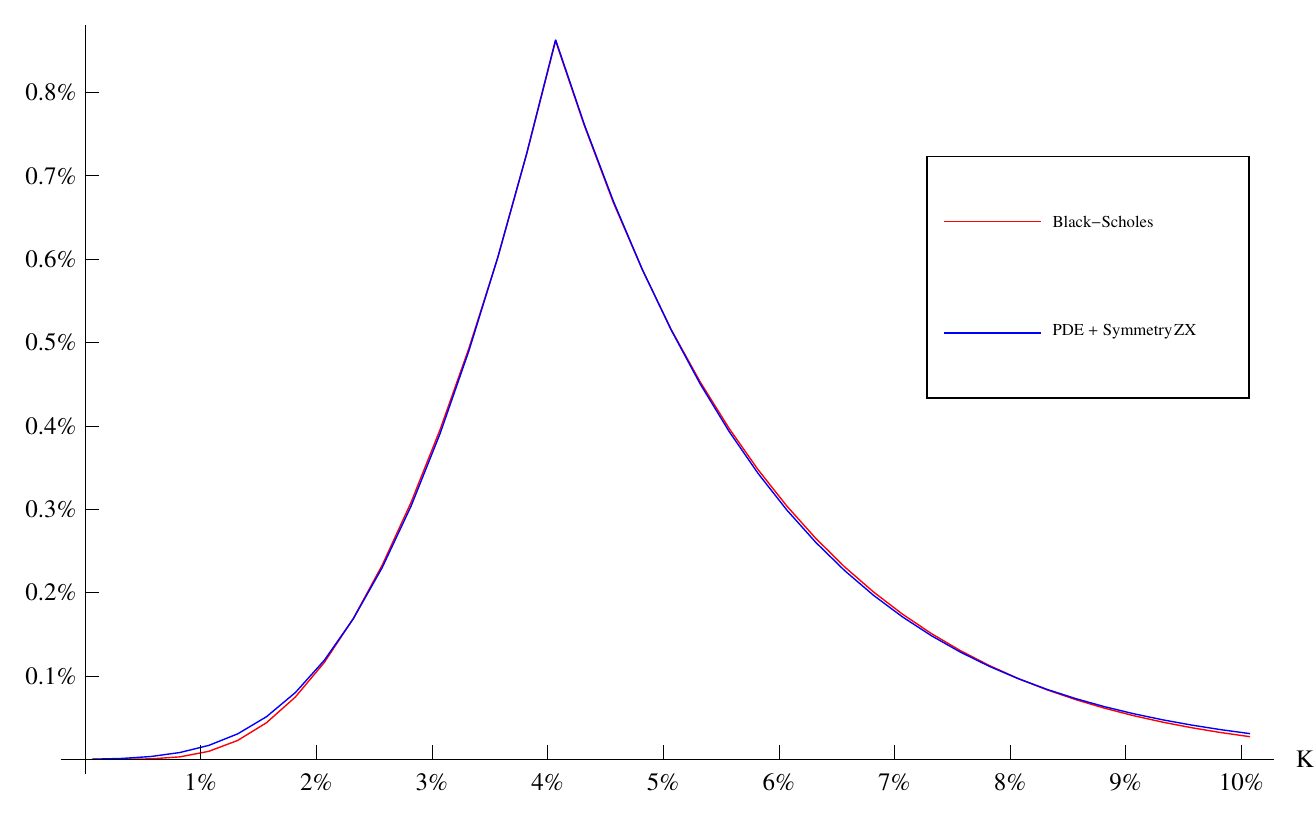}
\caption{\label{fig:USDCalibration.20131029.5Y5Y}Calibration results for USD 5Y5Y swaptions on October 29th, 2013.
The time values of swaptions expressed in the unit of 
forward swap annuity were computed and graphed
as a function of $K$. 
The red line was obtained by the Black-Scholes with linearly interpolated implied volatilities.
The blue line was computed using the ``PDE + Symmetry'' method with variables $Z_t$ and $X_t$. 
The SABR model parameters in \cref{tab:SABRparameters.20131029} were used.}  
\end{figure}
\begin{table}[htbp]
\centering
\caption{\label{tab:USDCalibration.20131029.5Y5Y}Calibration results for USD 5Y5Y swaptions on October 29th, 2013.
The time values of swaptions expressed in the unit of 
forward swap annuity were computed.
For Black-Scholes, the linearly interpolated implied volatilities were used.
For SABR, the ``PDE + Symmetry'' method with variables $Z_t$ and $X_t$ was used.
The SABR model parameters in \cref{tab:SABRparameters.20131029} were used.}
\begin{tabular}{|c|c|c|}
\hline 
Strike & Black-Scholes & SABR \\
\hline
2.07\% & 0.12\% & 0.12\%  \\
3.07\% & 0.40\% & 0.39\%  \\
3.57\% & 0.60\% & 0.60\%  \\
3.82\% & 0.73\% & 0.73\%  \\
4.07\% & 0.86\% & 0.86\%  \\
4.32\% & 0.76\% & 0.76\%  \\
4.57\% & 0.67\% & 0.67\%  \\
5.07\% & 0.52\% & 0.51\%  \\
6.07\% & 0.30\% & 0.30\%  \\
\hline
\end{tabular}
\end{table} 
\begin{figure}[htbp]
\centering
\includegraphics[width=1\columnwidth]{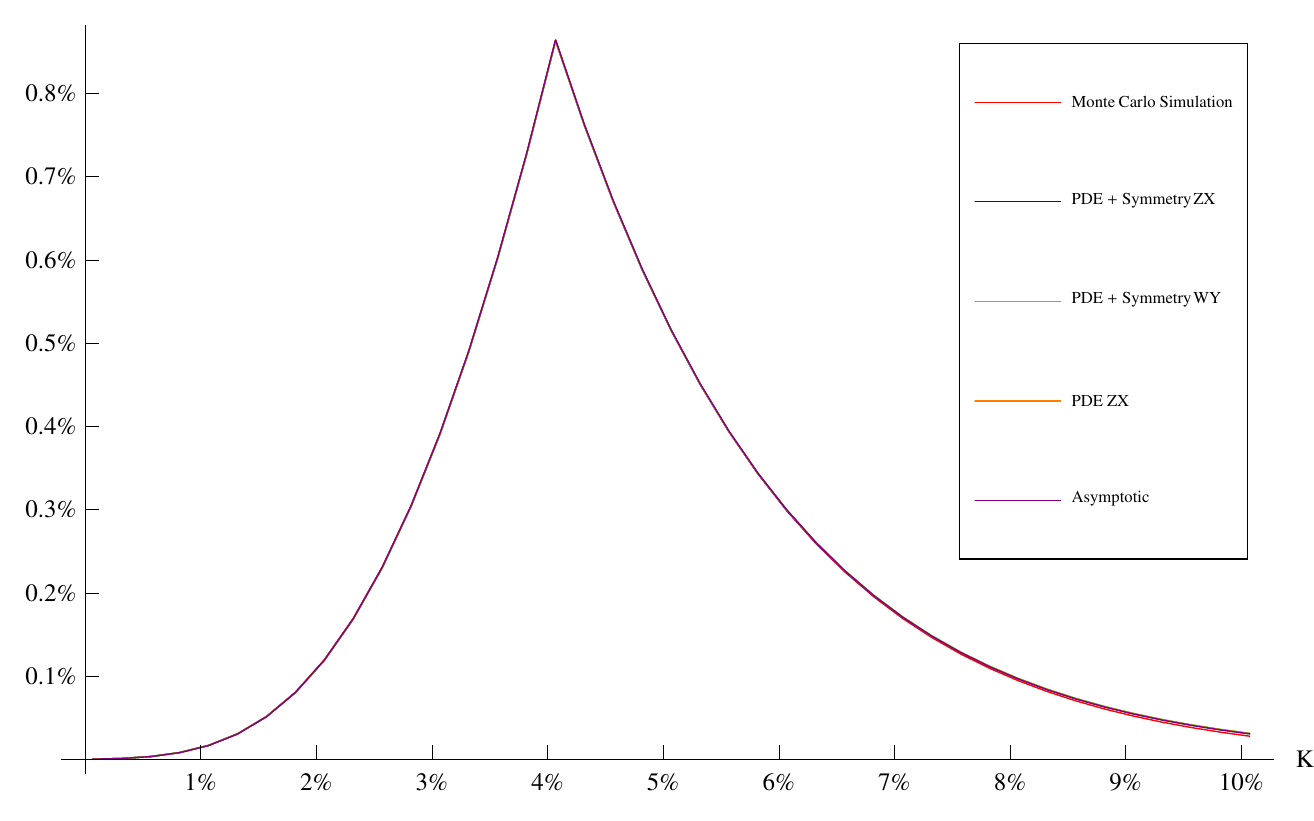}
\caption{\label{fig:comparison.20131029.5Y5Y}Pricing method comparison for USD 5Y5Y swaptions on October 29th, 2013. 
The time values of swaptions expressed in the unit of 
forward swap annuity were computed using various pricing methods.
The SABR model parameters in \cref{tab:SABRparameters.20131029} were used.}
\end{figure}
\begin{table}[htbp]
\centering
\caption{\label{tab:comparison.20131029.5Y5Y}Pricing method comparison for USD 5Y5Y swaptions on October 29th, 2013. 
The time values of swaptions expressed in the unit of 
forward swap annuity were computed using various pricing methods.
The SABR model parameters in \cref{tab:SABRparameters.20131029} were used.}
\resizebox{\columnwidth}{!}{%
\begin{tabular}{|*{6}{c|}}
\hline 
Strike & Monte Carlo simulation  & PDE + Symmetry ZX & PDE + Symmetry WY & PDE ZX & Asymptotic\\
\hline 
0.07\% & 0.00\% & 0.00\%  & 0.00\% & 0.00\% & 0.00\% \\
0.57\% & 0.00\% & 0.00\%  & 0.00\% & 0.00\% & 0.00\% \\
1.07\% & 0.02\% & 0.02\%  & 0.02\% & 0.02\% & 0.02\% \\
1.57\% & 0.05\% & 0.05\%  & 0.05\% & 0.05\% & 0.05\% \\
2.07\% & 0.12\% & 0.12\%  & 0.12\% & 0.12\% & 0.12\% \\
2.57\% & 0.23\% & 0.23\%  & 0.23\% & 0.23\% & 0.23\% \\
3.07\% & 0.39\% & 0.39\%  & 0.39\% & 0.39\% & 0.39\% \\
3.57\% & 0.60\% & 0.60\%  & 0.60\% & 0.60\% & 0.60\% \\
3.82\% & 0.73\% & 0.73\%  & 0.73\% & 0.73\% & 0.73\% \\
4.07\% & 0.86\% & 0.86\%  & 0.86\% & 0.86\% & 0.86\% \\
4.32\% & 0.76\% & 0.76\%  & 0.76\% & 0.76\% & 0.76\% \\
4.82\% & 0.59\% & 0.59\%  & 0.59\% & 0.59\% & 0.59\% \\
5.32\% & 0.45\% & 0.45\%  & 0.45\% & 0.45\% & 0.45\% \\
5.82\% & 0.34\% & 0.34\%  & 0.34\% & 0.34\% & 0.34\% \\
6.32\% & 0.26\% & 0.26\%  & 0.26\% & 0.26\% & 0.26\% \\
6.82\% & 0.20\% & 0.20\%  & 0.20\% & 0.20\% & 0.20\% \\
7.32\% & 0.15\% & 0.15\%  & 0.15\% & 0.15\% & 0.15\% \\
7.82\% & 0.11\% & 0.11\%  & 0.11\% & 0.11\% & 0.11\% \\
8.32\% & 0.08\% & 0.08\%  & 0.08\% & 0.08\% & 0.08\% \\
8.82\% & 0.06\% & 0.06\%  & 0.06\% & 0.06\% & 0.06\% \\
9.32\% & 0.04\% & 0.05\%  & 0.05\% & 0.05\% & 0.05\% \\
9.82\% & 0.03\% & 0.04\%  & 0.04\% & 0.04\% & 0.04\% \\
10.32\% & 0.02\% & 0.03\%  & 0.03\% & 0.03\% & 0.03\% \\
\hline
\end{tabular}%
}
\end{table}
\clearpage
\begin{figure}[htbp]
\centering
\includegraphics[width=1\columnwidth]{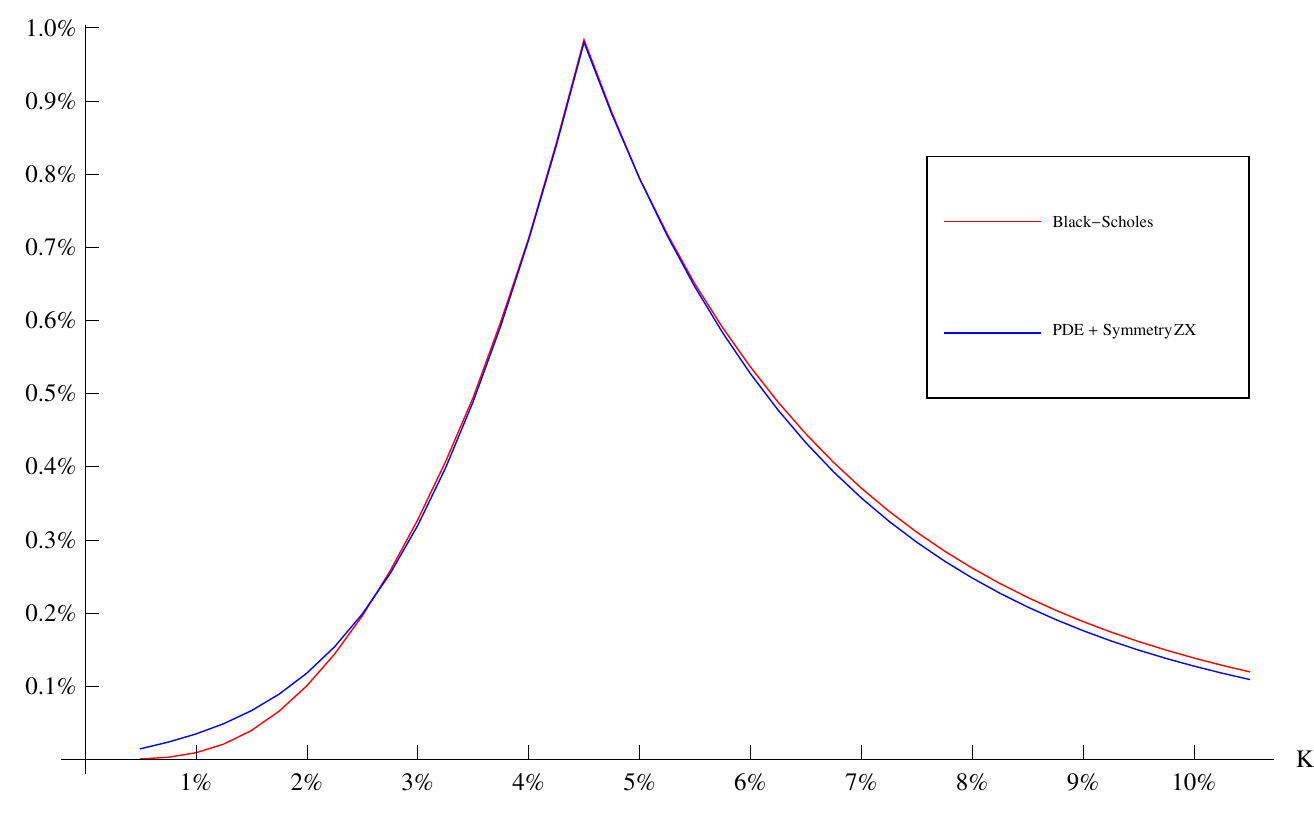}
\caption{\label{fig:USDCalibration.20131029.10Y10Y}Calibration results for USD 10Y10Y swaptions on October 29th, 2013.
The time values of swaptions expressed in the unit of 
forward swap annuity were computed and graphed
as a function of $K$. 
The red line was obtained by the Black-Scholes with linearly interpolated implied volatilities.
The blue line was computed using the ``PDE + Symmetry'' method with variables $Z_t$ and $X_t$. 
The SABR model parameters in \cref{tab:SABRparameters.20131029} were used.}  
\end{figure}
\begin{table}[htbp]
\centering
\caption{\label{tab:USDCalibration.20131029.10Y10Y}Calibration results for USD 10Y10Y swaptions on October 29th, 2013.
The time values of swaptions expressed in the unit of 
forward swap annuity were computed.
For Black-Scholes, the linearly interpolated implied volatilities were used.
For SABR, the ``PDE + Symmetry'' method with variables $Z_t$ and $X_t$ was used.
The SABR model parameters in \cref{tab:SABRparameters.20131029} were used.}
\begin{tabular}{|c|c|c|}
\hline 
Strike & Black-Scholes & SABR \\
\hline
2.50\% & 0.20\% & 0.20\%  \\
3.50\% & 0.49\% & 0.49\%  \\
4.00\% & 0.71\% & 0.71\%  \\
4.25\% & 0.84\% & 0.84\%  \\
4.50\% & 0.98\% & 0.98\%  \\
4.75\% & 0.88\% & 0.88\%  \\
5.00\% & 0.79\% & 0.79\%  \\
5.50\% & 0.65\% & 0.65\%  \\
6.50\% & 0.44\% & 0.43\%  \\
\hline
\end{tabular}
\end{table} 
\begin{figure}[htbp]
\centering
\includegraphics[width=1\columnwidth]{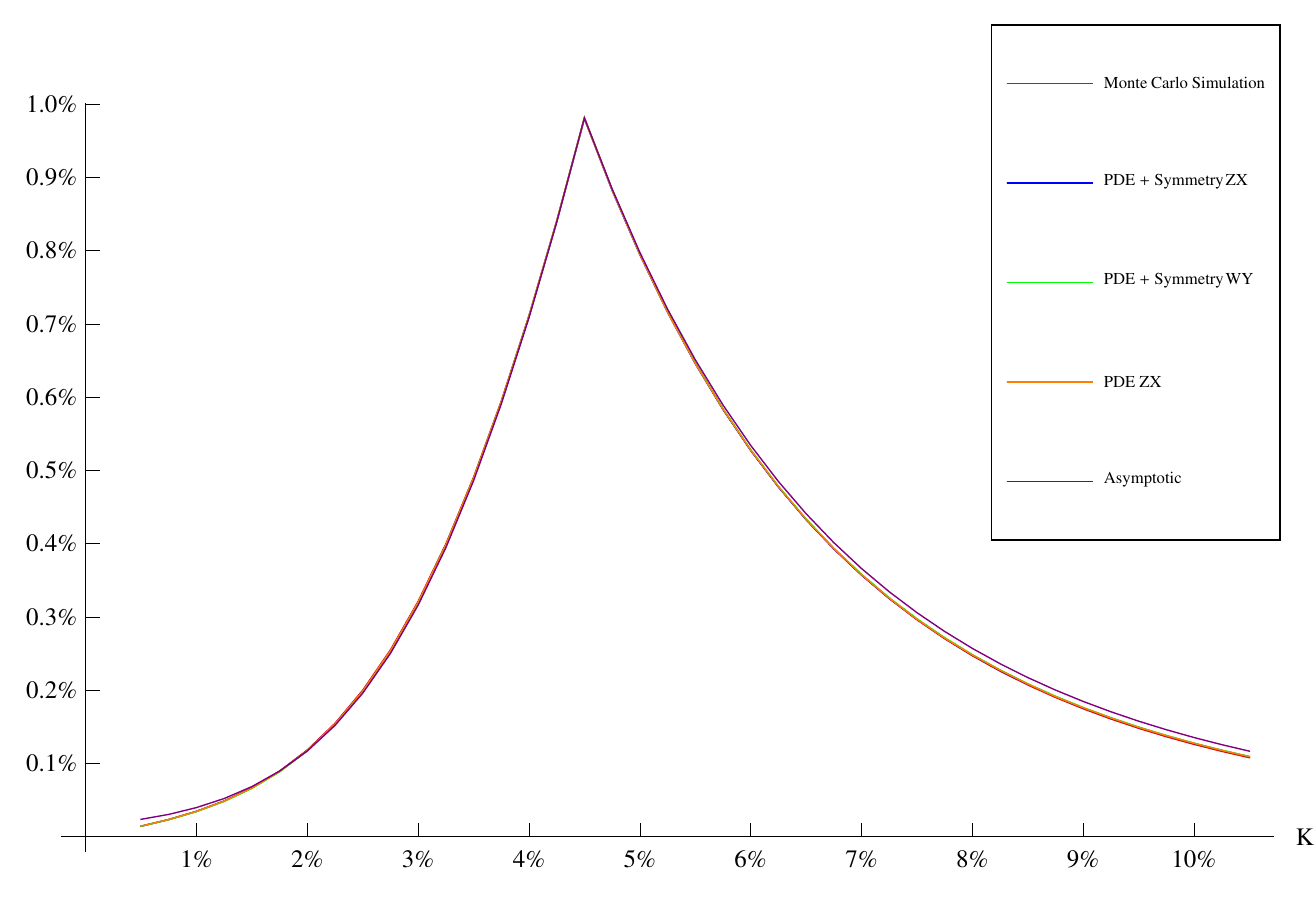}
\caption{\label{fig:comparison.20131029.10Y10Y}Pricing method comparison for USD 10Y10Y swaptions on October 29th, 2013. 
The time values of swaptions expressed in the unit of 
forward swap annuity were computed using various pricing methods.
The SABR model parameters in \cref{tab:SABRparameters.20131029} were used.}
\end{figure}
\begin{table}[htbp]
\centering
\caption{\label{tab:comparison.20131029.10Y10Y}Pricing method comparison for USD 10Y10Y swaptions on October 29th, 2013. 
The time values of swaptions expressed in the unit of 
forward swap annuity were computed using various pricing methods.
The SABR model parameters in \cref{tab:SABRparameters.20131029} were used.}
\resizebox{\columnwidth}{!}{%
\begin{tabular}{|*{6}{c|}}
\hline 
Strike & Monte Carlo simulation  & PDE + Symmetry ZX & PDE + Symmetry WY & PDE ZX & Asymptotic\\
\hline 
0.50\% & 0.01\% & 0.01\%  & 0.01\% & 0.01\% & 0.02\% \\
1.00\% & 0.03\% & 0.03\%  & 0.03\% & 0.03\% & 0.04\% \\
1.50\% & 0.07\% & 0.07\%  & 0.07\% & 0.07\% & 0.07\% \\
2.00\% & 0.12\% & 0.12\%  & 0.12\% & 0.12\% & 0.12\% \\
2.50\% & 0.20\% & 0.20\%  & 0.20\% & 0.20\% & 0.20\% \\
3.00\% & 0.32\% & 0.32\%  & 0.32\% & 0.32\% & 0.32\% \\
3.50\% & 0.49\% & 0.49\%  & 0.49\% & 0.49\% & 0.49\% \\
4.00\% & 0.71\% & 0.71\%  & 0.71\% & 0.71\% & 0.71\% \\
4.25\% & 0.84\% & 0.84\%  & 0.84\% & 0.84\% & 0.84\% \\
4.50\% & 0.98\% & 0.98\%  & 0.98\% & 0.98\% & 0.98\% \\
4.75\% & 0.88\% & 0.88\%  & 0.88\% & 0.88\% & 0.88\% \\
5.25\% & 0.72\% & 0.72\%  & 0.72\% & 0.72\% & 0.72\% \\
5.75\% & 0.58\% & 0.58\%  & 0.58\% & 0.58\% & 0.59\% \\
6.25\% & 0.48\% & 0.48\%  & 0.48\% & 0.48\% & 0.48\% \\
6.75\% & 0.39\% & 0.39\%  & 0.39\% & 0.39\% & 0.40\% \\
7.25\% & 0.32\% & 0.33\%  & 0.33\% & 0.33\% & 0.33\% \\
7.75\% & 0.27\% & 0.27\%  & 0.27\% & 0.27\% & 0.28\% \\
8.25\% & 0.23\% & 0.23\%  & 0.23\% & 0.23\% & 0.24\% \\
8.75\% & 0.19\% & 0.19\%  & 0.19\% & 0.19\% & 0.20\% \\
9.25\% & 0.16\% & 0.16\%  & 0.16\% & 0.16\% & 0.17\% \\
9.75\% & 0.14\% & 0.14\%  & 0.14\% & 0.14\% & 0.15\% \\
10.25\% & 0.12\% & 0.12\%  & 0.12\% & 0.12\% & 0.13\% \\
10.75\% & 0.10\% & 0.10\%  & 0.10\% & 0.10\% & 0.11\% \\
\hline
\end{tabular}%
}
\end{table}
\clearpage
\begin{figure}[htbp]
\centering
\includegraphics[width=1\columnwidth]{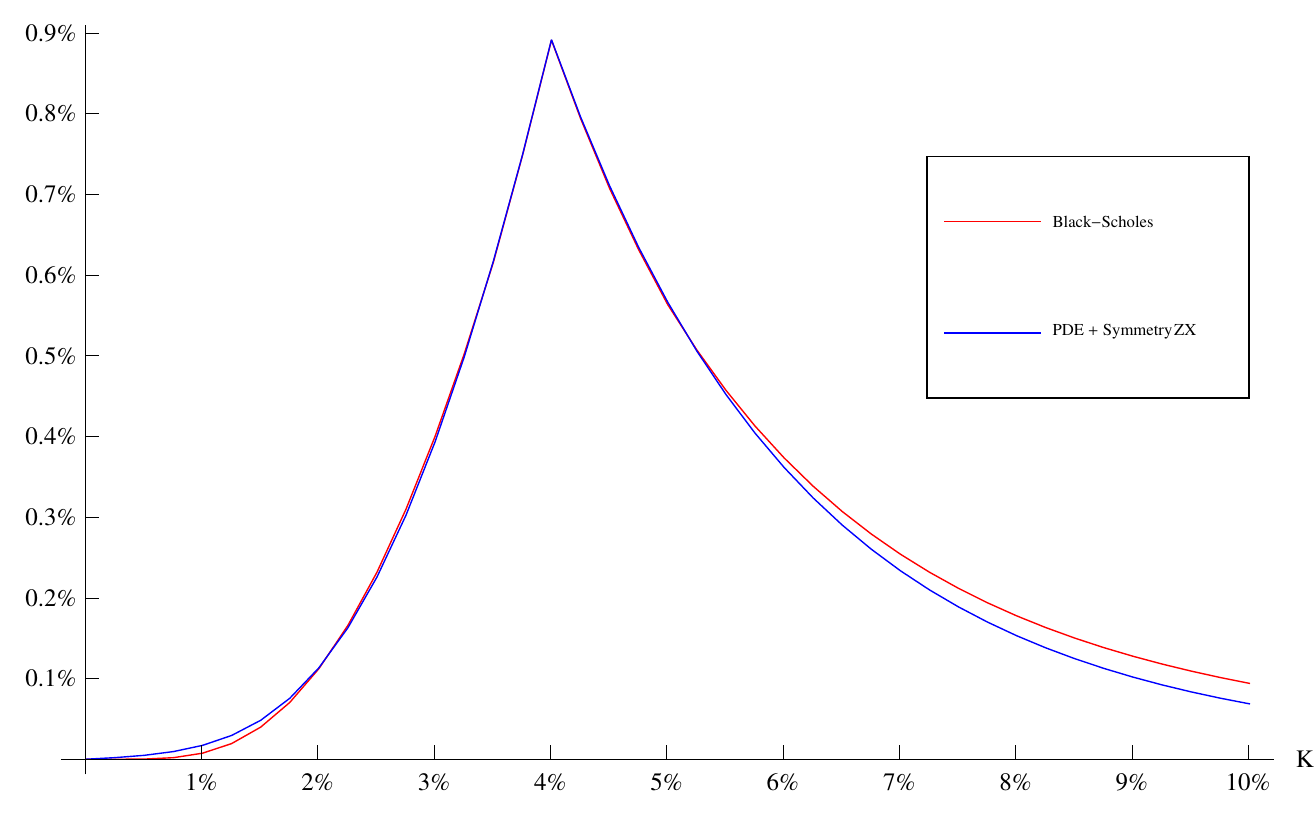}
\caption{\label{fig:USDCalibration.20131029.20Y20Y}Calibration results for USD 20Y20Y swaptions on October 29th, 2013.
The time values of swaptions expressed in the unit of 
forward swap annuity were computed and graphed
as a function of $K$. 
The red line was obtained by the Black-Scholes with linearly interpolated implied volatilities.
The blue line was computed using the ``PDE + Symmetry'' method with variables $Z_t$ and $X_t$. 
The SABR model parameters in \cref{tab:SABRparameters.20131029} were used.}  
\end{figure}
\begin{table}[htbp]
\centering
\caption{\label{tab:USDCalibration.20131029.20Y20Y}Calibration results for USD 20Y20Y swaptions on October 29th, 2013.
The time values of swaptions expressed in the unit of 
forward swap annuity were computed.
For Black-Scholes, the linearly interpolated implied volatilities were used.
For SABR, the ``PDE + Symmetry'' method with variables $Z_t$ and $X_t$ was used.
The SABR model parameters in \cref{tab:SABRparameters.20131029} were used.}
\begin{tabular}{|c|c|c|}
\hline 
Strike & Black-Scholes & SABR \\
\hline
2.01\% & 0.11\% & 0.11\%  \\
3.01\% & 0.40\% & 0.39\%  \\
3.51\% & 0.62\% & 0.62\%  \\
3.76\% & 0.75\% & 0.75\%  \\
4.01\% & 0.89\% & 0.89\%  \\
4.26\% & 0.79\% & 0.80\%  \\
4.51\% & 0.71\% & 0.71\%  \\
5.01\% & 0.56\% & 0.57\%  \\
6.01\% & 0.37\% & 0.36\%  \\
\hline
\end{tabular}
\end{table} 
\begin{figure}[htbp]
\centering
\includegraphics[width=1\columnwidth]{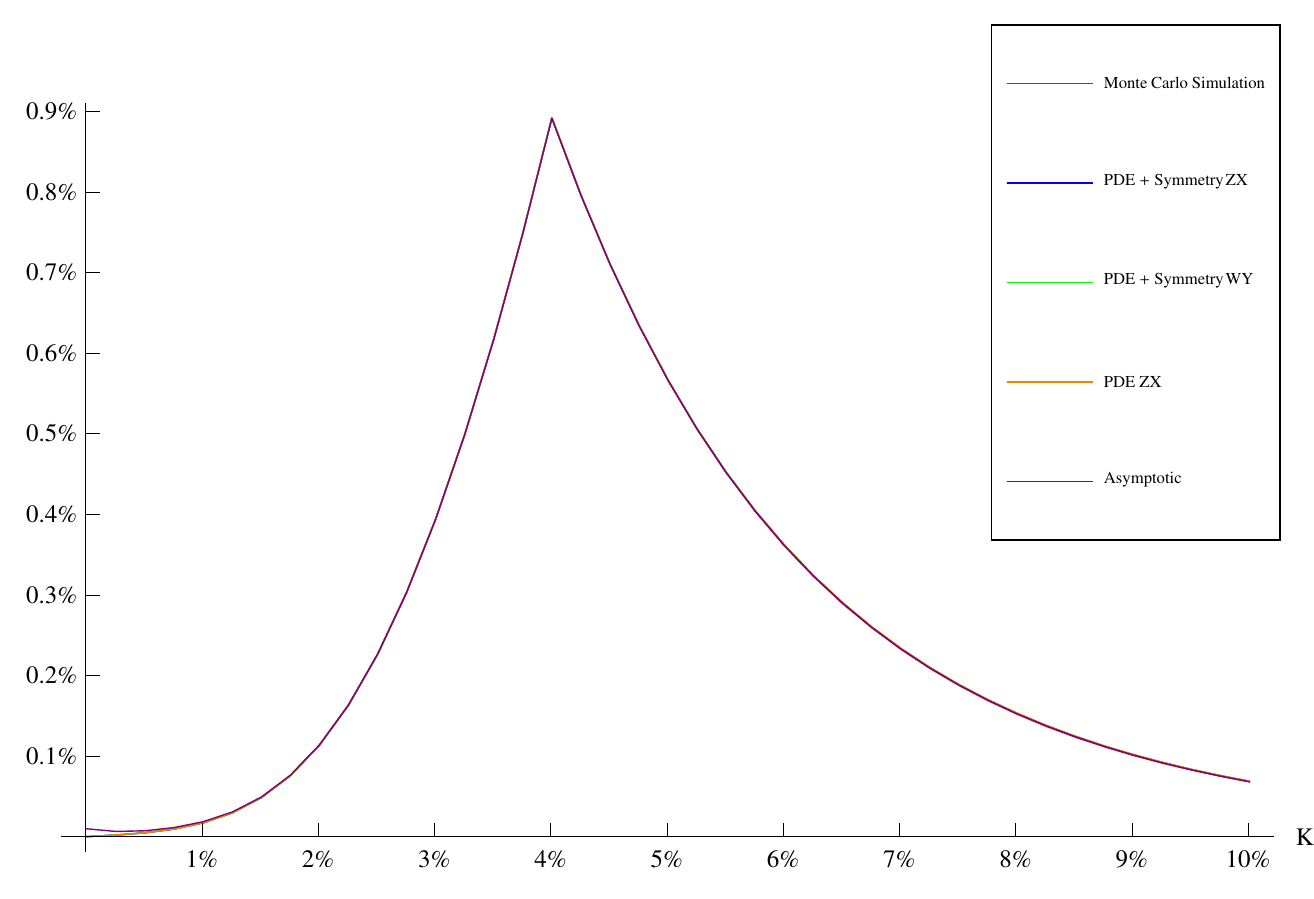}
\caption{\label{fig:comparison.20131029.20Y20Y}Pricing method comparison for USD 20Y20Y swaptions on October 29th, 2013. 
The time values of swaptions expressed in the unit of 
forward swap annuity were computed using various pricing methods.
The SABR model parameters in \cref{tab:SABRparameters.20131029} were used.}
\end{figure}
\begin{table}[htbp]
\centering
\caption{\label{tab:comparison.20131029.20Y20Y}Pricing method comparison for USD 20Y20Y swaptions on October 29th, 2013. 
The time values of swaptions expressed in the unit of 
forward swap annuity were computed using various pricing methods.
The SABR model parameters in \cref{tab:SABRparameters.20131029} were used.}
\resizebox{\columnwidth}{!}{%
\begin{tabular}{|*{6}{c|}}
\hline 
Strike & Monte Carlo simulation  & PDE + Symmetry ZX & PDE + Symmetry WY & PDE ZX & Asymptotic\\
\hline 
0.01\% & 0.00\% & 0.00\%  & 0.00\% & 0.00\% & 0.01\% \\
0.51\% & 0.00\% & 0.00\%  & 0.00\% & 0.00\% & 0.01\% \\
1.01\% & 0.02\% & 0.02\%  & 0.02\% & 0.02\% & 0.02\% \\
1.51\% & 0.05\% & 0.05\%  & 0.05\% & 0.05\% & 0.05\% \\
2.01\% & 0.11\% & 0.11\%  & 0.11\% & 0.11\% & 0.11\% \\
2.51\% & 0.23\% & 0.23\%  & 0.23\% & 0.23\% & 0.23\% \\
3.01\% & 0.39\% & 0.39\%  & 0.39\% & 0.39\% & 0.39\% \\
3.51\% & 0.62\% & 0.62\%  & 0.62\% & 0.62\% & 0.62\% \\
3.76\% & 0.75\% & 0.75\%  & 0.75\% & 0.75\% & 0.75\% \\
4.01\% & 0.89\% & 0.89\%  & 0.89\% & 0.89\% & 0.89\% \\
4.26\% & 0.80\% & 0.80\%  & 0.80\% & 0.80\% & 0.80\% \\
4.76\% & 0.63\% & 0.63\%  & 0.63\% & 0.63\% & 0.63\% \\
5.26\% & 0.51\% & 0.51\%  & 0.51\% & 0.51\% & 0.51\% \\
5.76\% & 0.40\% & 0.40\%  & 0.40\% & 0.40\% & 0.40\% \\
6.26\% & 0.32\% & 0.32\%  & 0.32\% & 0.32\% & 0.32\% \\
6.76\% & 0.26\% & 0.26\%  & 0.26\% & 0.26\% & 0.26\% \\
7.26\% & 0.21\% & 0.21\%  & 0.21\% & 0.21\% & 0.21\% \\
7.76\% & 0.17\% & 0.17\%  & 0.17\% & 0.17\% & 0.17\% \\
8.26\% & 0.14\% & 0.14\%  & 0.14\% & 0.14\% & 0.14\% \\
8.76\% & 0.11\% & 0.11\%  & 0.11\% & 0.11\% & 0.11\% \\
9.26\% & 0.09\% & 0.09\%  & 0.09\% & 0.09\% & 0.09\% \\
9.76\% & 0.08\% & 0.08\%  & 0.08\% & 0.08\% & 0.08\% \\
10.26\% & 0.06\% & 0.06\%  & 0.06\% & 0.06\% & 0.06\% \\
\hline
\end{tabular}%
}
\end{table}
\clearpage
\subsection{Final Comments}
After all tests in this section, we conclude that our ``PDE + Symmetry'' method, especially with 
variables $Z_t$ and $X_t$, produces
accurate swaption prices. This method requires significantly less computational effort than the conventional method of solving the SABR PDE.
We finish this article with a brief comment on performance. All numerical routines have a trade-off between
performance and accuracy. We can always speed up the routine at the expense of accuracy.  
That being said, we believe that the performance of
our implementation is good enough for commercial applications.
Our routine was implemented in Scala(version 2.10.2) and was tested on a 6-core hyper-threading enabled 
Intel\textsuperscript\textregistered  Xeon\textsuperscript\textregistered  processor E5-1650 running at 3.20GHz.
With reasonable accuracy(error $\le$ 1bp of the forward swap annuity), 10 year swaptions can be priced in as little as $25$ milliseconds with
$N_T = 240, N_U = 50, N_V = 256$. This should be fast enough to allow real time calibration and pricing.

\clearpage
\bibliographystyle{plainnat}
\bibliography{bibfile}
\newpage
\fancyhf{}
\renewcommand{\headrulewidth}{0pt} % remove lines as well
\cfoot{\thepage}  
 
\section*{Disclaimer}

\scriptsize The information herein has been prepared solely for informational purposes and is not an offer to buy or sell or a solicitation of an offer to buy or
sell any security or instrument or to participate in any trading strategy.  Any such offer would be made only after a prospective participant had completed its own 
independent investigation of the securities, instruments or transactions and received all information it required to make its own investment decision, including, 
where applicable, a review of any offering circular or memorandum describing such security or instrument, which would contain material information not contained 
herein and to which prospective participants are referred.  No representation or warranty can be given with respect to the accuracy or completeness of the information 
herein, or that any future offer of securities, instruments or transactions will conform to the terms hereof.  Morgan Stanley and its affiliates disclaim any and all 
liability relating to this information.  Morgan Stanley, its affiliates and others associated with it may have positions in, and may effect transactions in, 
securities and instruments of issuers mentioned herein and may also perform or seek to perform investment banking services for the issuers of such securities and instruments.

The information herein may contain general, summary discussions of certain tax, regulatory, accounting and/or legal issues relevant to the proposed transaction.  
Any such discussion is necessarily generic and may not be applicable to, or complete for, any particular recipient's specific facts and circumstances. 
Morgan Stanley is not offering and does not purport to offer tax, regulatory, accounting or legal advice and this information should not be relied upon as such.  
Prior to entering into any proposed transaction, recipients should determine, in consultation with their own legal, tax, regulatory and accounting advisors, 
the economic risks and merits, as well as the legal, tax, regulatory and accounting characteristics and consequences, of the transaction.

Notwithstanding any other express or implied agreement, arrangement, or understanding to the contrary, Morgan Stanley and each recipient hereof are deemed to agree 
that both Morgan Stanley and such recipient (and their respective employees, representatives, and other agents) may disclose to any and all persons, without limitation of 
any kind, the U.S. federal income tax treatment of the securities, instruments or transactions described herein and any fact relating to the structure of the securities, 
instruments or transactions that may be relevant to understanding such tax treatment, and all materials of any kind (including opinions or other tax analyses) 
that are provided to such person relating to such tax treatment and tax structure, except to the extent confidentiality is reasonably necessary to comply with 
securities laws (including, where applicable, confidentiality regarding the identity of an issuer of securities or its affiliates, agents and advisors).

The projections or other estimates in these materials (if any), including estimates of returns or performance, are forward-looking statements based upon 
certain assumptions and are preliminary in nature.  Any assumptions used in any such projection or estimate that were provided by a recipient are noted herein.  
Actual results are difficult to predict and may depend upon events outside the issuer's or Morgan Stanley's control.  
Actual events may differ from those assumed and changes to any assumptions may have a material impact on any projections or estimates.  
Other events not taken into account may occur and may significantly affect the analysis.  
Certain assumptions may have been made for modeling purposes only to simplify the presentation and/or calculation of any projections or estimates, 
and Morgan Stanley does not represent that any such assumptions will reflect actual future events.  
Accordingly, there can be no assurance that estimated returns or projections will be realized or that actual returns or performance results will not be materially 
different than those estimated herein.  Any such estimated returns and projections should be viewed as hypothetical.  
Recipients should conduct their own analysis, using such assumptions as they deem appropriate, and should fully consider other available information in making a decision 
regarding these securities, instruments or transactions.  
Past performance is not necessarily indicative of future results.  
Price and availability are subject to change without notice.
The offer or sale of securities, instruments or transactions may be restricted by law.  
Additionally, transfers of any such securities, instruments or transactions may be limited by law or the terms thereof.  
Unless specifically noted herein, neither Morgan Stanley nor any issuer of securities or instruments has taken or will take any action in any jurisdiction that would permit 
a public offering of securities or instruments, or possession or distribution of any offering material in relation thereto, in any country or jurisdiction where action for 
such purpose is required.  
Recipients are required to inform themselves of and comply with any legal or contractual restrictions on their purchase, holding, sale, exercise of rights or performance of 
obligations under any transaction.  
Morgan Stanley does not undertake or have any responsibility to notify you of any changes to the attached information. 

With respect to any recipient in the U.K., the information herein has been issued by Morgan Stanley \& Co. International Limited, regulated by the U.K. Financial Services Authority.  
THIS COMMUNICATION IS DIRECTED IN THE UK TO THOSE PERSONS WHO ARE MARKET COUNTERPARTIES OR 
INTERMEDIATE CUSTOMERS (AS DEFINED IN THE UK FINANCIAL SERVICES AUTHORITY'S RULES).

ADDITIONAL INFORMATION IS AVAILABLE UPON REQUEST.

\end{document}